\documentclass{article}

\usepackage[fleqn]{amsmath}
\usepackage{cite,latexsym,amssymb,tabularx,supertabular,bm,pbox}
\usepackage{bbm}
\numberwithin{equation}{section}

\usepackage{graphicx}
\newcommand{\diagram}[2][{}]{\pbox{\textwidth}{\includegraphics[#1]{{#2}}}}
\usepackage{xcolor}
\usepackage{framed}
\definecolor{shadecolor}{gray}{0.95}
\usepackage{float}
\usepackage{afterpage}
\usepackage[latin1]{inputenc}
\usepackage[rightcaption]{sidecap}
\usepackage[font=small,format=plain,labelfont=bf,up,textfont=sf,up]{caption}

\usepackage{subfig}
\usepackage{wrapfig}

\usepackage[pdftex,backref,colorlinks]{hyperref}
\usepackage[figure]{hypcap}
\usepackage[colorinlistoftodos,textsize=footnotesize]{todonotes}
\usepackage{setspace}
\usepackage[letterpaper]{geometry}
\newcommand\T{\rule{0pt}{2.6ex}}
\newcommand\B{\rule[-1.2ex]{0pt}{0pt}}

\newcommand{\tr}{{\rm tr}}

\newcommand{\xbj}{x_{\text{bj}}}
\newcommand{\xp}{{x_{\mathbb{P}}}}
\newcommand{\nat}{\text{ln}}
\newcommand{\gev}{\text{GeV}}
\newcommand{\mev}{\text{MeV}}
\newcommand{\sep}{s_{ep}}
\newcommand{\sgp}{s_{\gamma^{*}p}}

\newcommand{\gasp}{\gamma^{*}p}
\newcommand{\qqp}{q\bar {q}}
\newcommand{\vrr}{{\bm r}}
\newcommand{\vdl}{{\bm r}_1}
\newcommand{\vdr}{{\bm r}_2}

\def\lsi{\raise0.3ex\hbox{$<$\kern-0.75em\raise-1.1ex\hbox{$\sim$}}}
\def\gsi{\raise0.3ex\hbox{$>$\kern-0.75em\raise-1.1ex\hbox{$\sim$}}}

\newcommand{\msbar}{{\overline{{\rm MS}}}}

\newcommand{\cut}{\hspace{1.5mm}}

\begin{document}

\noindent{\Large\sf\textbf{HERA-data in the light of small $x$
    evolution with state of the art NLO input}}

\vspace{0.5cm} {\large { Janne Kuokkanen$^1$ , Kari Rummukainen$^2$ ,
    and Heribert Weigert$^3$ } }
\\\vspace{0.2cm}\\
{\footnotesize {$^1$ Department of Physical Sciences, University of Oulu,
    P.O. Box 3000, FI-90014 Oulu,
    Finland} \\
  {$^2$ Department of Physics and Helsinki Institute of Physics,
    P.O.Box 64, FI-00014 University of Helsinki, Finland
  } \\
  {$^3$ University of Cape Town,
    Dept. of Physics, Private Bag X3, Rondebosch 7701, South Africa} \\

}

\vspace{.2cm}
\noindent\begin{center}
\begin{minipage}{.92\textwidth}
  {\sf Both total and diffractive cross sections from HERA are
    successfully confronted with JIMWLK evolution equations in the
    asymptotic pseudo-scaling region. We present a consistent,
    simultaneous description of both types of cross sections that
    includes NLO corrections in the form of running coupling and
    energy conservation corrections. The inclusion of energy
    conservation corrections allows to match all available data with
    $\xbj \le .02$ i.e. up to $Q^2\le 1200\ \gev^2$. We discuss the
    effects of quark masses including charm, contrast asymptotic and
    pre-asymptotic fit strategies, and survey non-perturbative
    uncertainties related to impact parameter dependence.  }
\end{minipage}
\end{center}
\vspace{.5cm}

\textsf{HIP-2011-23/TH} 
\vspace{.5cm}

\section{Introduction}
\label{sec:introduction}

Much of the abundant particle production in modern collider
experiments at high energies is triggered by gluon channels, thus
imprinting the features of gluon phase space on many observables. This
is the basis of the importance of the Color Glass Condensate
(CGC)~\cite{Gribov:1984tu, Mueller:1986wy, Mueller:1994rr,
  Mueller:1994jq, Mueller:1995gb, McLerran:1993ka, McLerran:1994ni,
  McLerran:1994ka, McLerran:1994vd, Kovchegov:1996ty,
  Kovchegov:1997pc, Jalilian-Marian:1997xn, Jalilian-Marian:1997jx,
  Jalilian-Marian:1997gr, Jalilian-Marian:1997dw,
  Jalilian-Marian:1998cb, Kovner:2000pt, Weigert:2000gi, Iancu:2000hn,
  Ferreiro:2001qy, Kovchegov:1999yj, Kovchegov:1999ua,
  Balitsky:1996ub, Balitsky:1997mk, Balitsky:1998ya, Iancu:2003xm,
  Weigert:2005us, Jalilian-Marian:2005jf,Gelis:2010nm} for virtually
all current collider experiments be they designed to answer particle-
or and heavy-ion-physics questions.

The characteristic feature of enhanced gluon emission into the final
state at high energies is the emergence of an energy dependent
transverse correlation length $R_s(x)$. Its associated conjugate
momentum scale $Q_s(x)\sim 1/R_s(x)$ signifies the onset of gluon
saturation, hence the name saturation scale.  As gluon numbers rise
with energy, the correlation length of the dense gluon cloud shrinks,
the saturation scale increases.

The mere existence of such an energy driven scale has led to a large
body of phenomenological literature, often based on identifying
quantities that can be expected to crucially depend on the saturation
scale $Q_s$ to gain insight into the energy dependence of some
observable by applying a scaling argument.  The origin of this idea
predates the observation that perturbative QCD allows us to predict
$Q_s$-scaling in the context of the JIMWLK equation (or in an
independent scattering approximation the BK
equation~\cite{Balitsky:1996ub,
  Balitsky:1997mk,Balitsky:1998ya,Kovchegov:1999yj, Kovchegov:1999ua})
and has provided us with the discovery of what was called geometric
scaling in HERA data~\cite{Golec-Biernat:1998js, Golec-Biernat:1999qd,
  Stasto:2000er}: plotting the $e p$ cross sections measured at HERA
not as a function of rapidity $Y=\ln(1/x)$ and momentum transfer $Q^2$
independently, but instead as a function of the scaling variable
$Q^2/Q_s^2(x)$ reveals beautiful scaling features of the data for
$x\le 10^{-2}$ that extend even to diffractive measurements. This
scaling subsumes the complete energy dependence of the data at $x\le
10^{-2}$ in a single energy dependent function, $Q_s(x)$.

The standard approach to cross correlate HERA data in the $Q^2$
direction using Dokshitzer-Gribov-Lipatov-Altarelli-Parisi (DGLAP)
evolution equations reproduces this scaling feature only accidentally,
in the sense that they match data which themselves exhibit
scaling. There is no {\em intrinsic} reason why such a scaling feature
should emerge {\em within} the domain of validity of the linear DGLAP
formalism\footnote{The derivation of scaling ``within'' DGLAP
  in~\cite{Caola:2008xr} skirts the region of applicability of the
  argument: the scaling solutions shown there push into the region of
  large gluon densities where nonlinear/higher twist corrections are
  bound to become important.}. By contrast, in the CGC context,
scaling is a natural byproduct of nonlinear features of gluon emission
and saturation, which adds a particular interest to a comparison of
CGC results with HERA data.

The derivation of the JIMWLK evolution equation relies on the
prerequisite that there exists a frame in which the gluon field of the
target becomes strong, while the gluon field of the projectile is
weak. The resummation of the large target field then induces
nonlinearities that capture the effects of gluon saturation and,
through evolution, driven by perturbative gluon emission in the weak
field projectile, induce saturation and the appearance of a saturation
scale $Q_s(x)$ (its inverse $R_s\sim 1/Q_s$ has the interpretation of
a transverse correlation length in the dense gluon cloud).

While the mere presence of such an energy dependent scale will impose
its mark on any observable, strict mathematical scaling, i.e. the
notion that \emph{all} the cross sections and correlators of the
theory share an energy dependence that can solely be expressed in term
of that of the saturation scale $Q_s(x)$ is too naive an expectation:
It is more natural that different n-point functions have their own
associated scales unless tied together by JIMWLK evolution by
belonging to the same Balitsky-hierarchy for which a strong
factorization feature holds that allows to express these generically
independent correlators to be expressed in terms of a single one, at
least to good approximation. The scaling observed in the HERA total
cross section should be interpreted as a signal for such a correlator
factorization and the possibility to indeed truncate the Balitsky
hierarchy of the dipole amplitude, the correlator dominating the total
cross section.

Since the same correlators dominate both total and diffractive cross
sections at HERA, the same $x$-dependent scale should also be in effect
there, a fact established early in the Golec-Biernat--Wüsthoff (GB-W)
model~\cite{Golec-Biernat:1998js, Golec-Biernat:1999qd,
  Stasto:2000er}.

A second caveat arises from the QCD scale anomaly: even for
observables which are predominantly determined by only a single
correlator, true scaling is only to be expected at leading order
(LO). At next to leading order (NLO) exact scaling receives {\em
  small} corrections [which manifest themselves as a slow drift of
correlator shapes] purely induced by the scale anomaly, i.e. the
running of the coupling. We will refer to this regime as the
asymptotic or (pseudo-) scaling regime.

Despite the phenomenological success of models that assume (pseudo-)
scaling of the cross section, it is by no means clear that the data
would show clear evidence of strict or near scaling in HERA data or even for
the HERA total cross section alone.

The question if the scaling observed in the $\gamma^* p$ cross
section~\cite{Breitweg:2000yn, Breitweg:1998dz, Chekanov:2001qu} and
the rapidity gap events~\cite{Derrick:1995wv, Breitweg:1997aa,
  Adloff:1997sc, Breitweg:1998gc} (see~\cite{Chekanov:2008cw} for the
most recent combined update of all ZEUS data) at HERA affects all
correlators, i.e. represents true (pseudo-) scaling, or only signifies
the presence of an intrinsic scale, indicative merely of the presence
of nonlinearities, with scaling on a correlator level only apparent
and potentially limited to a very narrow kinematic range is
surprisingly hard to settle.

Let us emphasize that a survey of the literature presents us with a
very ambiguous picture: All models from GB-W to the BK inspired
parametrization of the dipole cross section by Iancu, Itakura and
Munier~\cite{Iancu:2003ge} (IIM) to the BFKL+saturation boundary model
of Mueller and Triantafyllopoulos~\cite{Mueller:2002zm,
  Triantafyllopoulos:2002nz} (MT) exhibit either strict or
pseudo-scaling of correlators, since scaling in this sense is one of
the main constraining features in the construction of such models.

However, the shapes of these various (pseudo-) scaling solutions are
notably different from each other and, as we will see below (see
Fig.~\ref{fig:rs_vs_lambda}), from the scaling solution imposed by the
evolution equation at NLO: the imposed scaling shapes in these models
do {\em not} resemble the scaling shapes that emerge as solutions of
the evolution equation.

Moreover, even within a CGC framework, strict or near scaling of
correlators does not seem to be required to obtain a successful fit to
the HERA total cross section: This was shown by Albacete, Armesto,
Milhano, and Salgado in~\cite{Albacete:2009fh, Albacete:2009ps}. They
omit all NLO contributions beyond running coupling, but start
evolution with a shape close to the GB-W parametrization. In this
treatment, correlators only approach the scaling shape imposed by the
evolution equation at the far end of the $x$ range covered by the HERA
experiments.

Our own
simulations~\cite{Weigert:2007hk,Weigert:2008zz,Weigert:2009zz} add in
energy conservation corrections to cover NLO corrections beyond the
running coupling contributions and provide an equally convincing fit
in the pseudo-scaling region. These same NLO corrections restrict our
treatment to the region near (pseudo-) scaling due to stability and
self consistency considerations that only emerge once all these NLO
corrections are taken into account (see
Sec.~\ref{sec:scal-behav-self-cons-NLO}).

We will attempt to summarize the status of the theoretical tools
presently available (state of the art are NLO evolution combined with
LO impact factors, for a more in depth discussion see below) and
explore what kind of tension available data pose on our theoretical
analysis by considering both the HERA total cross section and the
rapidity gap events.

As our first set of results emerges from fits to the total cross
section: we will argue that simulations that include NLO corrections
in the form of running coupling corrections and energy conservation
corrections (the most complete set of NLO corrections currently
available) favor fits in the pseudo-scaling region based on fits of
the total cross section. The treatment of quark masses has proven to
be somewhat problematic in earlier fits, which have generically used
constituent like values of around $140\ \mev$ for light quarks and
$1.4\ \gev$ for charm. Conceptually, quark masses are subleading in
the small $x$ limit where factors of the form
$(\alpha_s)^{n+m}(\ln(1/x))^n$ are used to sort contributions by
importance and hence should not pose a serious difficulty. Choosing
current quark masses instead of constituent quark masses we find that
fits are indeed feasible with the quark masses having their largest
effect in the nonperturbative range with $Q^2 < 1\ \gev^2$. We find
that the idea of~\cite{Golec-Biernat:1998js} to address the situation
by replacing $x$ by some $x_{\text{eff}}$ to modify the small $Q^2$
limit can improve the fit, although the specific form introduced
in~\cite{Golec-Biernat:1998js} proves unusable. Such resummations are
by construction nonperturbative and should at this stage be taken to
merely indicate the relevance of nonperturbative input in this range
of phase space.

A second set of insights emerge from a study of rapidity gap events:
Our analysis is based on the fit parameters extracted from total cross
sections and clearly shows us the limits of a fit with incomplete NLO
input, despite the quite satisfactory fit quality. A comparison with
data clearly requires a $q\Bar q g$-component in the impact factor for
which at present we only have a rough substitute which is only
reliable in the large $Q^2$ limit and was already devised long ago in
the context of the GB-W model~\cite{Golec-Biernat:1999qd}.  This
prevents us from using observables beyond the total cross section to
get a closer look at the details of JIMWLK evolution -- this crude
treatment is not suitable for such a precision study. It turns out
that non-perturbative aspects which enter through the impact parameter
dependence adds additional uncertainties as one steps beyond the total
cross section. It affects the relative normalization of individual
Fock-space components as soon as NLO impact factors start to play an
important role in resolving the structure of the cross section.

This leads us to conclude that to address questions such as precision
fits on initial conditions, the reliability of truncations of the full
JIMWLK framework, or the size and nature of subleading
$N_c$-corrections as advocated in~\cite{Marquet:2010cf} with any
definiteness, we need full knowledge of all NLO contributions
including the appropriate impact factors for the observable in
question as well as an improved understanding of the non-perturbative
aspects of the impact parameter dependence.

Any progress in this respect will also improve the utility of HERA
fits as an input for fits to RHIC and LHC data and will be one of the
most important tasks for the near future.
 
The structure of the paper is as follows: We begin by recapitulating
the theoretical ingredients necessary to define the underlying
observables at zeroth order in Sec.~\ref{sec:gener-disc-total}
and~\ref{sec:rapid-gaps-zeroth-order}, putting some emphasis on the
approximations underlying the expressions usually given in the
literature.  We review our present knowledge of the NLO corrections
and how to use truncations of JIMWLK evolution to efficiently
implement them in Sec.~\ref{sec:beyond-zeroth-order}.

Sec.~\ref{sec:lessons-from-total-cross-section} is devoted to the
systematics of a fit to the total cross section in the asymptotic
pseudo-scaling region. We discuss general features of the asymptotic
fit such as evolution speeds, $Q_s$ in HERA phase space and correlator
properties in Sec.~\ref{sec:gen-feat}. This is followed by a thorough
study of the role of the energy conservation correction
(Sec.~\ref{sec:syst-from-fits-light-quarks-energy-cons}), and the
effect of quark masses (Sec.~\ref{sec:quark-masses}).

The fit obtained in Sec.~\ref{sec:lessons-from-total-cross-section} is
then applied to diffractive data in
Sec.~\ref{sec:lessons-from-diffr-data}. We begin our discussion in
Sec.~\ref{sec:nonp-infl-b-dep} with estimates of the non-perturbative
uncertainties induced by our lack of knowledge of impact parameter-
($b$-) dependence of the eikonal correlators in a proton or nuclear
target.

In Sec.~\ref{sec:pre-as-examples} we compare our asymptotic approach
that includes both running coupling corrections and the energy
conservation correction with more conventional fits that only use the
running coupling effect, but leave out the energy conservation
correction and use the pre-asymptotic regime of small $x$
evolution. We show that features of the solutions that are recovered
perturbatively in the asymptotic region must be imprinted at least
partially via the initial condition in the pre-asymptotic approach.

Sec.~\ref{sec:pre-as-examples} collects our main results in attempt to
provide a synthesis.

Several appendices provide a number of generic expressions
(Sec.~\ref{sec:kin-approx}), ancillary results partly indispensable
to reconstruct our numerical simulations (Sec.~\ref{sec:definitions}
through~\ref{sec:tools-diffr-cross}), as well as number of consistency
checks (Sec.~\ref{sec:con-checks}).

\section{Deep inelastic scattering at small \texorpdfstring{$x$}{x}}

\subsection{Total \texorpdfstring{$\gamma^* A$}{gamma* A} cross sections at zeroth order}
\label{sec:gener-disc-total}

There are many phenomenological applications to various physical
processes and differential cross sections that are based on the idea
of obtaining energy dependence from the scaling behavior in terms of
the saturation scale. However, the observable directly addressed by
JIMWLK evolution is simply the total cross section in very asymmetric
collisions such as $e A$ or $p A$ experiments.

\begin{figure}[H]  \centering
\includegraphics[width=\textwidth]{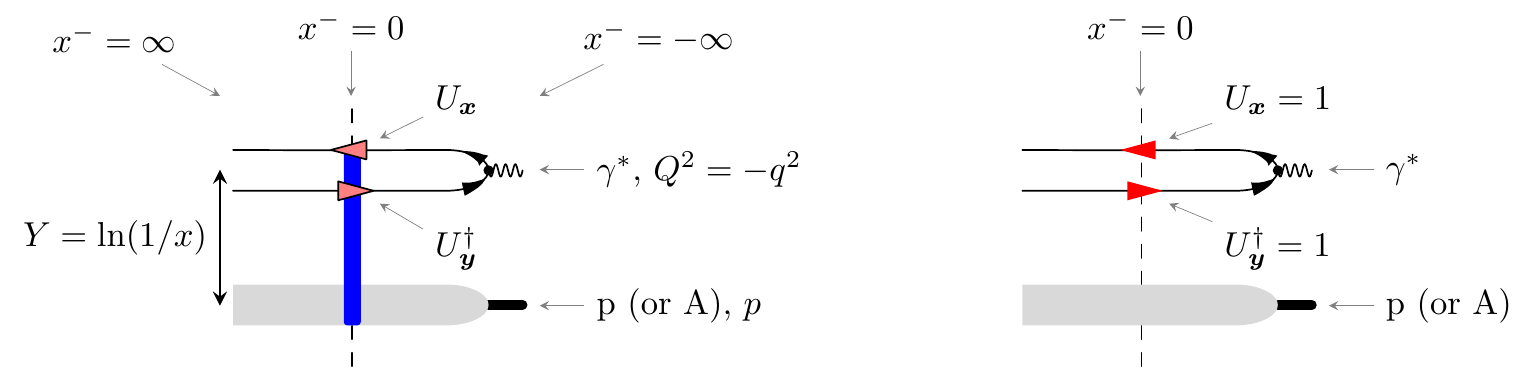}

\caption{Diagrammatic representation of the amplitude for $\gamma^* A$
  scattering at small $x$ at momentum transfer $Q^2=-q^2$. Light cone ``time''
  $x^-$ runs from right to left. The interacting ``out-state'' (left) contain
  nontrivial interaction between projectile and target, which is marked by a
  vertical bar (blue online) at $x^-=0$ that indicates the interaction region
  and markers for the Wilson lines picked up by the projectile constituents.
  The non-interacting ``in-state'' (right) instead has no interactions and
  correspondingly trivial Wilson line factors at $x^-=0$.}
  \label{fig:gamma*-A-diagr-notation}
\end{figure}

The strong asymmetry between projectile and target serves to justify the
notion that the gluon field of the nuclear target with atomic number $A$ can
be thought of as much larger than that of the much simpler electron or proton
projectile. The asymmetry is used to describe the projectile in terms of a
very simple wave function with only a very few (valence) partons, which then
scatter on the large target field with a very large longitudinal
momentum. This justifies the description of the interaction of such a
projectile constituent with the large target field in a no recoil
approximation.  The no recoil approximation fixes the projectile constituent
onto a worldline at a transverse position unaltered during the interaction
with the target. Multiple interactions with the target field then build a
non-Abelian eikonal factor, a path ordered exponential that captures the
interaction of the constituent with the target field. This way the interaction
of a quark in the projectile with the target is represented as an
$SU(N_c)$-valued field in the transverse plane $U_{\bm x}$, an antiquark
analogously enters as a $U_{\bm x}^\dagger$ and a gluon in the projectile
interacts as $\Tilde U_{\bm x}$ (the tilde denotes the adjoint
representation). The graphical notation used to represent this is shown inf
Fig.~\ref{fig:gamma*-A-diagr-notation} for the example of the zeroth order
contribution to the $\gamma^* A$ scattering amplitude.

The corresponding zeroth order total cross section arises as the absolute value
squared of the difference of this interacting diagram with its noninteracting
counterpart\footnote{Note that we keep the dashed vertical line that marks
  $x^-=0$ also in the noninteracting case. We will need this below to
  distinguish where gluon vertices connect with respect to $x^-=0$.}
\begin{subequations}
  \label{eq:zeroeth-order-cross}
\begin{equation}
  \label{eq:zeroeth-order-cross-a}
  \left\vert \diagram[width=1.2cm]{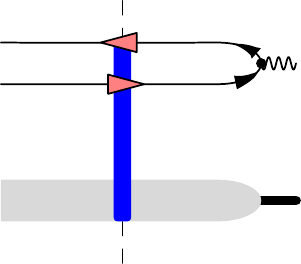}\right. - 
   \left. \diagram[width=1.2cm]{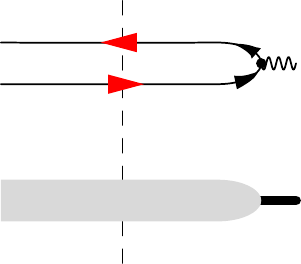} \right\vert^2
=\, 
\left( \diagram[width=1.2cm]{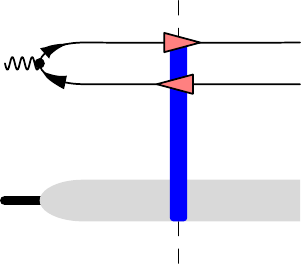} - \diagram[width=1.2cm]{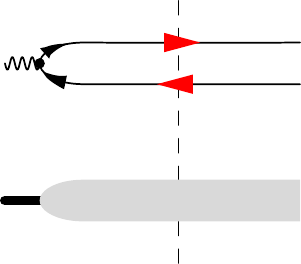} \right)
\left( \diagram[width=1.2cm]{qqbint-R} - \diagram[width=1.2cm]{qqbnoint-R}
\right)
\ .
\end{equation}
If the transverse momentum integrals are unrestricted, they will identify the
transverse coordinates left and right of the cut, so that the the $U$-content
of these diagrams is partially simplified:
\begin{equation}
  \label{eq:zeroeth-order-cross-b}
\eqref{eq:zeroeth-order-cross-a} =\, 
  \underset{\tr U_{\bm y'}U^\dagger_{\bm x'} U_{\bm x}U^\dagger_{\bm
      y}\xrightarrow[\bm y'\to\bm y]{\bm x'\to\bm x}\tr 1}{
    \diagram[width=1.2cm]{qqbint-L}\cut\diagram[width=1.2cm]{qqbint-R}}
  - \underset{\tr U_{\bm x}U^\dagger_{\bm y}}{
    \diagram[width=1.2cm]{qqbnoint-L}\cut\diagram[width=1.2cm]{qqbint-R}}
  - \underset{\tr U_{\bm y}U^\dagger_{\bm x}}{
    \diagram[width=1.2cm]{qqbint-L}\cut\diagram[width=1.2cm]{qqbnoint-R}}
  + \underset{\tr 1\vphantom{U^\dagger_{\bm y}}}{
    \diagram[width=1.2cm]{qqbnoint-L}\cut\diagram[width=1.2cm]{qqbnoint-R}}
\end{equation}
\end{subequations}
Note in particular that, with this assumption, the $U$-factors left and right of the cut in the
$\overline{\text{out}}$-out overlap cancel against each
other. Diagrammatically, we have
\begin{align}
  \label{eq:U-left-right-cancel}
          \diagram[width=1.2cm]{qqbint-L}\cut\diagram[width=1.2cm]{qqbint-R}
      \xrightarrow[\bm y'\to\bm y]{\bm x'\to\bm x}   \diagram[width=1.2cm]{qqbnoint-L}\cut\diagram[width=1.2cm]{qqbnoint-R}
\end{align}
While this is assumed universally in the literature without discussion, one
should be clear that this is an approximation: phase space integrals over the
transverse momenta $\bm k_i$ of the final state quarks are limited by
$W^2 :=s_{\gamma* A}$ (the invariant mass of the $\gamma^* A$-system) both in
$t :=-(\bm k_1+\bm k_2)^2$ and the invariant mass of the produced
pair. Consequently, also the momentum integrals over the transverse momenta
are \emph{not} unrestricted and thus the primed and unprimed coordinates are
not \emph{quite} the same in an exact treatment: even the total cross section
will have a contribution from a four Wilson line operator. At small
$Q_s^2/W^2$, and presently achievable theoretical accuracy (see our discussion
of NLO contributions below and the discussion of coincidence limits given
in~\cite{Kovchegov:2008mk,Marquet:2010cf}), it is fully justified to follow
custom and take the limit shown in~\eqref{eq:U-left-right-cancel} and to
incorporate this four point function only in a coincidence limit in which it
becomes trivial.

With this caveat, it is the $U$-content of the two remaining diagrams, namely
\begin{equation}
\label{eq:hatSqqbardef}
  \Hat S_{\bm{x y}}^{q\Bar q} := \frac{\tr\left(U_{\bm x} U_{\bm
        y}^\dagger\right)}{N_c}
\end{equation}
and its complex conjugate that capture the interaction of the $q\Bar q$ pair
with the target. 

The presence of the target wave function induces an energy dependent
averaging process, that for the applications below will be strictly
real, so that we may (and will) not distinguish $S_Y^{q\Bar q}(\bm
x,\bm y) := \langle \Hat S_{\bm{x y}}^{q\Bar q} \rangle(Y)$ from its
complex conjugate $\langle \Hat S_{\bm{y x}}^{\Bar q q} \rangle(Y)$ in
the following. For the total cross section, the target interaction of
Eq.~(\ref{eq:zeroeth-order-cross}) can be fully summarized by the
dipole operator
\begin{align}
  \label{eq:dipole-amplitude}
  \Hat N^{q\Bar q}_{\bm x \bm y} :=   \frac{1}{N_c} 
  \tr(1-U_{\bm{x}} U^\dagger_{\bm{y}})
\end{align}
(and its complex conjugate), and the average $N^{q\Bar q}_Y(\bm x, \bm
y) := \langle \Hat N^{q\Bar q}_{\bm x \bm y} \rangle(Y)$, the dipole
amplitude.  It is the average over the target wave function, an
operation that involves both perturbative and non-perturbative
information, that proves the most difficult part of this calculation
and induces the energy (or $Y$-) dependence of the cross section. The
tool to extract this energy dependence is the JIMWLK equation, and its
associated framework. At zeroth and leading order (LO) in
$\alpha_s\ln(1/x)$ this allows us to describe the $\gamma^* A$ cross
section at a given energy entirely in terms of this simple amplitude
without reference of ``higher'' Fock-space components of the
projectile, simply by subsuming (in the sense of a renormalization
group procedure) all other strongly interacting components into the
averaging procedure.

The last ingredient of Eq.~(\ref{eq:zeroeth-order-cross}) not yet
spelled out analytically, the wave function of the virtual photon, is
known exactly [see for example~\cite{Nikolaev:1990ja} and
Eqs.~\eqref{eq:dipole-wf-ovelap}], with both longitudinal and
transverse polarizations contributing additively to the total cross
section, $\sigma^{\gamma^* p}_{\text{tot}}(Y,Q^2) = \sigma^{\gamma^*
  p}_T(Y,Q^2) +\sigma^{\gamma^* p}_L(Y,Q^2)$.

This leads to an expression for the cross section in the form of a convolution
in terms of the transverse coordinates which characterize the eikonal
scattering position of the $q\Bar q$ pair. Using $\bm{r}=\bm{x}-\bm{y}$ and
$\bm b =z\bm x+\Bar z\bm y$ to denote dipole size and impact
parameter respectively, one obtains (notations inspired
by~\cite{Marquet:2007nf})
\begin{equation}
  \label{eq:dipole-cross}
  \sigma^{\gamma^* p}_{T,L}(Y,Q^2) 
  =\sum\limits_f\int\!\!d^2 \bm{r}
  \int\limits^1_0\!dz\
  \Phi_{T,L}^f(z,\bm r,{\bm r}',Q^2)
  \int d^2\bm{b} \ 2 \ N^{q\Bar q}_Y(\bm x,\bm y)
  \ ,
\end{equation}
where $z$ and $\Bar z:=1-z$ denote the longitudinal momentum fractions
carried by the quark and antiquark respectively.  For a fixed $Q^2$,
polarization and flavor $f$, the photon wave function product
$\Phi_{T,L}^f(z,\bm r,{\bm r}',Q^2)$\footnote{The notation for the
  expressions shown here and in Eq.~\eqref{eq:dipole-wf-ovelap}
  anticipate the diffractive case. There the structure of the wave
  function overlaps remains the same, only the dipole sizes on both
  sides of the final state cut must be distinguished.}  encodes the
probability to find a $q\Bar q$ pair of size $|\bm{r}|$, polarization
$T$ or $L$, and longitudinal momentum fraction $z$ inside the virtual
photon:
\begin{subequations}
  \label{eq:dipole-wf-ovelap}
\begin{align}
\label{eq:Phi-trans}
\Phi^f_T(z,\bm{r},\bm{r}';Q^2)=
\frac{\alpha_{em}N_c}{2\pi^2}e_f^2
\Bigl(
& (z^2+\Bar z^2)Q_f^2
\frac{\bm{r}\cdot\bm{r}'}{|\bm{r}||\bm{r}'|}
K_1(Q_f|\bm{r}|)K_1(Q_f|\bm{r}'|)
\notag \\ &
+m_f^2 K_0(Q_f|\bm{r}|)
K_0(Q_f|\bm{r}'|)\Bigr)
\ ,
\end{align}
\begin{align}
\label{eq:Phi-long}
\Phi^f_L(z,\bm{r},\bm{r}';Q^2)=
\frac{\alpha_{em}N_c}{2\pi^2}e_f^2
4Q^2 z^2\Bar z^2 K_0(Q_f|\bm{r}|)
K_0(Q_f|\bm{r}'|)\ .
\end{align}
\end{subequations}
In the above, $e_f$ and $m_f$ denote the charge and mass of the quark with 
flavor $f$ and 
\begin{align}
\label{eq:Q_f}
Q_f^2\!=\! z\Bar z Q^2\!+\!m_f^2\ .
\end{align}

The impact parameter ($b$) integrated dipole amplitude has the
interpretation of a $q\Bar q$-dipole cross section on the target:
\begin{align}
  \label{eq:dipole-cross-def}
  \sigma_{q\Bar q}(Y,(\bm x-\bm y)^2) := 2 \int d^2\bm{b} \ N^{q\Bar
    q}_Y(\bm x,\bm y) \ .
\end{align}
It carries the energy dependence of the cross section in terms of $1/x =
e^{Y}$, the relative boost factor between the projectile and the target. The
separation into wave-function factors (generically called impact factors) and
dipole cross section (more generically Wilson line $n$-point functions) is
prototypical to all observables in the high energy limit and extends to higher
order in perturbation theory.

A description in terms of structure functions,
$F_2$, $F_{T,L}$  corresponds to a purely kinematical reparametrization
according to the standard relation
\begin{align}\label{eq:short_cross}
\sigma_{\text{tot}}^{\gamma^* p}(\xbj,Q^2) 
= 
\frac{4\pi^2\ \alpha_{em}}{Q^2}F_2(\xbj,Q^2)
=
\frac{4\pi^2\ \alpha_{em}}{Q^2}\left(F_T(\xbj,Q^2)+F_L(\xbj,Q^2)\right)
\ . 
\end{align}
Please note that this does not imply a general link with particle
distributions outside the parton gas region with $Q^2 \gg Q_s^2(x)$ where a
twist expansion becomes valid.

\subsection{Rapidity gaps in \texorpdfstring{$\gamma^* A$}{gamma* A} at zeroth order}
\label{sec:rapid-gaps-zeroth-order}

\begin{figure}[H]
  \centering
\includegraphics[width=\textwidth]{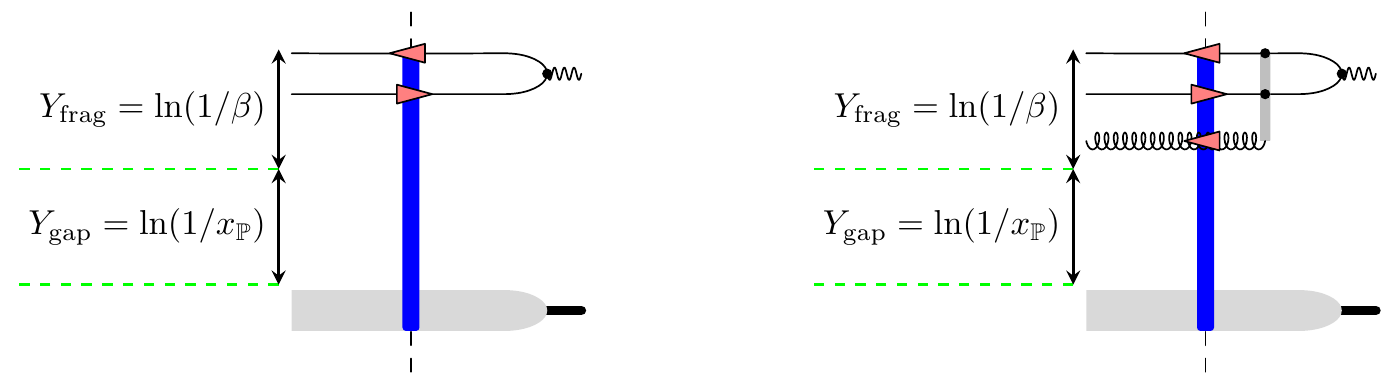}

\caption{Rapidity gap events differ from generic events contributing
  to the total cross section by a target side rapidity gap of size
  $Y_{\rm{gap}}=\ln(1/x_{\mathbb{P}})$ into which no gluons are
  emitted. This gap is complemented by a projectile fragmentation
  range of size $Y_{\rm{frag}}=\ln(1/\beta)$, such that
  $Y=Y_{\rm{gap}}+Y_{\rm{frag}}$.}
  \label{fig:gap-kinematics}
\end{figure}

Apart from the total cross section, large complementary data set is
available for rapidity gap events, in which the virtual photon
fragments into (predominantly) a $q\Bar q$-pair accompanied by a gluon
shower (which then hadronizes before it reaches the detector) that
remains well separated from the target fragmentation region by a large
rapidity gap. The kinematical setting for rapidity gap events is
sketched in Fig.~\ref{fig:gap-kinematics}.
The experimental situation is characterized by a quite strong similarity of
the energy dependence of both total and diffractive cross sections.

Theoretically, all differences with the expression for the total cross
section Eq.~(\ref{eq:zeroeth-order-cross}) arise from restrictions on
the final state: With a rapidity gap on the target side, the target
stays intact. Therefore, in the final state, the target is projected
back onto its wave-function in each amplitude factor. This leads to
separate target averages $\langle\ldots\rangle(Y)$ in each
amplitude. Since no net color can be exchanged across the gap, the
$q\Bar q$-final state is necessarily projected onto a singlet. Once
perturbative corrections are taken into account, and partons added to
the projectile fragmentation region, they remain in an overall
singlet.

At zeroth order in $\alpha_s\ln(1/x)$, the projectile only contains a
$q\Bar q$ pair, the cross section is given by
\begin{align}
  \label{eq:zeroth-order-restricted}
  & \left( \diagram[width=1.2cm]{qqbint-L} - \diagram[width=1.2cm]{qqbnoint-L}
  \right)
\diagram[height=1.1cm]{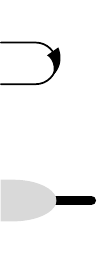}\frac{1}{N_c}
\diagram[height=1.1cm]{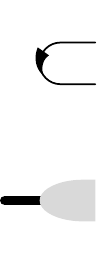}
\left( \diagram[width=1.2cm]{qqbint-R} - \diagram[width=1.2cm]{qqbnoint-R}
\right)
\end{align}
in close analogy with~\eqref{eq:zeroeth-order-cross}.  

We note that each of these restrictions individually (the separate averages as
well as the singlet projection of the projectile constituents) will prevent
the simplification of the $U$-content that takes place in the
$\overline{\text{out}}$-out overlap for the total cross section due to
Eq.~(\ref{eq:U-left-right-cancel}). Instead
\begin{align}
  \label{eq:U-left-right--dont-cancel}
       \diagram[width=1.2cm]{qqbint-L}\diagram[height=1.1cm]{qqbs-t-Rc}
\diagram[height=1.1cm]{qqbs-t-L}\diagram[width=1.2cm]{qqbint-R}
   \neq 
  \diagram[width=1.2cm]{qqbnoint-L}\diagram[height=1.1cm]{qqbs-t-Rc}
\diagram[height=1.1cm]{qqbs-t-L}\diagram[width=1.2cm]{qqbnoint-R}
\end{align}
As an immediate consequence, not only the mixed ($\overline{\text{in}}$-out)
overlaps but also the $\overline{\text{out}}$-out overlap will acquire
nontrivial energy dependence.

What is left to understand are the restrictions on the transverse
coordinates of Wilson lines in this expression as imposed by phase
space integrals -- again such a restriction alone would be sufficient
to induce a nontrivial Wilson-line four-point function. To this end,
note that $\beta$ and $Q^2$ together determine the invariant mass of
the projectile fragments:
\begin{align}
  \label{eq:beta-M_X}
  \beta = \frac{Q^2}{Q^2+ M_X^2}\ .
\end{align}
This restricts the integration over transverse momenta. To be
specific, assume $n$ projectile constituents, denote longitudinal
momentum fractions by $z_i$ (with $\sum\limits_i^n z_i = 1$) and final
state transverse momenta by $\bm k_i := \Tilde{\bm k}_i+z_i\bm\Delta$
(where $\bm\Delta := \sum\limits_i^n \bm k_i$). Then the invariant
mass of the $n$-particle final state is
\begin{align}
  \label{eq:M_Xn-k_i}
  M_{X,n}^2 = \sum_i^n \frac{\Tilde{\bm k}_i^2+m_i^2}{z_i}
\end{align}
so that a restriction on $M_X$ imposes one linear constraint on the
$\Tilde{\bm k}_i^2$. To understand what this implies for the transverse
coordinates, define
\begin{align}
  \label{eq:bandr}
&  \bm r_i := \bm x_i-\bm x_n; \hspace{1cm} 
  \bm r'_i := \bm x'_i-\bm x'_n; \hspace{1cm} 
  \bm b := \sum\limits_i z_i \bm  r_i; \hspace{1cm} 
  \bm b' := \sum\limits_i z_i \bm  r'_i; 
\end{align}
and rewrite the exponents of the transverse momentum phase factors as
\begin{align}
  \label{eq:phase-factors}
  \sum\limits_i^n \bm k_i \cdot (\bm x'_i-\bm x_i)= \bm
  \Delta\cdot\Bigl(\bm b'-\bm b \Bigr)
  +\sum\limits_{i=1}^{n-1} {\Tilde{\bm k}}_i \cdot 
  \Bigl(\bm r'_i-\bm r_i\Bigr)
\   .
\end{align}
This implies that integration over momentum transfer $t =
-\bm\Delta^2$ will identify the light cone c.m. coordinates $\bm b$
and $\bm b'$ if one ignores the kinematical upper limit on $t$ as for
the total cross section. One of the $n-1$ $\Tilde{\bm k_i}$
integrations on the other hand is restricted by~(\ref{eq:M_Xn-k_i}),
so that one of the $n-1$ independent distance pairings $\bm r_i$ and
$\bm r_i'$ will remain independent after all unconstrained integrals
are carried out. For the $q\Bar q$ final state encountered at leading
order, there is only one independent momentum variable available to
begin with $\bm\kappa = \Tilde{\bm k}_1 = - \Tilde{\bm k}_2$. It is,
therefore, tied directly to $M_X$ via
\begin{align}
  \label{eq:M_X2-kappa}
  M_{X,2}^2 = \frac{\bm \kappa^2+m_i^2}{z \Bar z}
\end{align}
and leaves behind phase factor containing $(\bm r'-\bm
r)\cdot\bm\kappa$ where the dipole sizes $\bm r$ and $\bm r'$ in
amplitude and complex conjugate amplitude remain independent. For each
flavor independently, the length of $\bm\kappa$ is fixed in terms of
$\beta$ and the quark mass via~\eqref{eq:M_X2-kappa}
and~\eqref{eq:beta-M_X} as
\begin{align}
  \label{eq:kappa_f}
  \bm\kappa_f^2 := z\Bar z Q^2\frac{\Bar\beta}{\beta} - m_f^2\ .
\end{align}
One obtains (see again~\cite{Marquet:2007nf})
\begin{align}
  \label{eq:dsigmadbeta}
  \frac{d\sigma^{\gamma^*A\rightarrow Xp}_{T,L}}{d\beta} (x,Q^2) = &
  \frac1\pi \frac{Q^2}{4\beta^2}\sum_f \int d^2\bm{r} \int d^2\bm{r}'
  \int\limits_0^{2\pi} \frac{d\varphi_{\bm\kappa}}{2\pi} \int_0^1 dz\
  z\Bar z\ \Theta(\bm{\kappa}_f^2)\
  e^{i\bm{\kappa}_f\cdot(\bm{r}'-\bm{r})}\\ & \hspace{1cm}\times
  \Phi_{T,L}^f(z,\bm{r},\bm{r}';Q^2) \int d^2\bm b\ (N^{q\Bar q}_Y(\bm
  r',\bm b'))^* N^{q\Bar q}_Y(\bm r,\bm b) \notag
\end{align}
where $Y=\ln(1/x)$.  Integrating this result over $\beta$ leads to the
total diffractive cross section. This step identifies $\bm r$ with
$\bm r'$ if one extends the upper phase space boundary in $M_X$ from
$W^2$ to $\infty$ as discussed earlier. The result in turn maps back
onto the total cross section if one removes the singlet projection in
the final state which replaces $(N^{q\Bar q}_Y(\bm r',\bm b'))^*
N^{q\Bar q}_Y(\bm r,\bm b)$ by $(N^{q\Bar q}_Y(\bm r',\bm b'))^* +
N^{q\Bar q}_Y(\bm r,\bm b)$.

Note that this relationship of cross sections uniquely identifies the
$x$ arguments of the averages in the diffractive cross sections to be
the overall Bjorken $x$ of the process, not $x_{\mathbbm P}$ as
usually assumed in the literature. The numerical effect of such a
replacement is, however, not large enough to affect the quality of any
diffractive fits with state of the art expressions. These expressions
suffer from more serious defects: incomplete NLO impact factors and
nonperturbative normalization effects associated with the
corresponding impact parameter averages as will be discusses below.

\subsection{Beyond zeroth order}
\label{sec:beyond-zeroth-order}

Leading order (LO) corrections to the above resum contributions
proportional to $(\alpha_s\ln(1/x))^n$ and are fully taken into
account by solving the LO-JIMWLK equation. The impact factors receive
no corrections to their zeroth order form. At next to leading order,
when contributions proportional to $\alpha_s(\alpha_s\ln(1/x))^n$ are
taken into account, both JIMWLK evolution \emph{and} the impact
factors receive corrections.

Running coupling corrections to the evolution of Wilson line $n$-point
functions have been calculated in full
generality~\cite{Gardi:2006rp,Kovchegov:2006vj,Balitsky:2006wa}. The remaining
conformal corrections to evolution have been obtained by Balitsky and
Chirilli\cite{Balitsky:2008zz}, who presently work on the expression for NLO
impact factors. A generalization of both aspects for arbitrary Wilson line
$n$-point functions is yet to be devised. These ingredients would be required
to extend the only existing treatment of
JIMWLK-evolution~\cite{Rummukainen:2003ns} beyond leading order.

Fortunately truncations of JIMWLK evolution to finite sets of
evolution equations, such as the Balitsky-Kovchegov (BK) equation or
its more general GT (Gaussian truncation) counterpart allow us to use
the available information to implement evolution at NLO. Accuracy on
the impact factor side at present remains at LO. NLO accuracy for the
impact factors introduces terms including $q\Bar q g$-correlators into
the expressions for both the total cross section
Eq.~(\ref{eq:short_cross}) and the diffractive cross
section~(\ref{eq:dsigmadbeta}).  We will see below that fits to the
total cross section are not affected strongly, but that already the
description of rapidity gap events suffers noticeably from this
limitation.

Even with more modest goals in mind, such as the description of the total
cross section it is mandatory to include NLO effects at least on the level of
evolution equations. Here NLO corrections induce
qualitatively new effects like scale breaking and a quantitatively important
reduction in evolution speed compared to the LO situation.  With the impact
factors remaining at LO it is sufficient to know the $Y$-dependence of the
dipole cross sections entering Eqs.~(\ref{eq:short_cross})
and~(\ref{eq:dsigmadbeta}), but once the NLO impact factors are known this is
no longer sufficient ($q\Bar q g$ operators will require consistent treatment)
and one is forced to either use the full JIMWLK-evolution
framework or choose a ``suitable'' truncation. To appreciate what is involved,
we briefly recapitulate the necessary tools.

\subsubsection{JIMWLK and its truncations at LO}
\label{sec:jimwlk-its-trunc-LO}

The JIMWLK evolution equation provides a means to calculate the
energy- (or rapidity-) dependence of arbitrary $U$-correlators by
first introducing an energy- (or rapidity-) dependent statistical
weight $Z_Y[U]$ for the configurations of the $U$-fields. The dipole
correlator of Eq.~(\ref{eq:dipole-amplitude}) is then expressed as a
functional integral of the form
\begin{align}
  \label{eq:Z-for-N}
  \frac{1}{N_c} 
  \left\langle\tr(1-U_{\bm{x}} U^\dagger_{\bm{y}})
  \right\rangle(Y) = \int \Hat D[U] \frac{1}{N_c} 
  \tr(1-U_{\bm{x}} U^\dagger_{\bm{y}}) Z_Y[U]
\end{align}
where $\Hat D[U]$ is a functional Haar measure. This is meaningful in
the sense that it allows to calculate the average of \emph{any}
operators, if it is possible to describe the evolution of \emph{all}
averages in terms of the evolution of the weight $Z_Y[U]$ defining the
averaging procedure. This is the main content of the JIMWLK equation:
it abstracts the energy dependence of the average from the operator
being averaged by describing it as a functional evolution equation for
$Z_Y[U]$. At LO, the equation takes the form of a functional
Fokker-Planck equation
\begin{align}
  \label{eq:JIMWLK}
  \frac{d}{d Y} Z_Y[U] = -H_{\text{JIMWLK}} Z_Y[U]
\end{align}
that traces how additional gluons are added to the phase space of the
projectile as one increases the energy of the collision.  To arrive
at~(\ref{eq:JIMWLK}) one needs to prove~\cite{Weigert:2000gi} that
this equation indeed allows to find the energy dependence of arbitrary
$U$-correlators, not just the simple $q\Bar q$-operator entering the
dipole cross section. For such generic operators $\Hat O[U]$, $Z_Y[U]$
defines a target average via
\begin{align}
  \label{eq:O-U-average}
  \langle \Hat O[U]\rangle(Y)  := \int \Hat D[U] \Hat O[U] Z_Y[U]
\end{align}
so that ~(\ref{eq:JIMWLK}) implies an evolution equation for each such
operator $\Hat O[U]$ that takes the form
\begin{align}
  \label{eq:JIMWLK-op}
  \frac{d}{d Y} \langle \Hat O[U]\rangle(Y)  
  = -\langle H_{\text{JIMWLK}}\Hat O[U]\rangle(Y)
\ .
\end{align}
One of the features of JIMWLK evolution is that non-singlet correlators are
exponentially suppressed by infrared divergent contributions, only singlet
correlators survive.

A key feature of $Y$-evolution is that
Eq.~(\ref{eq:JIMWLK}) gives rise to coupled hierarchies of evolution
equations for $U$-correlators, known as Balitsky hierarchies. This already
becomes manifest in the evolution equation for the $q\Bar q$ operator at LO:
It can be written as
  \begin{align}
    \label{eq:preBKU}
    \frac{d}{dY} 
 \langle 
 \tr(
  U_{\bm{x}}
  U^\dagger_{\bm{y}}) 
\rangle(Y)    
=\frac{\alpha_s}{\pi^2}\int d^2z\ {\cal K}_{\bm{x z y}} 
  \left(
  \langle
\big[\Tilde U_{\bm{z}}\big]^{a b}\
    \tr(
  t^a 
  U_{\bm{x}}
  t^b 
  U^\dagger_{\bm{y}}) 
  \rangle(Y)
  -C_f  \langle 
 \tr(
  U_{\bm{x}}
  U^\dagger_{\bm{y}})  \rangle(Y)    
  \right)
  \end{align}
or, using~(\ref{eq:hatSqqbardef}) and the Fierz identity
\begin{equation}
  \label{eq:Fierz}
  \big[\Tilde U_{\bm{z}}\big]^{a b}
  2 \tr(t^a U_{\bm{x}}t^b U^\dagger_{\bm{y}}) =
  \tr( U_{\bm{x}}U^\dagger_{\bm{z}}) \
           \tr( U_{\bm{z}}U^\dagger_{\bm{y}})
           -\frac{1}{N_c}\tr( U_{\bm{x}}U^\dagger_{\bm{y}})
\end{equation}
as
\begin{align}
  \label{eq:preBKS}
 \frac{d}{d Y} \langle \Hat S_{\bm{x y}} \rangle(Y)
  =\frac{\alpha_s N_c}{2\pi^2}\int d^2z\,
  \ {\cal K}_{\bm{x z y}} \
 \langle \Hat S_{\bm{x z}} 
\Hat S_{\bm{z y}} 
 -
\Hat S_{\bm{x y}} \rangle(Y)
\ .
\end{align}
The integral kernel in both~(\ref{eq:preBKU}) and~(\ref{eq:preBKS})
is given by~\cite{Mueller:1994rr,Kovchegov:1999yj}
\begin{align}\label{K}
  {\cal K}_{\bm{x z y}} \, := \, \frac{({\bm x} - {\bm y})^2}{({\bm x}
    - {\bm z})^2 \ ({\bm z} - {\bm y})^2}\ .
\end{align}
Eqs.~(\ref{eq:preBKU}) and~(\ref{eq:preBKS}) do not represent closed
equations since the evolution of $\langle \tr( U_{\bm{x}}
U^\dagger_{\bm{y}}) \rangle(Y)$ depends on an operator with an
additional gluon operator insertion made manifest by the $\tilde U$
appearing in the first term on the right hand side of
Eq.~(\ref{eq:preBKU}). The evolution equation of that new operator,
$\langle \big[\Tilde U_{\bm{z}}\big]^{a b}\ \tr( t^a U_{\bm{x}} t^b
U^\dagger_{\bm{y}}) \rangle(Y)$, in turn will involve yet one more
insertion of a gluon operator $\tilde U$, iteratively creating an
infinite coupled hierarchy of evolution equations, the Balitsky
hierarchy of the quark dipole operator~(\ref{eq:hatSqqbardef})
\cite{Balitsky:1997mk, Balitsky:1996ub}. JIMWLK evolution summarizes
the totality of all such hierarchies, based on any (gauge invariant)
combination of multipole operators but can only be solved
numerically~\cite{Rummukainen:2003ns,Kovchegov:2008mk,Lappi:2011ju} at
considerable computational cost.

One may note that within such hierarchies one finds numerous cross- and
self-referencing patterns that can be exposed by looking at coincidence limits
of coordinates in the operators involved. Taking $\langle \big[\Tilde
U_{\bm{z}}\big]^{a b}\ \tr( t^a U_{\bm{x}} t^b U^\dagger_{\bm{y}}) \rangle(Y)$
as an example, one finds its evolution equation linked with that of $q\Bar q$-
and $g g$-dipole operator averages to which it reduces in the limits $\bm
x\to\bm y$ and $\bm z\to\bm x$ (or $\bm y$) respectively. More generically,
generalizing from $q\Bar q$ dipoles to ${\cal R}\Bar{\cal R}$-dipoles (where
${\cal R}$ refers to an arbitrary representation ${\cal R}$) one finds their
evolution equation to contain the operator $\big[\Tilde U_{\bm{z}}\big]^{a b}\
\overset{{\scriptscriptstyle\cal R}}\tr( \overset{{\scriptscriptstyle\cal R}}
t^a \overset{{\scriptscriptstyle\cal R}} U_{\bm{x}}
\overset{{\scriptscriptstyle\cal R}} t^b \overset{{\scriptscriptstyle\cal R}}
U^\dagger_{\bm{y}})$ (see~\cite{Kovchegov:2008mk}). The evolution equations of
this operator are mapped onto those of ${\cal R}\Bar{\cal R}$-dipoles or $g
g$-dipoles in the two coincidence limits according to
\begin{subequations}
  \label{eq:3-point-generic-coincidence}
\begin{align}
  \label{eq:generic-x=y-limit}
\lim\limits_{y\to x}     \big[\Tilde U_{\bm{z}}\big]^{a b}\
   \overset{{\scriptscriptstyle\cal R}}\tr(
\overset{{\scriptscriptstyle\cal R}} t^a 
\overset{{\scriptscriptstyle\cal R}} U_{\bm{x}}
\overset{{\scriptscriptstyle\cal R}} t^b 
\overset{{\scriptscriptstyle\cal R}}
  U^\dagger_{\bm{y}})
& =  C_{\cal R} \frac{d_{\cal R}}{d_A}
\Tilde\tr\left(\Tilde U_{\bm z} \Tilde U_{\bm x}^\dagger \right)
\ ,
\\
  \label{eq:z=x,y-limit}
\lim\limits_{
      \bm z\to \bm y\ \text{or}\ \bm x}    
\big[\Tilde U_{\bm{z}}\big]^{a b}\
   \overset{{\scriptscriptstyle\cal R}}\tr(
\overset{{\scriptscriptstyle\cal R}} t^a 
\overset{{\scriptscriptstyle\cal R}} U_{\bm{x}}
\overset{{\scriptscriptstyle\cal R}} t^b 
\overset{{\scriptscriptstyle\cal R}}
  U^\dagger_{\bm{y}}) 
& = C_{\cal R}   \,
\overset{{\scriptscriptstyle\cal R}}\tr(
\overset{{\scriptscriptstyle\cal R}} U_{\bm{x}}
\overset{{\scriptscriptstyle\cal R}} U_{\bm{y}}^\dagger
) 
\ .
\end{align}
\end{subequations}
The more $U$-fields involved, the more constraints are imposed by
coincidence limits and one may construct whole towers of operators
linked downwards by coincidence limits.  All these structures and
relationships are automatically preserved and maintained in full
JIMWLK evolution which, at least at one loop accuracy, can be
simulated numerically.\footnote{Limitations are imposed only by
  available computational resources with limits on evolution ranges
  and initial conditions that can be accommodated without losing
  numerical accuracy.}

The theoretical picture can be simplified and the numerical effort required to
solve the evolution equation can be reduced significantly by truncating the
hierarchies. Any such truncation comes at the price of introducing an
additional approximation. The most widely used truncation of JIMWLK evolution
is known as the BK approximation. It assumes the factorization
\begin{align}
  \label{eq:BKfact}
  \langle \Hat S_{\bm{x z}} 
\Hat S_{\bm{z y}} \rangle(Y) \to 
 \langle \Hat S_{\bm{x z}} \rangle(Y)\ \langle
\Hat S_{\bm{z y}} \rangle(Y)
\ ,
\end{align}
which turns Eq.~\eqref{eq:preBKS} into a closed equation in terms of
$\langle\Hat S_{\bm{x y}} \rangle(Y)$ only and thus decouples the rest of the
Balitsky hierarchy. The BK truncation is valid and is parametrically justified
in the large-$N_c$ limit for scattering on a large dilute nuclear
target. Using \eqref{eq:BKfact} in \eqref{eq:preBKS} we obtain the BK
evolution equation
\begin{align}
  \label{eq:BK}
  \frac{d}{d Y} \langle \Hat S_{\bm{x y}} \rangle(Y) =\frac{\alpha_s
    N_c}{2\pi^2}\int d^2z\, \ {\cal K}_{\bm{x z y}} \ \left[ \langle
    \Hat S_{\bm{x z}} \rangle (Y) \ \langle \Hat S_{\bm{z y}} \rangle
    (Y) - \langle \Hat S_{\bm{x y}} \rangle(Y) \right]\ .
\end{align}
An alternative truncation has been discussed
in~\cite{Kovner:2001vi,Weigert:2005us,Kovchegov:2008mk} and dubbed the
Gaussian truncation (GT) in\cite{Kovchegov:2008mk}. In spirit, it approximates
the JIMWLK average with Glauber-iterated two gluon t-channel exchange  with
the target.

This leads to explicit expressions for multi-$U$-correlators, in terms of a
two point function ${\cal G}_{Y,\bm x\bm y}$, for example
\begin{subequations}
  \label{eq:simpleGcorr}
\begin{align}
  \label{eq:UUdaggersol}
  \langle \overset{{\scriptscriptstyle\cal R}}\tr(
\overset{{\scriptscriptstyle\cal R}} U_{\bm{x}}
\overset{{\scriptscriptstyle\cal R}}
  U^\dagger_{\bm{y}})
\rangle(Y) = &\, \ d_{\cal R}\ e^{-C_{\cal R}{\cal G}_{Y,\bm{x y}}}
\ , 
\\
  \label{eq:UtrtUtUdagger}
\langle 
\big[\Tilde U_{\bm{z}}\big]^{a b} 
\overset{{\scriptscriptstyle\cal R}}\tr(
\overset{{\scriptscriptstyle\cal R}} t^a 
\overset{{\scriptscriptstyle\cal R}} U_{\bm{x}}
\overset{{\scriptscriptstyle\cal R}} t^b 
\overset{{\scriptscriptstyle\cal R}}
  U^\dagger_{\bm{y}})
\rangle(Y) 
= & \,
 C_{\cal R} d_{\cal R}\ e^{
  -\frac{N_c}2\left(
  {\cal G}_{Y,\bm{x z}} + {\cal G}_{Y,\bm{z y}}- {\cal G}_{Y,\bm{x y}}  
\right)
-C_{\cal R} {\cal G}_{Y,\bm{x y}} 
}  
\end{align}
\end{subequations}
and a consistent description of evolution for dipoles in all those arbitrary
representations in terms of a single equation for ${\cal G}$
(see~\cite{Kovchegov:2008mk})
\begin{equation}
  \label{eq:tilde-G-evo-short}
 \frac{d}{d Y} {\cal G}_{Y,\bm{x y}}  =  \frac{\alpha_s}{\pi^2} \int\!\!d^2z\  
 {\cal K}_{\bm{x z y}} \biggl(
 1-  e^{-\frac{ N_c}{2} \bigl(
 {\cal G}_{Y,{\bm{x z}}} +{\cal G}_{Y,{\bm{y z}}}
 - {\cal G}_{Y,{\bm{x y}}}\bigr)}
\biggr)
\ ,
\end{equation}
irrespective of the representation ${\cal R}$. One may think of GT as
a truncation that, compared to BK, includes the minimal subset of
$1/N_c$ suppressed contributions needed to restore the group
theoretical coincidence limits (\ref{eq:3-point-generic-coincidence})
which are automatically satisfied by full JIMWLK evolution. It can be
shown\cite{Kovchegov:2008mk} that the dynamical content of GT and BK
are in fact the same in the sense that replacing $q\Bar q$- and $g
g$-dipoles appearing in Eqs.~(\ref{eq:simpleGcorr}) (directly or in
certain limits) by their large-$N_c$ counterparts (proportional to
$e^{-\frac{N_c}2{\cal G}_{Y,\bm{x y}}}$ and $e^{-N_c {\cal G}_{Y,\bm{x
      y}}}$ respectively) maps the BK equation~(\ref{eq:BK})
onto~(\ref{eq:tilde-G-evo-short}) and vice versa.  GT improves over BK
not in terms of dynamical content, but in the way this content is
mapped onto different correlators of the theory.

It was shown in~\cite{Kovchegov:2008mk} at one loop accuracy, that the
modification of the truncation encoded in Eqs.~(\ref{eq:simpleGcorr})
and~(\ref{eq:tilde-G-evo-short}) under the name Gaussian truncation
leads to slightly better agreement with full JIMWLK evolution than the
BK truncation. We will use both BK and GT at NLO accuracy to compare
to data below and will see that GT results in a slight improvement of
the fit in keeping with the slightly better match of GT with JIMWLK
evolution.  The main advantage of the Gaussian truncation is that it
allows to consistently describe general $n$-point functions such as
that on the left of~(\ref{eq:U-left-right-cancel}) before the local
limit is taken. Unlike the BK approximation it is versatile enough to
allow us to test the reliability of the phase space approximations
that are built into~(\ref{eq:short_cross}) without being hampered by
${\cal O}(1/N_c^2)$ corrections.

\subsubsection{Evolution at NLO}
\label{sec:evolution-at-nlo}

At NLO the picture gets even more complicated. At this accuracy not
only single gluon Wilson lines are added to the original dipoles at
the level of Eq.~(\ref{eq:preBKS}), also insertions of nonlocal
operators such as $\tr(t^a U_{\bm z_2} t^b U_{\bm z_2}^\dagger)$ (in
the case of quark contributions) appear on the right hand side. It is
clear that a full JIMWLK treatment becomes more and more costly and
suitable truncations more and more of a necessity. In the light of the
increasingly complicated insertions one would expect a strict leading
$1/N_c$ BK approximation to become rather crude. The Gaussian
truncation on the other hand should remain a viable candidate for a
useful truncation. Here we choose an NLO treatment that, for the total
cross section~(\ref{eq:short_cross}) and the diffractive cross
section~(\ref{eq:dsigmadbeta}) allows us to use both BK and GT with
comparable accuracy, and only keep in mind that once accuracy is high
enough to consider testing the validity
of~(\ref{eq:U-left-right-cancel}) for the total cross section or more
differential observables as discussed in~\cite{Marquet:2010cf}, the
Gaussian truncation becomes the tool to choose. Our main reason not to
use the full dipole evolution as presented in~\cite{Balitsky:2008zz}
is numerical efficiency. While it is easy to include running coupling
corrections according
to~\cite{Gardi:2006rp,Kovchegov:2006vj,Balitsky:2006wa}, an
implementation of the conformal corrections is impractical. Instead we
adopt to substitute the conformal corrections with an energy (or
rather longitudinal momentum) conservation correction as suggested by
Gotsman, Levin, Maor and Naftali~\cite{Gotsman:2004xb}. This is an
attempt to resum DGLAP type corrections that enter small-$x$ evolution
at NLO that resum collinear contributions to all orders but should not
lead to any double counting conflicts with the resummation of running
coupling corrections.

In the BK-truncation the NLO equation to solve takes the form
\begin{align}\label{eq:bks_modified}
\frac{d}{d Y} S_{Y;\bm{xy}}=\frac{N_{c}}{2\pi^{2}} \int d^{2} z \ 
\mathcal{M}_{\bm{xzy}}\Bigl(1-\frac{d}{d Y}\Bigr)
(S_{Y;\bm{xz}}S_{Y;\bm{zy}}-S_{Y;\bm{xy}})
\end{align}
while its GT counterpart reads
\begin{align}\label{eq:GT_modified}
\frac{d}{d Y} e^{-C_f{\cal G}_{Y;\bm{xy}}} 
=\frac{N_{c}}{2\pi^{2}} \int d^{2} z \ \mathcal{M}_{\bm{xzy}}
\Bigl(1-\frac{d}{d Y}\Bigr)
\biggl(
 1-  e^{-\frac{ N_c}{2} \bigl(
 {\cal G}_{Y,{\bm{x z}}} +{\cal G}_{Y,{\bm{y z}}}
 - {\cal G}_{Y,{\bm{x y}}}\bigr)}
\biggr) e^{-C_f{\cal G}_{Y;\bm{xy}}}\ .
\end{align}
The kernel function $\mathcal{M}_{\bm{xzy}}\equiv {\cal
  K}_{\bm{xzy}}R^{\text{eff}}_{\bm{xzy}}$ is a product of the leading
order BFKL/BK kernel ${\cal K}_{\bm{xzy}}$ of~(\ref{K}) and what one
may call the effective running strong coupling
$R^{\text{eff}}_{\bm{xzy}}$. The energy conservation corrections are
represented by the derivative term on the right hand sides. Without
it, both equations can be solved by a single step of numerical
integrations based on the input of $S_Y$ or ${\cal G}_Y$ alone. To
access the derivative one needs to know these functions at two $Y$
values, $Y$ and $Y+\Delta Y$, which forces us to use a much more
costly iterative procedure described in
App.\ref{sec:energy-conservation}.

To arrive at a precise form for $R^{\text{eff}}_{\bm{xzy}}$ one should
be aware that there exists no canonical way to separate running
coupling corrections from the conformal contributions at NLO. Only
their sum is unambiguously defined, to split them apart one is forced
to introduce a separation scheme as discussed
in~\cite{Gardi:2006rp,Kovchegov:2006vj,Balitsky:2006wa,Kovchegov:2006wf}.

\begin{figure}[htb]
 \centering
 \includegraphics[width=.33\linewidth]{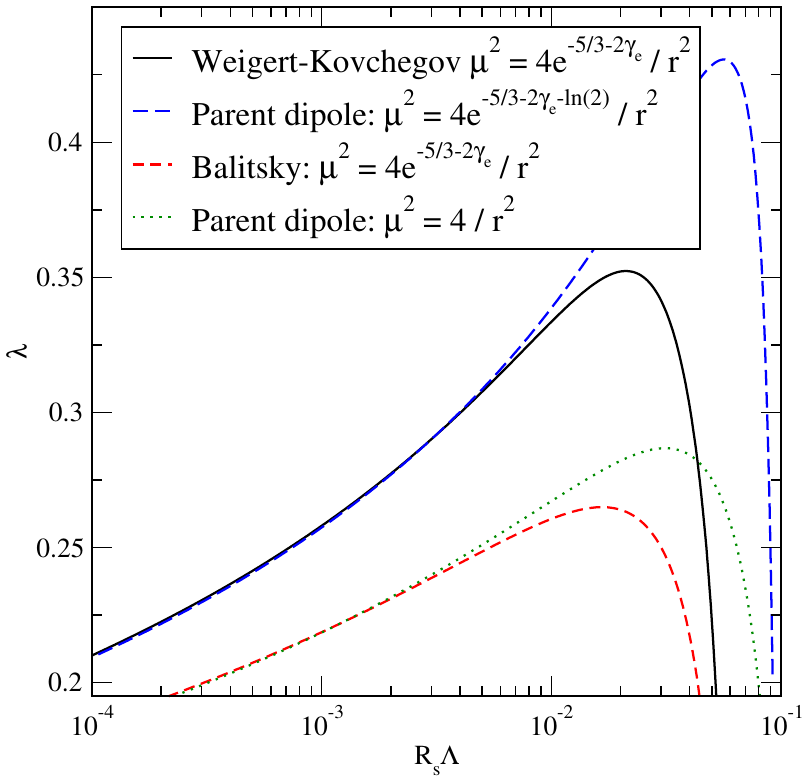}
 \hfill
 \includegraphics[width=.324\linewidth]{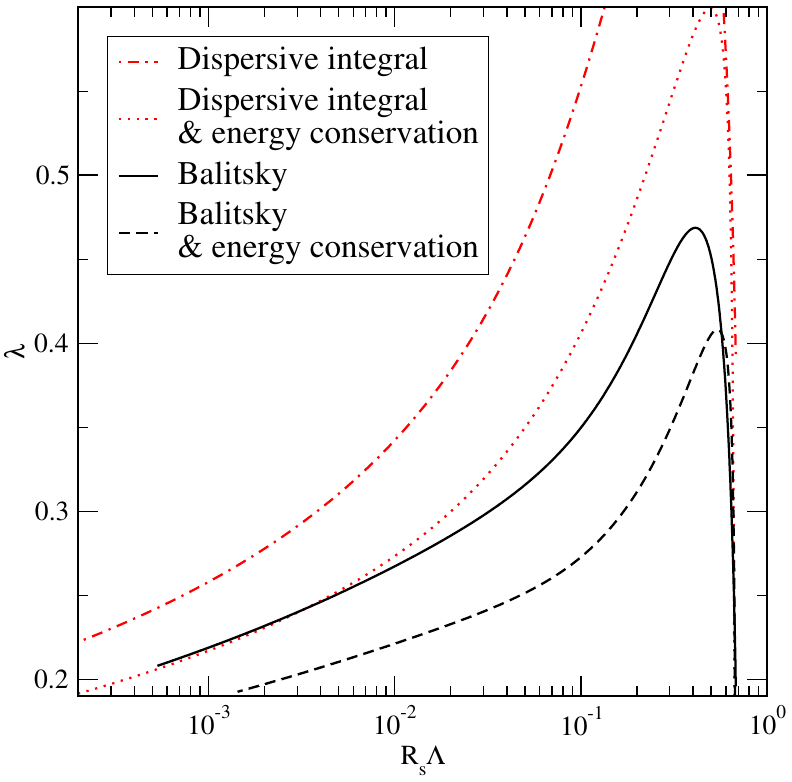}
 \hfill
 \includegraphics[width=.33\linewidth]{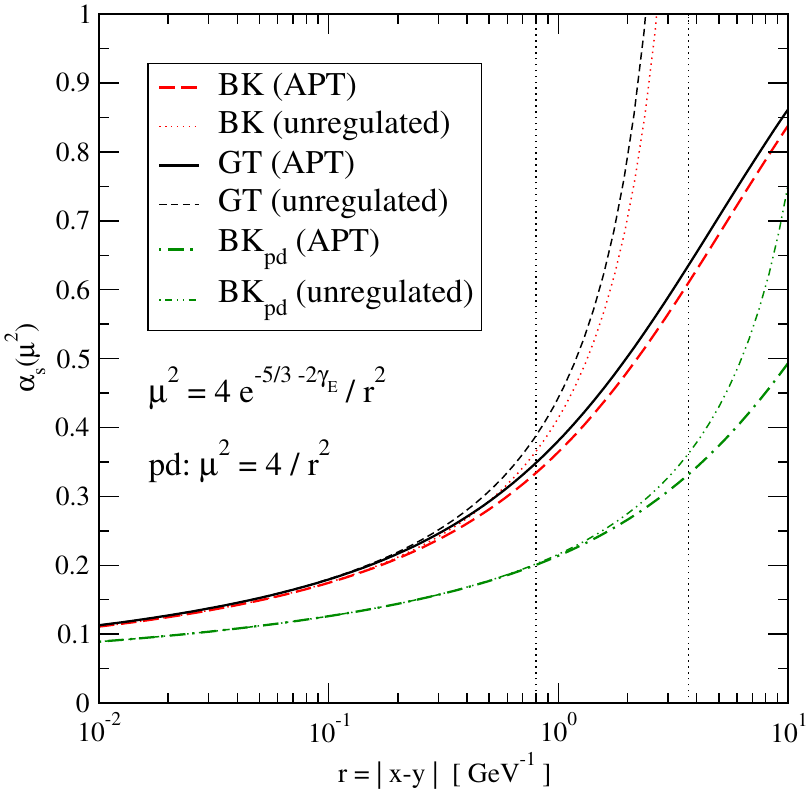}
 \caption{\textbf{Left:} The evolution speed $\lambda$ as a function of
   dimensionless units $R_s \Lambda$ for two different approaches
   mentioned in text. \textbf{Middle:} Evolution speed after adding in
   the energy conservation correction on top of running coupling {\bf
     Right:} The running couplings (regulated vs. unregulated) used in
   different BK schemes plotted against the parent dipole size $r$.
   The vertical lines bracket approximately the region of saturation
   corresponding to HERA, $0.8 < R_s < 3.7$ for $5\times 10^{-7} <
   \xbj < 0.02$.}
 \label{fig:running_coups-updated}
\end{figure}

For data comparison we will adopt the separation scheme that subsumes
most of the known NLO corrections into the running coupling
contribution as suggested by Balitsky~\cite{Balitsky:2006wa} instead
of the scheme originally suggested
in~\cite{Gardi:2006rp,Kovchegov:2006vj}.  We introduce the shorthand
notations
\begin{equation}
  \label{eq:distshort}
  r=\vert {\vrr}\vert=\vert\bm{x}-\bm{y}\vert\ ,\quad
  r_1 = \vert{\vdl}\vert=\vert \bm{x}-\bm{z}\vert\ ,\quad
  r_2=\vert {\vdr}\vert=\vert \bm{y}-\bm{z}\vert\ ,
\end{equation}
to refer to coordinate differences\footnote{$r$ is the size of the
  parent dipole $(\qqp\ \text{dipole})$ and $r_1,r_2$ refer to
  daughter dipoles $(qg\ \text{and}\ \bar{q}g\ \text{dipoles})$.  All
  expressions encountered are fully symmetric in interchange $r_1
  \leftrightarrow r_2$.} and
\begin{align}\label{eq:dipoles}
\mu^2(r) = \frac{{\cal C}^2}{r^2}
;\quad {\cal C}^2=4e^{-5/3-2\gamma_E}\ ,
\end{align}
to refer to the scales in the coupling
constants\footnote{$\gamma_E=0.5772\dots$ is the Euler-Mascheroni
  constant. The scale factor ${\cal C}^2 = 4 e^{-2\gamma_E-5/3}$ is
  specific to the $\msbar$ scheme, it would be replaced by $4
  e^{-2\gamma_E}$ in the $V$-scheme.}.  Now we may write the running
coupling to be used in Eq.~\eqref{eq:bks_modified} as
\begin{align} \label{eq:quark_run}
 R^{\text{eff}}_{\bm x\bm y\bm z} = 
 R^{\text{eff}}(r,{r_1},{r_2})=
 \alpha_{s}(\mu(r))\left[
   1+\frac{r_1^2}{r^2}\left(
     \frac{\alpha_{s}(\mu(r_2))}{\alpha_{s}(\mu(r_1))}-1
   \right)
   +\frac{r_2^2}{r^2}\left(
     \frac{\alpha_{s}(\mu(r_1))}{\alpha_{s}(\mu(r_2))}-1
   \right)
 \right]\
  ,
\end{align}
which is a combination of three individual running couplings given by the
standard perturbative one loop running
\begin{align}\label{eq:standard_run}
\alpha_{s}(\mu)=\frac{4\pi}{\beta_0}\frac{1}{\nat(\mu^2 /\Lambda^2)}\ ;
\quad \beta_{0} = (11 N_c-2 N_{f})/3\ ;
\quad N_{f},N_c=3\ \Rightarrow \beta_{0}=9\ .
\end{align}
The choice to subsume as large a part of the NLO corrections into the
running coupling has a large effect if all other NLO corrections are
ignored. Comparing evolution the separation schemes
of~\cite{Gardi:2006rp,Kovchegov:2006vj} and~\cite{Balitsky:2006wa}
under these conditions has a large effect on the evolution speed as
shown in Fig.~\ref{fig:running_coups-updated}, left panel. For
completeness this figure also includes the effect of parent dipole
running used widely in the literature. This procedure postulates an
effective coupling $R_{\text{eff}}:= \alpha_s(\mu(r))$ with the scale
factor ${\cal C}^2 = 4$.  This choice leads to almost the same
evolution speed than Eq.~\eqref{eq:quark_run} as can be seen from the
leftmost of Figs.~\ref{fig:running_coups-updated}.

The idea behind the energy conservation correction is to include an
all orders resummation of collinear corrections that start to
contribute at NLO -- they should make up the bulk of contributions not
yet included. Unfortunately, we lack a derivation that would allow us
to interlink its treatment with the separation schemes used to define
the running coupling corrections: double counting of contributions can
not be excluded without this information. Correspondingly, the energy
conservation corrections do not alleviate the difference between
evolution speeds induced by the different running coupling schemes
once included in the calculation. This is shown in
Fig.~\ref{fig:running_coups-updated}, middle panel and should be
considered a major uncertainty in the present approach. However, this
is of little consequence for data fits. Once we allow $\Lambda$ to
become a fit parameter, the difference of these treatments can be
reabsorbed by a rescaling of that factor: after rescaling, the shapes
of dipole amplitudes differ very little and allow fits of equal
quality. We will find, however, that the energy conservation
correction is a prerequisite to obtain a good fit in the
pseudo-scaling region. A treatment that explicitly removes double
counting from the outset would be much preferable.

One unavoidable complication remains in the region where
$R_s(Y)\Lambda$ is near one, the region where the IR safety arguments
that render JIMWLK evolution a self consistent procedure do not
apply. JIMWLK evolution is justified where $R_s(Y)\Lambda \ll 1$,
since at distances larger than the correlation length $R_s(Y)$, the
correlator part of the evolution equations~\eqref{eq:bks_modified}
or~\eqref{eq:GT_modified} very quickly approaches zero, so that for
small enough $R_s$ any sensitivity on how one regulates the Landau
pole in the running coupling modified kernel ${\cal M}_{\bm{x z y}}$
disappears -- IR uncertainties are effectively eliminated.

If, however, $R_s(Y)\Lambda$ is near one, regulator effects become
visible. They affect mostly the large $r$ part of the dipole amplitude
and integrated quantities such as the evolution speed, as defined
in~\eqref{eq:lambda-def}.

Since, as discussed below, the data lie in a range where the scale
separation between $\Lambda$ and $Q_s$ is not safely established, we
have to face uncertainties induced by our choice of regulator, our
model for the IR behavior of the QCD coupling. The uncertainties
encountered are shown in the panel on the right of
Fig.~\ref{fig:running_coups-updated} and involve an APT
regulator~\cite{Shirkov:1997wi,Solovtsov:1999in} and the extreme case
of a practically unregulated coupling (corresponding to a cutoff
treatment where $\alpha_s$ is the running coupling frozen only after
it reaches 30.).  This serves to illustrate the potential size and
\emph{range} of influence of the uncertainty induced by the presence
of the Landau pole.

\subsubsection{Scaling behavior and self-consistency at NLO}
\label{sec:scal-behav-self-cons-NLO}

JIMWLK evolution as well as both its BK and GT truncations are IR safe
in the sense that in a transversely infinite medium and an initial
condition with short enough initial correlation length $R_s(Y_0)$ all
further evolution is governed by contributions on perturbative scales
with most contributions arising near $R_s(Y) < R_s(Y_0)$. Under these
conditions the integrand vanishes exponentially near the Landau pole
and one may argue on physics grounds that contributions to evolution
at the Landau pole may be neglected. In practice this is achieved by
introducing some form of regulator. While at asymptotically high
energies where $Q_s \gg 1\ \gev$ the precise choice of regulator
\emph{cannot} affect evolution, real data are far from that region and
the choice of regulator will impact any comparison with
experiment. For definiteness we choose what is known as an APT
regulator, see for
example~\cite{Shirkov:1997wi,Solovtsov:1999in}\footnote{APT stands for
  Analytic Perturbation Theory. We use only its treatment of the
  Landau pole a practical way to regulate the running coupling.}, and
simply subtract the Landau pole from the expression for the running
coupling. The advantage of using an APT regulator over the other
possibilities is that it produces a very smooth behavior of the
effective running coupling $R_{\text{eff}}$ even if it depends on all
three scales $\mu_r$, $\mu_{r_1}$ and $\mu_{r_2}$. If one were to
simply freeze the individual couplings below some fixed scale in
$R_{\text{eff}}$ of Eq.~(\ref{eq:quark_run}) the result would be
discontinuous. This is partially compensated by a small ${\cal K}$ in
the evolution equation as long as energy conservation corrections are
ignored. The stability of the iteration procedure used to implement
the energy conservation correction described in
App~\ref{sec:energy-conservation}, however, is strongly reduced by
such discontinuities. The restriction to transversely infinite media
is conceptually more serious: The JIMWLK framework does not correctly
describe the transverse growth of a finite target. Near the edge,
gluon densities become small and the evolution equations match up with
the BFKL equation, unphysical Coulomb tails are no longer shielded by
a density induced correlation length and nonperturbative contribution
start to dominate. For this reason one usually uses JIMWLK equations
to describe the in medium behavior and models any size effects. The
simplest possible treatment for dipole cross section would be to
postulate a separation of $\bm r$- and $b$-dependence in simple
correlators like the dipole amplitude of Eq.~(\ref{eq:dipole-cross})
and introduce a profile function $T(\bm b)$ to model the dipole cross
section as
\begin{align}
  \label{eq:fact-dipole-cross}
  \sigma_{q\Bar q}(Y,(\bm x-\bm y)^2) = \sigma_0 \ N^{q\Bar q}_{Y; \bm x \bm y}
\hspace{.4cm}\text{with}\hspace{.4cm}
   N^{q\Bar q}_{Y; \bm x \bm y} = N^{q\Bar q}_{Y; \bm x-\bm y}
\hspace{.4cm}\text{and}\hspace{.4cm}
\sigma_0 := 2 \int d^2\bm{b}\ T(\bm b) 
\ .
\end{align}
With this assumption, all the nonperturbative information is encoded
in a single constant $\sigma_0$, which is then matched to data.  While
such a treatment is adequate for a LO treatment of the total cross
section, we will see below that such an approach induces uncertainties
already at leading order once more sophisticated observables are
considered, rapidity gap events and the cross sections
of~\cite{Marquet:2010cf} are among those. See
Sec.~\ref{sec:nonp-infl-b-dep} and the discussion leading up to it.

The numerical impact of NLO corrections is best illustrated by
studying its impact on evolution speed $\lambda$, defined as
\begin{align}
  \label{eq:lambda-def}
  \lambda(Y):= -\frac1{\pi} \int \frac{d^2\bm{r}}{\bm r^2}\ \frac{d}{d
    Y} S_{Y; \bm x\bm y}^{q\bar{q}} \ .
\end{align}
This definition is equivalent to the ``naive'' definition of $\lambda$
as the rate of change of the saturation scale, $\lambda(Y)=\frac{d}{d
  Y} \ln(Q_s^2(Y))$, wherever strict scaling
holds~\cite{Iancu:2002tr}, but provides a useful generalization
wherever strict scaling does not hold: This includes the
pseudo-scaling case encountered as one steps beyond
LO~\cite{Rummukainen:2003ns}. Where it becomes necessary to plot
energy dependent quantities we often choose to plot against $R_s(Y)$
as the intrinsic $Y$-dependent scale of our simulations. For
simplicity we take $R_s(Y)$ from
\begin{align}
  \label{eq:RsYdef}
  S_{Y,\bm{xy}}^{q\bar{q}}(|\bm{r}|=R_s(Y))=\frac{1}{2}\ .
\end{align}

\begin{figure}[htb]
 \centering
 \includegraphics[width=.325\linewidth]{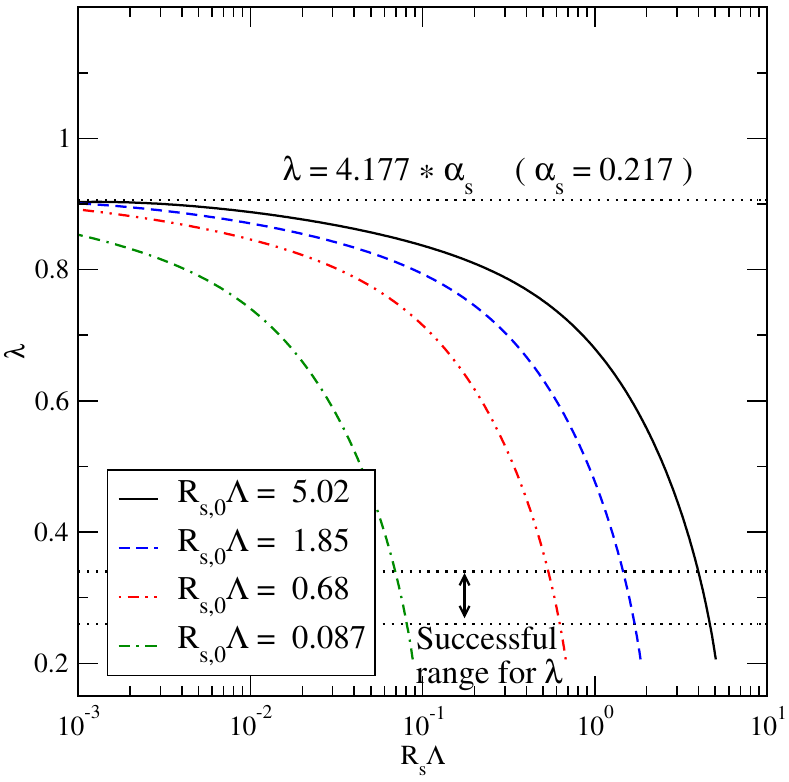}
 \hfill
 \includegraphics[width=.325\linewidth]{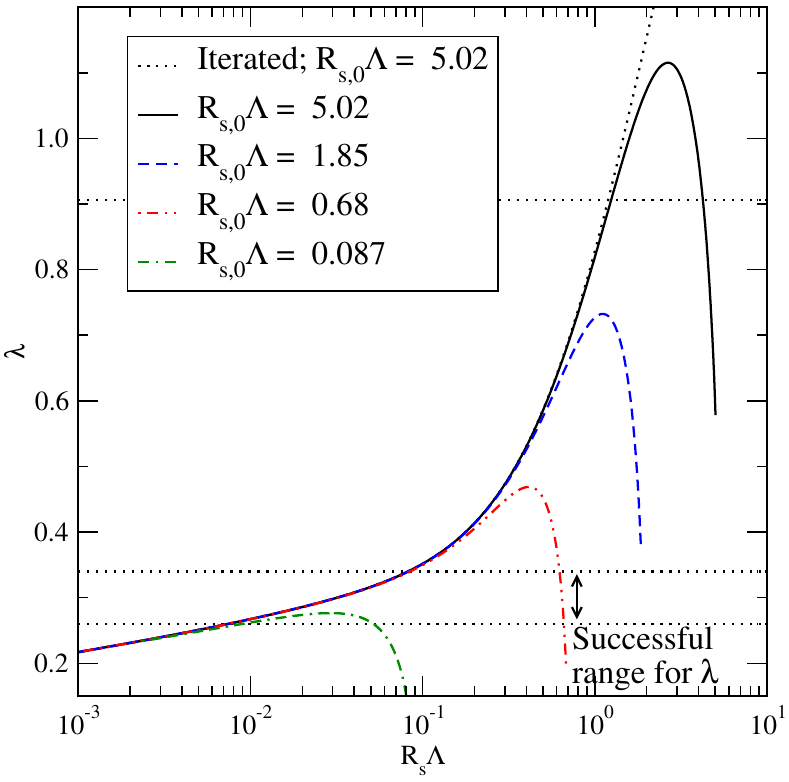}
 \hfill
 \includegraphics[width=.325\linewidth]{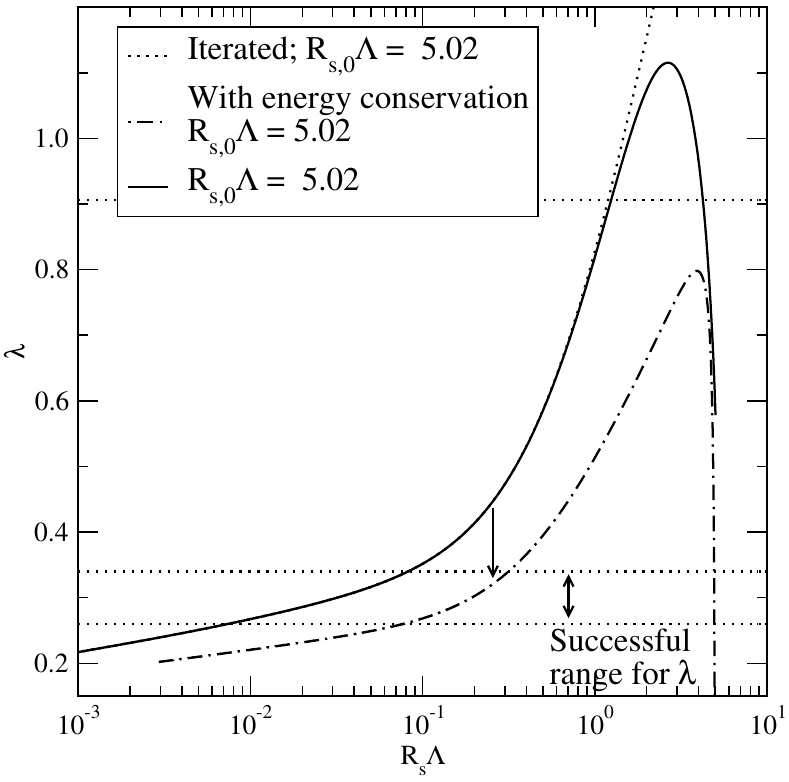}
 \caption{\textbf{Left:} LO i.e. fixed coupling BK evolution:
   asymptotically the evolution speed $\lambda \rightarrow
   \text{\bf{const}}\times\alpha_s$ for any relevant initial state.
   \textbf{Middle:} NLO BK evolution: the running coupling slows down the
   evolution. After the initial state effects are erased, the
   evolution speed settles on the same asymptotic line. \textbf{Right:}
   DGLAP type NLO corrections slow down the evolution further.}
 \label{fig:nlo_illustrated}
\end{figure}

Fig.~\ref{fig:nlo_illustrated} shows the behavior of the evolution
speed as one incorporates NLO corrections: At leading order, i.e. with
fixed coupling and without the energy conservation correction
(Fig.~\ref{fig:nlo_illustrated}, left) different initial conditions in
the course of evolution towards higher $Y$ (moving along the curves
from right to left in the figures, from large $R_s(Y)$ to small
$R_s(Y)$) all merge up with the asymptotic scaling regime in which
$\lambda$ becomes a constant. As running coupling is turned on
(Fig.~\ref{fig:nlo_illustrated}, middle), true scaling turns into
pseudo-scaling: even the asymptotic speed remains $Y$-dependent. For
$R_s(Y)\Lambda < 1$, evolution is slowed down drastically. Adding our
last NLO ingredient (the energy conservation correction) leads to
further slowdown as shown in Fig.~\ref{fig:nlo_illustrated} on the
right. For comparison, all the panels in
Fig.~\ref{fig:nlo_illustrated} show the $\lambda$ range that leads to
successful fits in our own fits and the scaling models of GB-W and
IIM. One concludes that NLO corrections to evolution (running coupling
contributions in particular) are an essential ingredient to a
successful fit to HERA data.

In preparation to the discussion of the fit procedure below let us
note that we have intentionally plotted evolution speeds against
$R_s(Y) \Lambda$, where, at this point, $\Lambda$ is the QCD scale in
the coupling. In the fits below, however, we will allow $\Lambda$ to
vary.  The reason for this is twofold: we have observed already that
such a change of scale can absorb most of the differences between
different coupling separation schemes, beyond that one may also think
of this a resummation of nonperturbative effects in the initial
condition that affects both the values of $R_s(Y)$ at all $Y$ and the
associated evolution speeds $\lambda(Y)$ (through the size of the
coupling). If the energy conservation correction is omitted as
in~\cite{Albacete:2009fh}, the latter is main justification to treat
the scale factor as a fit parameter. In any case, it is through this
fit procedure that we set the overall scale of $R_s(Y)$ in physical
units.\footnote{\cite{Albacete:2009fh} fix the value of $\Lambda$ and
  introduce a scale factor c to achieve the same goal.}

\begin{figure}[htb]
  \centering
  \includegraphics[width=.4\textwidth]{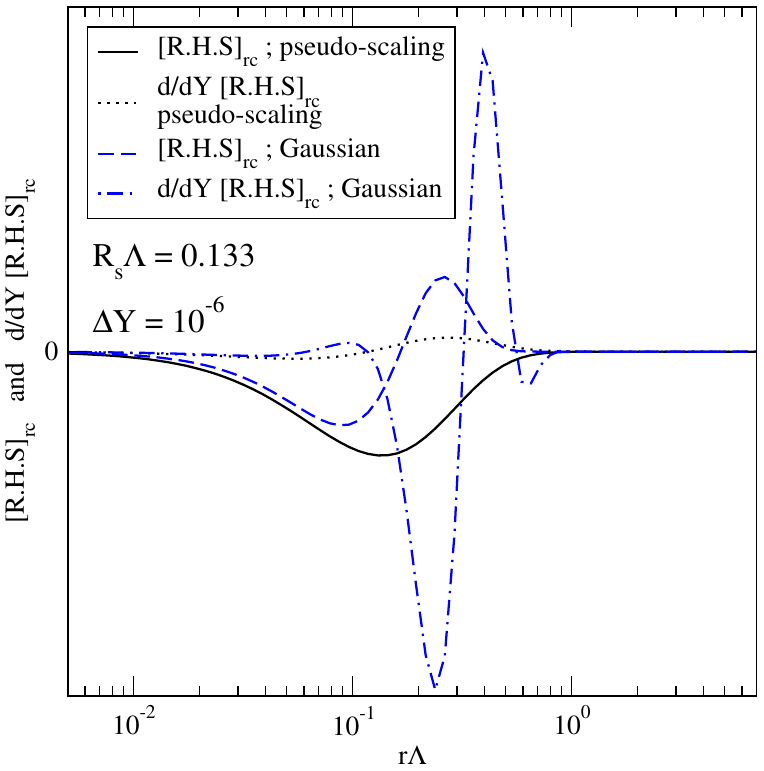}
  \hfill 
  \includegraphics[width=.4\textwidth]{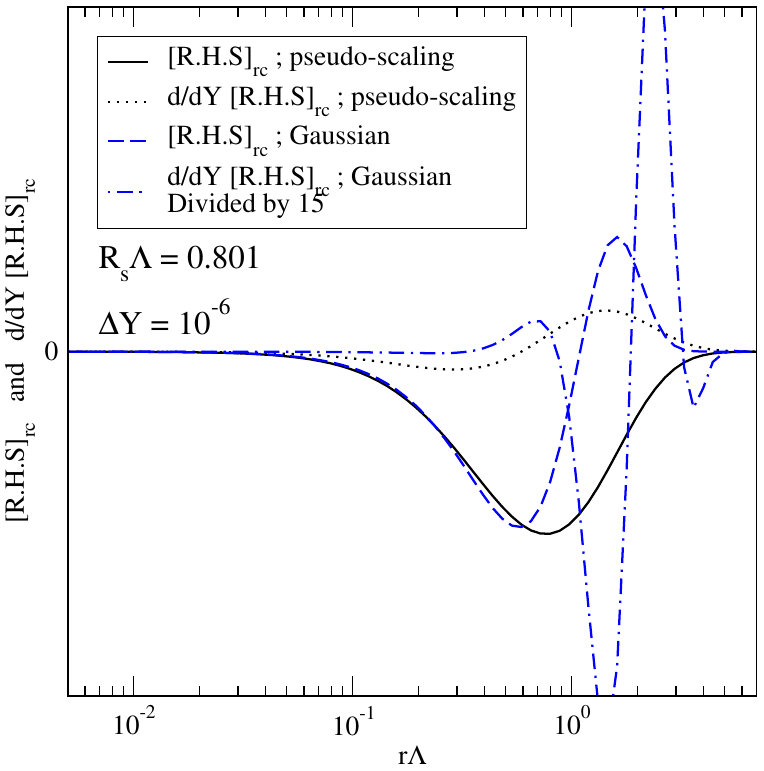}
  \caption{ Relative size of the energy conservation correction: the
    contributions to the r.h.s. of Eq.~\eqref{eq:GT_modified} at a
    fixed $R_s(Y)$ (left: $1/R_s\ll \Lambda$, right: $R_s$ near
    $\Lambda$) are split up into the contribution without the energy
    conservation correction, labeled $[\text{R.H.S.}]_{\text{rc}}$ (it
    contains \emph{only} the running coupling corrections) and the
    energy conservation correction
    $\frac{d}{dY}[\text{R.H.S.}]_{\text{rc}}$. In the scaling regime,
    the the energy conservation correction is subleading. Away from
    the scaling regime (exemplified by a Gaussian correlator shape at
    the same correlation length) the energy conservation correction
    dominates.}
  \label{fig:energy-cons-corr}
\end{figure}

Evolution with NLO effects included allows us an additional cross
check on the self consistency of our tools: For the calculation to be
self consistent, there should be a clear hierarchy of size between
leading and subleading contributions, unless the subleading
contributions introduce a qualitatively new feature. This does apply
to our discussion of evolution speed near the scaling regime: a new
qualitative feature (scale breaking and the appearance of running
coupling effects) change evolution speeds dramatically, while the
effect of energy conservation corrections induce only minor
corrections. This, however, is not universal: the relative size of the
energy conservation corrections compared to evolution without it
strongly depends on the shape of the solutions. Near or in the
pseudo-scaling regime the energy conservation corrections have a
relatively small effect. Far from the pseudo scaling regime, the
energy conservation correction dominates the r.h.s of the evolution
equation over a large range of $R_s(Y)$ scales. This difference in
behavior is shown in Fig.~\ref{fig:energy-cons-corr}.
It is accompanied by severe numerical stability problems away from the
pseudo-scaling region: We have employed iterative methods to obtain
the derivative term as well as backwards difference methods (once two
or more adjacent time steps are known with sufficient accuracy) and
both require step sizes $\Delta Y$ beyond anything even remotely
practical to stabilize the numerical results\footnote{This is also the
  reason why we have chosen an approximate iteration procedure
  outlined in App.~\ref{sec:energy-conservation} to construct the
  solution in the pseudo-scaling region.}.

Both these observations, the dominance of the energy conservation
correction and the numerical stability, lead us to believe that the
evolution equation in its present form is less reliable far away from
the scaling regime. It would appear that additional resummations are
necessary to reliably address the region far from scaling. The nature
of such corrections is not known at present. We therefore advocate the
use of the pseudo-scaling regime in data comparisons as long as fits
in this region are at all possible.

\section{Lessons from the total cross section}
\label{sec:lessons-from-total-cross-section}

\subsection{General features}
\label{sec:gen-feat}

In the following we will first confront the data with evolution
assuming three light quarks with current quark masses $\lesssim 5\
\mev$. This is sufficient to discuss the main features of the fit and
any fit tensions. Fits with only a single mass are much faster to do
and hence a more efficient tool to clarify such systematic questions.

After scouting the terrain in this manner we verify that the inclusion of
quark masses will not change our conclusions and explore which phase space
ranges are most affected by the inclusion of mass effects.

\begin{figure}[htb]
 \centering
 \includegraphics[width=0.40\linewidth]{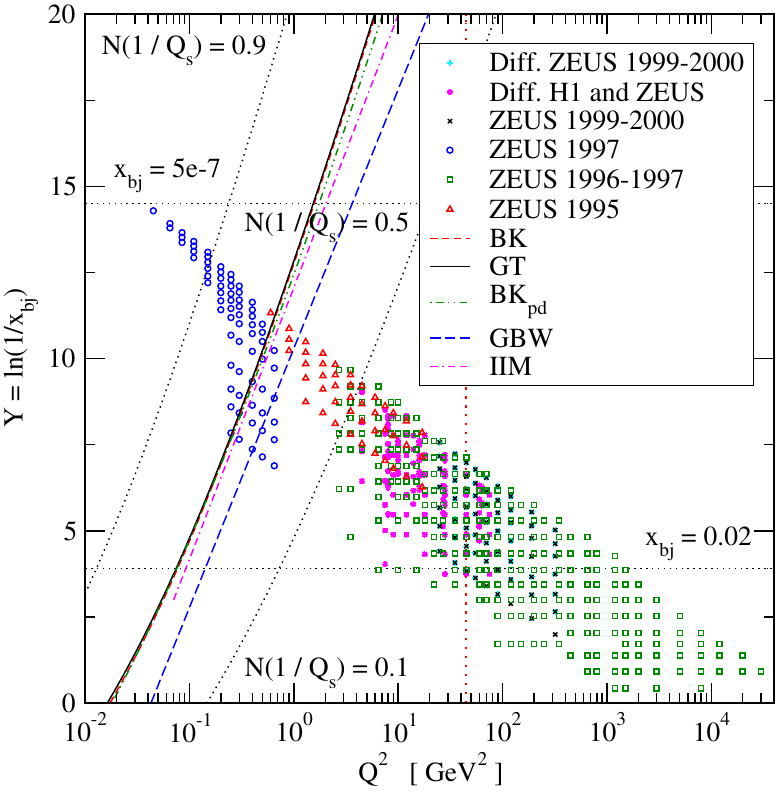}
 \hfill
 \includegraphics[width=0.40\linewidth]{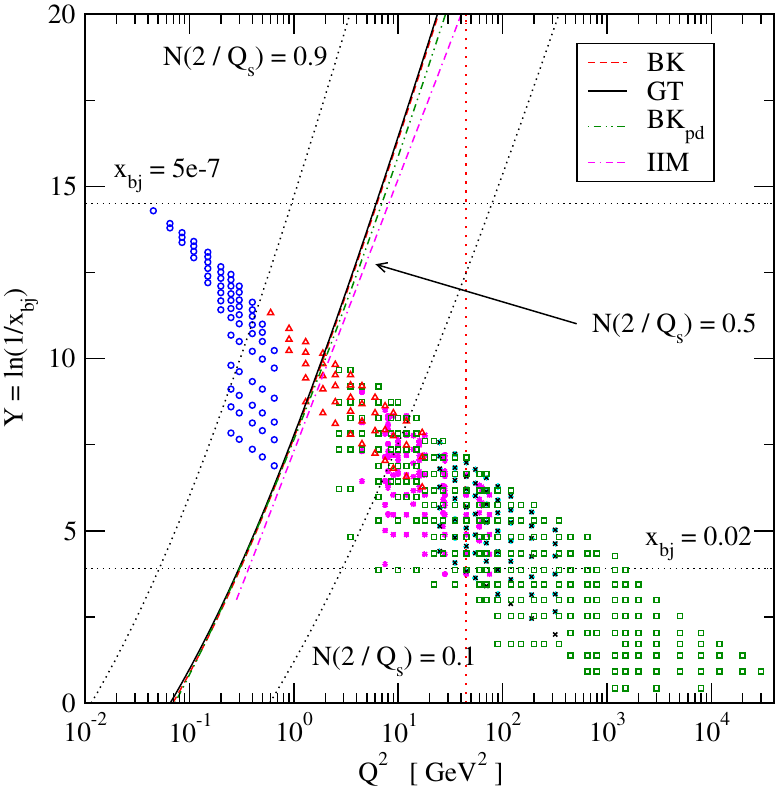}
 \caption{\textbf{Left:} The phase phase of all (inclusive and diffractive) HERA
   data included in fits together with the saturation momenta $Q_s^2(Y) =
   1/R_s^2(Y)$ obtained from different fit approaches. The horizontal lines at
   $x=5\times10^{-7}$ and $x=0.02$ indicate the small $x$ range used in the
   fits. In diffractive fits no restrictions were used. \textbf{Right:} To
   illustrate that slight changes of scale definitions strongly affect the
   appearance of this plot we replace $Q_s^2(Y) = 1/R_s^2(Y)$ by
   $Q_s^2(Y) = 4/R_s^2(Y)$.}
 \label{fig:qs_on_hera_phase}
\end{figure}

One generic feature is shared by \emph{all} fits we have undertaken:
the values of $R_s$ appear to be well constrained by data, an
impression of this is shown in Fig.~\ref{fig:qs_on_hera_phase} which
shows alternatively $Q_s(Y)=1/R_s(Y)$ and $Q_s(Y) = 4/R_s(Y)$ as they
emerge from the fits overlaid on the data used. Sloped dotted lines
indicate where the $q\Bar q$-dipole amplitude crosses $0.1$ and $0.9$
respectively. Details on these fits will be given
below. With $1/R_s(Y)$ of the order of $1\ \gev$ one would expect that
nonperturbative contributions to evolution are inevitable, but to
judge nonperturbative influence in a meaningful way, we need a more
detailed analysis. Below we will comment on two aspects: IR effects in
the evolution and phase space features in the case of mass effects.

The second feature that is quite well constrained by data,
irrespective of details of the theoretical input, is the slope of
$R_s(Y)$ in Fig.~\ref{fig:qs_on_hera_phase}, or alternatively the
evolution speed $\lambda(Y)$. Experience from GB-W and IIM model fits
with $\lambda = .31$ and $\lambda = .29$ respectively as well as our
fit experience with solutions from the evolution equations all
establish a viable range for $\lambda(Y)$ that falls into the
successful fit range indicated in Fig.~\ref{fig:nlo_illustrated}. If
this is generic, fits without energy conservation corrections will
necessarily have smaller $R_s \Lambda$ values than fits that include
such a contribution to evolution (Fig.~\ref{fig:nlo_illustrated},
right). Since the physical values of $R_s(Y)$ are very well
constrained this implies that fits without the energy conservation
correction require a noticeably smaller value for $\Lambda$ that fits
that include energy conservation corrections. The former get the
contributions from a region with smaller coupling and are less
sensitive on the choice of IR regulator for the Landau pole. The
considerably smaller $\Lambda$ values necessary without the energy
conservation corrections, however, would indicate a larger
nonperturbative resummation entering the evolution equation: none of
the alternatives is free of nonperturbative effects.

In an actual fit with solutions of an evolution equation, one needs to choose
an initial condition, in our case always of the form of the GB-W model,
specify the initial correlation length in units of $\Lambda$ and evolve this
initial condition in $Y$. As we proceed solving the equation the correlator
shapes change away from the GB-W form and approach the asymptotic or
pseudo-scaling regime. $R_s(Y)$ shrinks and $\lambda$ traces a trajectory as
indicated in Fig.~\ref{fig:nlo_illustrated}. For any given trajectory one then
needs to relate the rapidity variable of the simulation to the physical
rapidity by determining what we call $Y_{\text{offset}}$ or $Y_{\text{off}}$
to pinpoint $Y=0$.\footnote{Since we only start compare to data for $x \leq
  2\cdot 10^{-2}$ we may well obtain negative values of $Y_{\text{off}}$ and
  still have dipole cross sections to cover all the data range
  considered. We have no ambitions to extend or parametrizations to larger
  $x$.}  On a given trajectory, $Y_{\text{off}}$ is most closely linked to
evolution speed.  The fit of $Y_{\text{off}}$ is done simultaneously with a
fit of the physical units on $R_s(Y_0=-\ln(2\cdot 10^{-2}))$ by varying
$\Lambda$, and a fit the overall normalization $\sigma_0$ from
Eq.~(\ref{eq:fact-dipole-cross}).

A fit in the pseudo-scaling region, where the shape of the dipole
amplitude is fully determined by the nonlinearity with details of the
initial condition erased everywhere but in the extreme UV, the fit is
therefore a three parameter fit in terms of $Y_{\text{off}}$,
$\Lambda$, and $\sigma_0$.

Away from the asymptotic pseudo-scaling region many additional
features of the initial condition survive and may affect the quality
of the fit. At present we have no systematic tools to scan the space
of initial conditions and very little constraints from theory. Our
efforts below are meant to shed some first light on the issue using a
very hands on attitude that is limited in scope mostly by the cost of
creating a single trajectory on the one hand and the search of fit
parameters for a given trajectory on the other.

\subsection{Systematics from fits with light quarks, energy conservation
  included}
\label{sec:syst-from-fits-light-quarks-energy-cons}

\begin{figure}[htb]
 \centering
 \includegraphics[width=.328\linewidth]{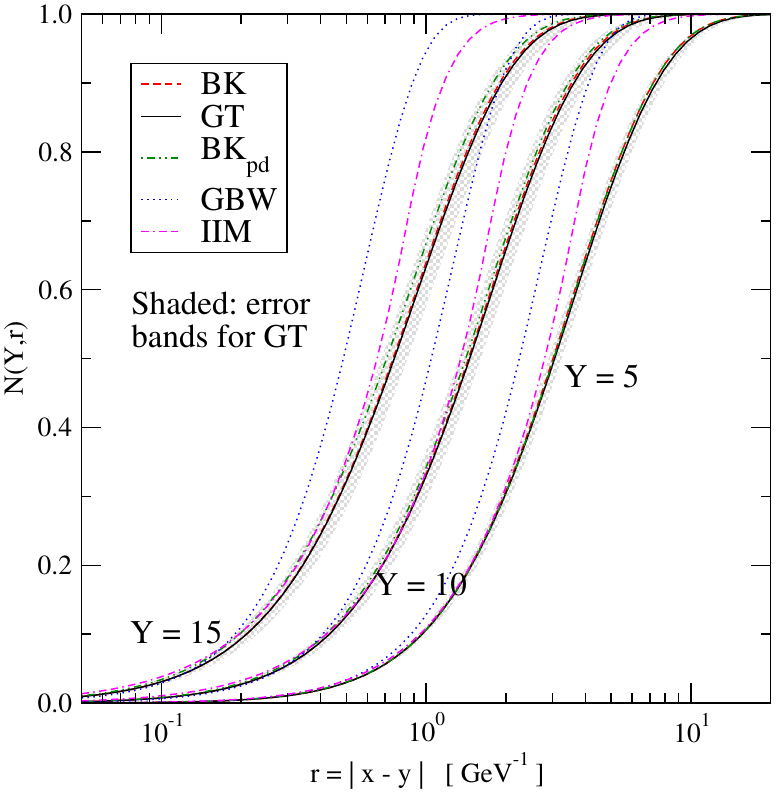}
 \hfill
 \includegraphics[width=.328\linewidth]{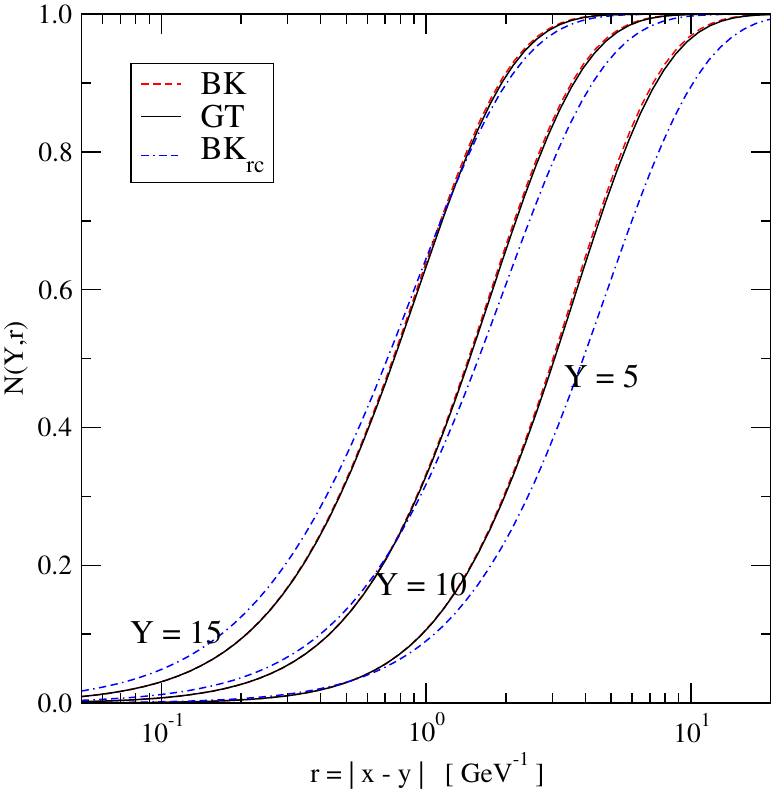}
 \hfill
 \includegraphics[width=.328\linewidth]{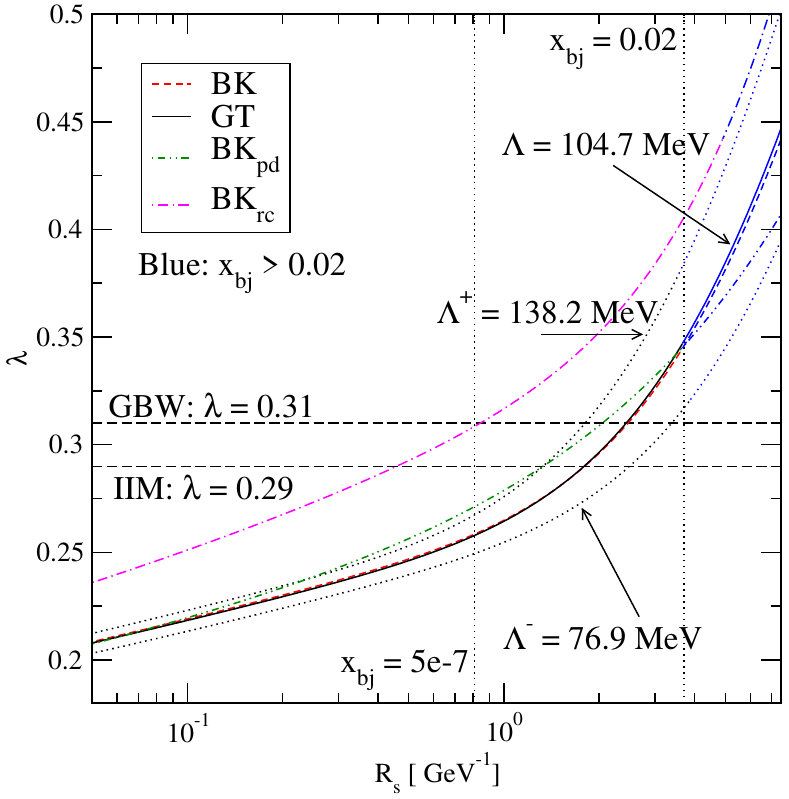}
 
 \caption{\textbf{Left:} The evolution of $N^{q\Bar q}_Y(r)$ in different
   approximations and models. The data align JIMWLK based descriptions
   and the IIM model in the region $N^{q\Bar q}_Y(r)\lesssim 0.4$ --
   the GB-W model deviates significantly.  \textbf{Middle:} The
   asymptotic (perturbatively imposed) solutions of the evolution
   equations are steeper after the energy conservation corrections are
   included. This benefits the fit quality overall \emph{and} at large
   $Q^2$ in particular.  \textbf{Right:} The evolution speed $\lambda$
   for all JIMWLK approximations considered as a function of
   correlation length $R_s$.  Error bands are extracted from the
   condition $\chi^{2}/\text{dof} = 1$, see
   Appendix~\ref{sec:definitions}. }
 \label{fig:rs_vs_lambda}
\end{figure}

Before we turn to fit quality, let us first collect the main features
the data fits impose on correlator shapes and evolution speed in
Fig.~\ref{fig:rs_vs_lambda}.

We have already stressed at the outset of
Sec.~\ref{sec:lessons-from-total-cross-section}, the most tightly
constrained feature in all the fits is $R_s(Y)$ in physical
units. This also manifests itself in the fact that after the fit is
performed, the shapes of dipole amplitudes agree very well within an
order of magnitude below $R_s(Y)$ no matter if the calculation is
based on an evolution equation or a model. This is shown in
Fig.~\ref{fig:rs_vs_lambda}, on the left panel for BK, GT, BK with
parent dipole running (labeled $\text{BK}_{\text{pd}}$) and the two
models GB-W and IIM. The models show noticeable deviations from the
fits based on the pseudo-scaling solutions from evolution equations at
$|\bm r| \gtrsim R_s(Y)$.

We had already seen in Fig.~\ref{fig:nlo_illustrated} that the main
effect of the energy conservation correction is a further slowdown of
evolution above that induced by the running of the coupling. The
middle panel in Fig.~\ref{fig:rs_vs_lambda} illustrates the second
major impact by comparing BK and GT fits (which include
the energy conservation correction) and an asymptotic BK fit without
this correction labeled $[\text{BK}]_{\text{rc}}$: correlator shapes
in the asymptotic regime stay \emph{steeper} than in the corresponding
evolution with the energy conservation correction omitted. 

Evolution speeds corresponding to the fits are shown in the rightmost
panel in Fig.~\ref{fig:rs_vs_lambda}. We indicate an associated error
band only for the GT fit. It encloses all fits that include the energy
conservation correction, but is clearly separated from the fit without
the energy conservation corrections $[BK]_{\text{rc}}$, complementing
the shape deviation already observed. The two models, by construction,
have constant evolution speeds that fall near the average of evolution
speeds obtained from the BK and GT fits, confirming our expectations.

Table~\ref{tab:inclusive_results} provides an assessment of the
quality of the fits illustrated previously in terms of dipole
amplitudes and evolution speeds in Fig.~\ref{fig:rs_vs_lambda}. The
table reflects our fit strategy of first obtaining a fit at low
$Q^2\le 45\ \gev^2$ which we then attempt to extend to a larger $Q^2$
range up to $1200\ \gev^2$, monitoring any change in fit parameters
required in the process. In the larger $Q^2$ range the models (GB-W
and IIM) are known to fail, they were designed only with small $Q^2$
values in mind, but also the small $x$ approximation employed in the
evolution equations will have to break down eventually -- even with
partial resummations like the energy conservation correction built
in. The success of such an extension is a measure as to how
efficiently the resummations recapture large $Q^2$ effects.
\begin{table}[!thb]
 \centering
\resizebox{\textwidth}{!}{
  \begin{tabular}{@{}l>{\hspace{1mm}}lllllll@{}}
    $\xbj\leq 0.02$ &   & $\text{BK}_{\text{pd}}$ & $\text{BK}$  & $\text{GT}$ & $\text{GB-W}$ & $\text{IIM}$ & $[\text{BK}]_{\text{rc}}$\\ [4mm]
        & \begin{minipage}[t]{2.2cm}
      $\lambda(\xbj)$\\
      \scriptsize $\xbj\in [\frac{5}{10^{7}},0.02])$
    \end{minipage}
    & 0.27-0.34 & 0.26-0.35 & 0.26-0.35 & 0.31 & 0.29& 0.31-0.44  \\[5mm] 
    &  $\Lambda$ $[ \mev ]$  & $82.4_{-24.4}^{+31.7}$ & $93.7_{-25.1}^{+30.7}$  & $104.7_{-27.8}^{+33.5}$ & $x_0=\frac{1.1}{10^{3}}$ & $x_0=\frac{1.1}{10^{4}}$ & 50.4 \\ [3mm]
    $Q^2 \leq$     &  $\chi^{2}/\text{224}$ & 0.818 & 0.811 & 0.810 & 1.401  & 0.828 & 1.760 \\ 
    $45\ \gev^2$   &  $\sigma_{0}$ $[\gev^{-2}]$ & 54.01 & 55.05 & 55.33 & 44.59  & 51.47 & 56.84  \\ [3mm]
    $Q^2 \leq$     &  $\chi^{2}/\text{295}$ & 0.979 & 0.812 & 0.805 & 1.978 & 1.037 & 4.783  \\ 
    $1200\ \gev^2$  &  $\sigma_{0}$ $[ \gev^{-2} ]$ & 53.69 & 55.01 & 55.35 & 44.48 & 51.08 & 54.79 \\ 
  \end{tabular}
}
\caption{Fit results to inclusive data
  from~\cite{Breitweg:2000yn,Breitweg:1998dz,Chekanov:2001qu,Chekanov:2008cw}. $\chi^2/\text{dof}$ is below one for a wide range of $\Lambda$ values indicated by the errors listed, see Fig.~\ref{fig:LAMBDA_vs_chi}.} 
\label{tab:inclusive_results}
\end{table}

The best fits are obtained when we use the NLO evolution equations
once energy conservation is included (columns labeled BK and GT as well as
BK$_{\text{pd}}$):\footnote{All these fits are performed with light
  quarks only, the role of physical quark masses is discussed in
  Sec.~\ref{sec:quark-masses}.} The fit quality is excellent
over both the small and the large $Q^2$ range. Already the low $Q^2$
range ($Q^2 < 45\ \gev^2$) determines both the physical units of
$R_s(Y)$ and the ideal evolution speed $\lambda(Y)$ (via
$Y_{\text{off}}$). To extend the fit to the full $Q^2$ range covered
by the data below $x=2\cdot 10^{-2}$ only $\sigma_0$ needs to be
readjusted. 
Note the \emph{excellent} $\chi^2/\text{dof}$ values over the whole
$Q^2$-range for both BK and GT truncations in the pseudo-scaling region.

An asymptotic fit with the energy conservation omitted is clearly
unworkable: its $\chi^2$-value is barely acceptable already for
$Q^2\le 45\ \gev^2$ and indicates an outright failure in the broader
$Q^2$ region up to $1200\ \gev^2$ -- it fares worse than the models
(if extrapolated into this region) by far. The origin for this is the
shallower shape of the dipole amplitudes observed in
Fig.~\ref{fig:rs_vs_lambda} (middle panel). The data require steeper
correlators and lower evolution speeds. Evolution speeds do slow down
as evolution proceeds to smaller $R_s(Y)\Lambda$, however, at the same
time correlators flatten out further to asymptotically approach the
shapes of the fixed coupling case as the running of the coupling slows
down with shrinking $R_s(Y)\Lambda$ values. This leaves a very small
window for $\Lambda$ (viewed as a fit parameter) in which the
correlators are still tolerably steep, but evolution speed is already
small enough. As a result all features of this fit deviate from those
shown in the left panel of Fig.~\ref{fig:rs_vs_lambda}: evolution
speed is still quite large, and the match of the dipole correlators
below $R_s$ observed in the left panel of Fig.~\ref{fig:rs_vs_lambda}
is lost as well.

The success of the asymptotic fit with energy conservation included
could be interpreted as an indication that the inclusion of the energy
conservation correction gives a better match to the perturbative
anomalous dimensions that govern the large $Q^2$-behavior. This is in
contrast to the preasymptotic fit
of~\cite{Albacete:2009ps,Albacete:2009fh} (which omits energy
conservation corrections) where relics of the steepness of the initial
condition --non-perturbative in nature-- allow for a good fit quality
at all $Q^2$. We will provide more details on this comparison in
Sec.~\ref{sec:pre-as-examples}.

The models, GB-W and IIM, are clearly limited to a smaller $Q^2$
region. This holds even for the IIM model, which does incorporate
additional perturbative information in the form of BFKL anomalous
dimensions.

\begin{figure}[htb] \centering
 \includegraphics[width=.342\linewidth]{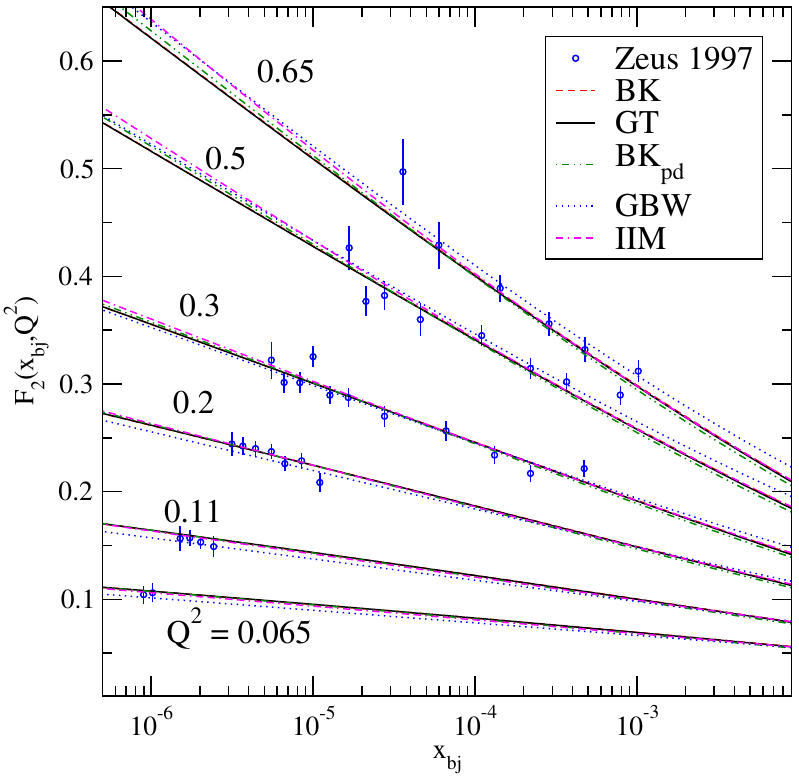}
 \hfill
 \includegraphics[width=.32\linewidth]{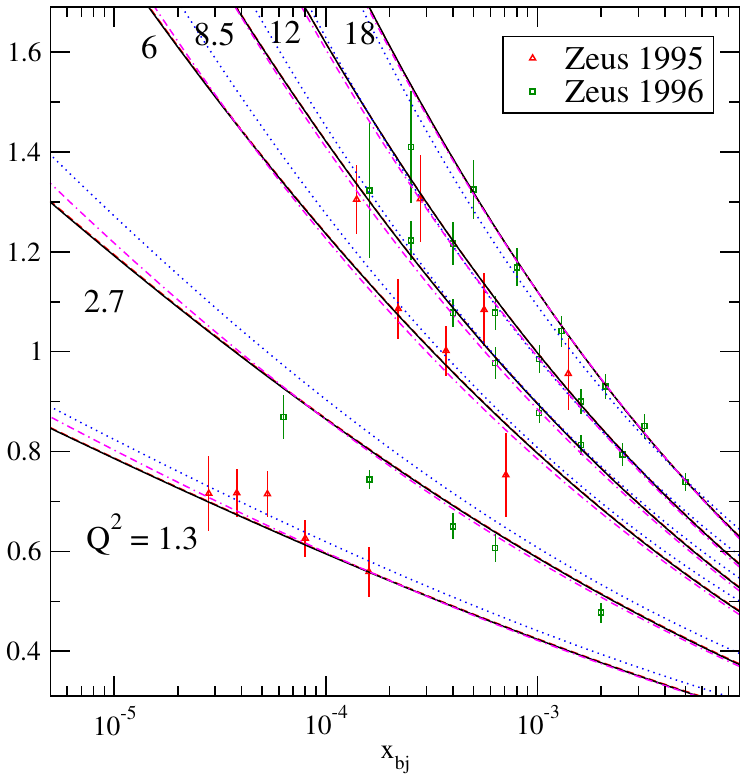}
 \hfill
 \includegraphics[width=.32\linewidth]{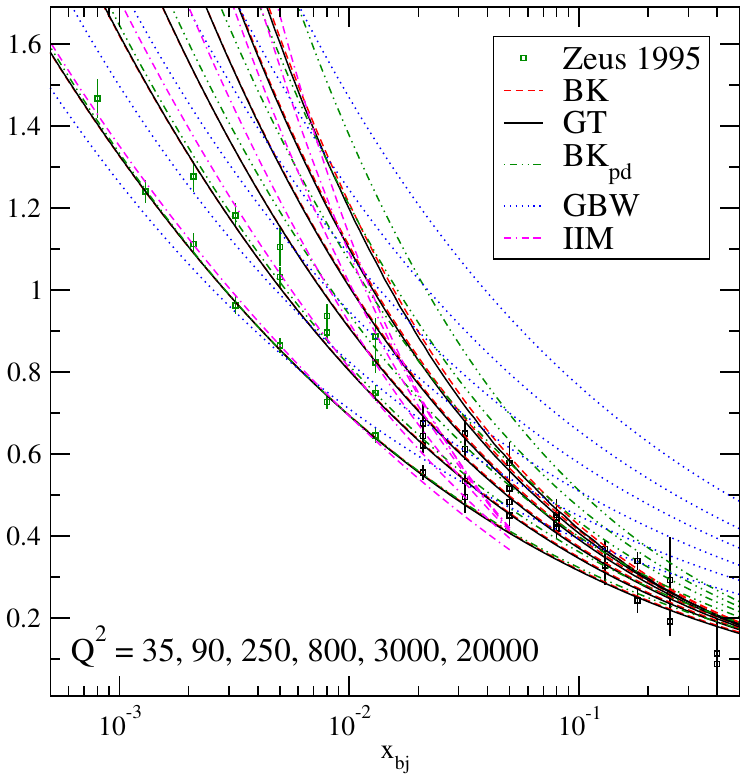}
 \caption{$F_2(x,Q^2)$ as a function of $x$ from a fit with light quarks
   compared to a subset of data .  Different descriptions are qualitatively
   the same at low and moderate values of $Q^2$ (on the left and middle
   respectively). At high $Q^2$ (on the right) only the BK and GT fits compare
   well to the data.}
 \label{fig:subsets_f2}
\end{figure}

\begin{figure}[htb]
 \centering
 \includegraphics[width=.342\linewidth]{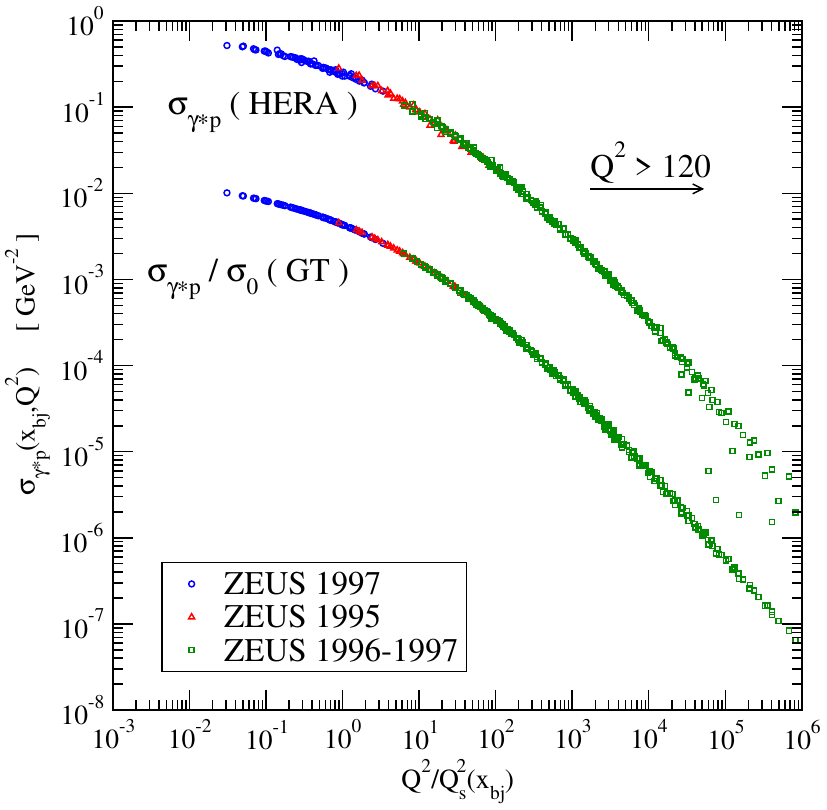}
 \hfill
 \includegraphics[width=.32\linewidth]{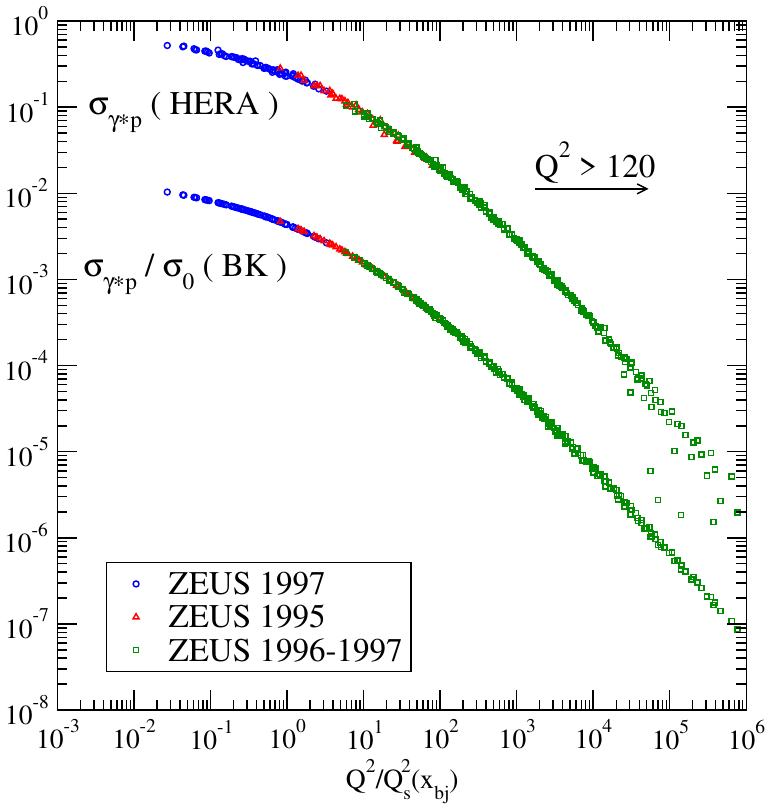}
 \hfill
 \includegraphics[width=.32\linewidth]{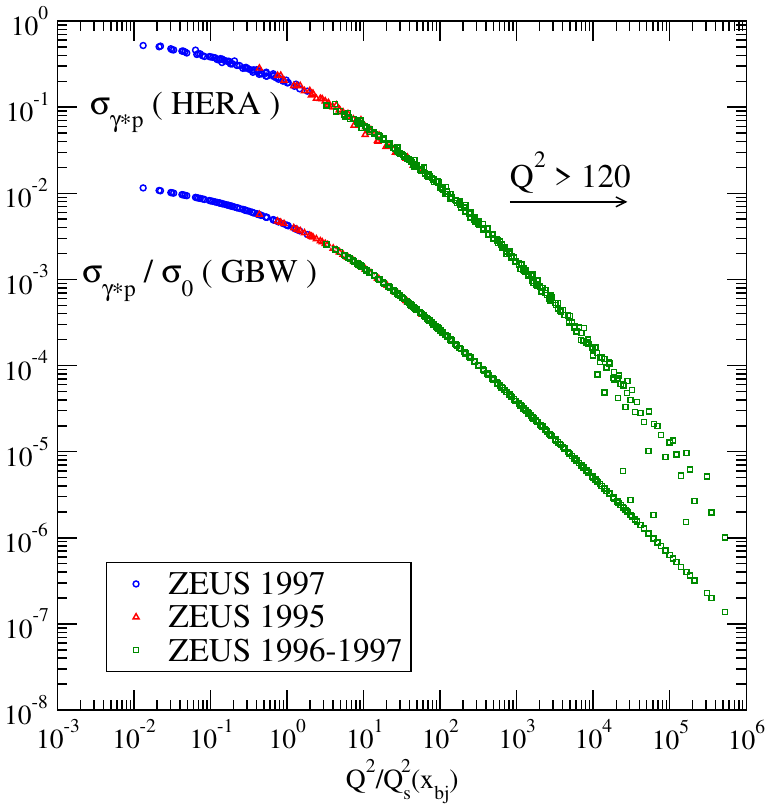}
 \caption{The $\gamma^{*}p$ cross section as a function of scaling
   variable $Q^{2}/Q_{s}^{2}$. The plots show both data and
   theoretical predictions (shifted downwards by $\sigma_0$ to avoid
   clutter).  \textbf{Left:} GT.  \textbf{Middle:} BK.  \textbf{Right:} The
   GB-W-model. Data
   from~\cite{Breitweg:2000yn,Breitweg:1998dz,Chekanov:2001qu}.}
 \label{fig:cross_sections}
\end{figure}

\begin{figure}[p]  \centering
 \includegraphics[width=0.939\linewidth]{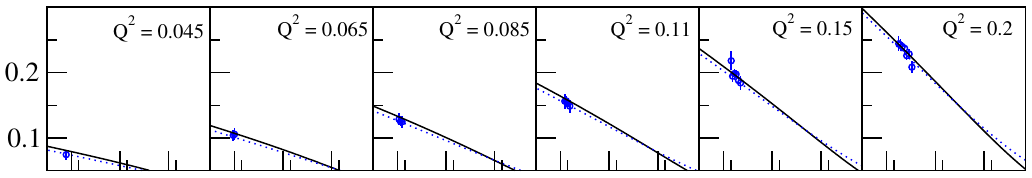}
 \vspace{-0.22cm}
\\
 \includegraphics[width=0.939\linewidth]{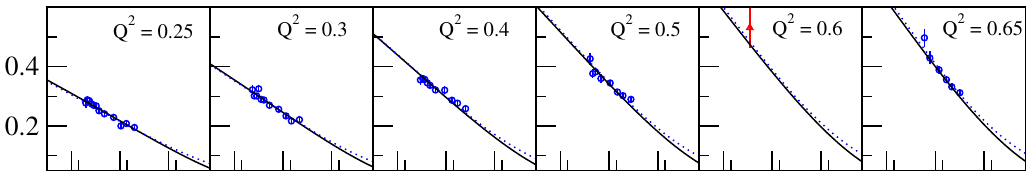}
 \vspace{-0.22cm}
\\
 \includegraphics[width=0.939\linewidth]{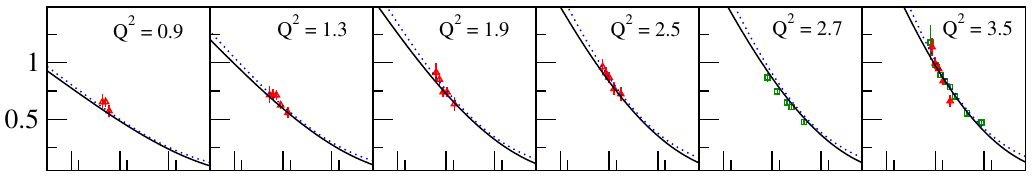}
 \vspace{-0.22cm}
\\
 \includegraphics[width=0.939\linewidth]{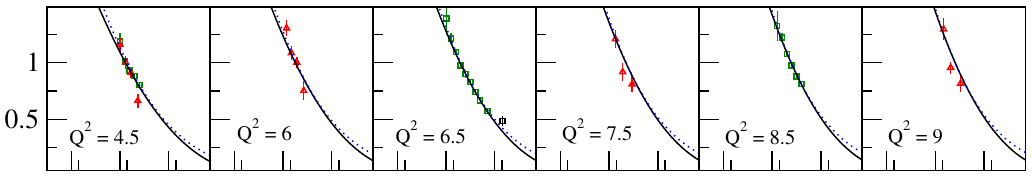}
 \vspace{-0.22cm}
\\
 \includegraphics[width=0.939\linewidth]{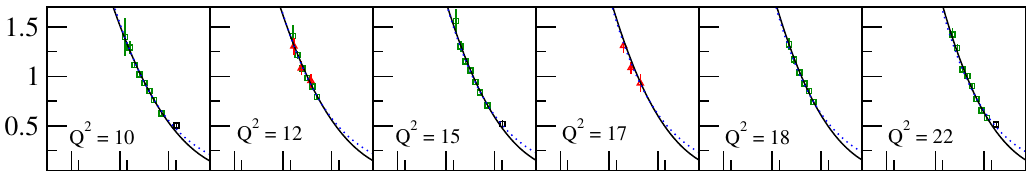}
 \vspace{-0.22cm}
\\
 \includegraphics[width=0.939\linewidth]{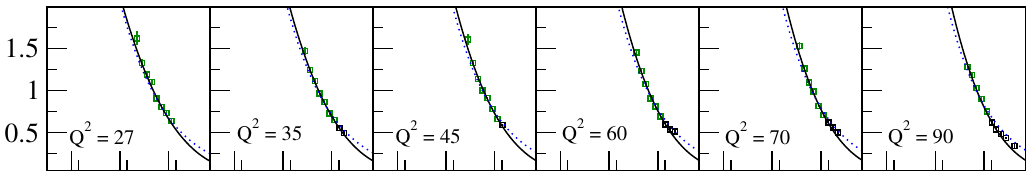}
 \vspace{-0.22cm}
\\
 \includegraphics[width=0.939\linewidth]{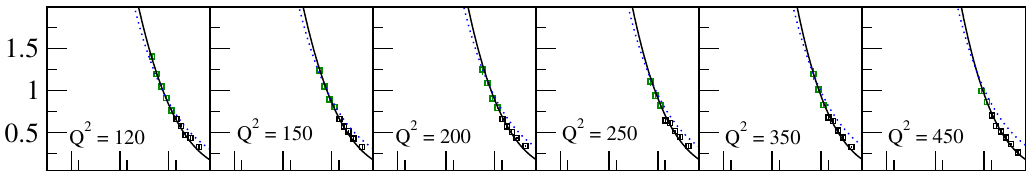}
 \vspace{-0.22cm}
\\
 \includegraphics[width=0.939\linewidth]{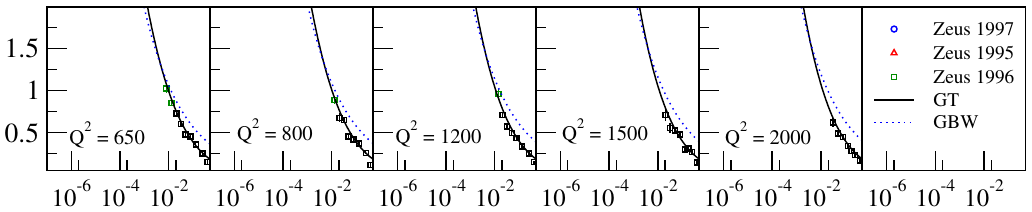}
 \vspace{-0.15cm}
 \vspace{+0.15cm}
 \caption{$F_{2}(\xbj,Q^2)$ as a function of $\xbj$. The data are from
   ZEUS~\cite{Breitweg:2000yn,Breitweg:1998dz,Chekanov:2001qu}.}
 \label{fig:inclusive_f2}
\end{figure}

Fig.~\ref{fig:subsets_f2} shows asymptotic solutions and models
against a subset of data to illustrate fit quality in different $Q^2$
ranges. The asymptotic fits with energy conservation included remain
valid to astonishingly large $Q^2$ values, exceeding the $1200\ \gev^2$
range over which we have kept track of $\chi^2$-values above.
This generic picture is reinforced if we contrast the scaling behavior
of theory and data as done in Fig.~\ref{fig:cross_sections} for GT
(left), BK (middle) GB-W fits (right). Shown are $\gasp$ cross
sections (see Eq.~\eqref{eq:short_cross} for the connection
$\sigma_{\gasp}\leftrightarrow F_2$) as a function of scaling variable
$Q^2 / Q_s^2(\xbj)$ for the three cases after the fit is
performed. The borders between moderate and high virtualities (above
$Q^2 = 120\ \gev^2$) are roughly indicated by the arrows overlaid on
the plots. Data sets from different running periods are plotted by
using different colors and symbols\footnote{The same notation for the
  inclusive data is applied throughout.}. To avoid overlapping, the
fit results are separated from the experimental data by dividing the
normalization factors $\sigma_0$ out. Note that the saturation scales
are different functions of $\xbj$ in each of the panels, although the
differences between the two truncations (GT, left and BK, middle) are
so small that no deviation can be discerned visually: both slopes
coincide excellently with the experimental data up to the largest
$Q^2/Q_s^2(\xbj)$. Contrary to that, the GB-W model shown on the right
can not resolve the high $Q^2$ data: the \emph{slopes} of data and
theoretical predictions start to deviate at large $Q^2/Q_s^2(\xbj)$.
A more detailed picture of the fit quality is given in
Fig.~\ref{fig:inclusive_f2}.

\subsection{Quark masses}
\label{sec:quark-masses}

As already mentioned, quark masses are formally a subleading effect
from the perspective of our small $x$ resummations, but they do impact
final state phase space and the width of wave functions in impact
factors quite severely. The use of constituent quark masses of $140\
\mev$ for light quarks and $1.4\ \gev$ for charm is quite widespread
in the context of the GB-W model and typically used in a restricted
$Q^2$-range (below $45\ \gev^2$). Since quark masses by nature are a
nonperturbative feature that should have its main effect at small
$Q^2$ one should expect that the inclusion of quark masses will not
spoil the excellent fit quality that was obtained in our light quark
fits for $Q^2$ above the heaviest quark included. From diffractive
measurements we know that charm quarks should contribute significantly
to the HERA cross sections while bottom quark contributions are
negligible. One should therefore complement the three light quarks
used above with a charm quark. This brings in a quark mass of
$1.2-1.4\ \gev$ depending on whether one considers current or
constituent quarks, both of which are of the same order as $1/R_s(Y)$
in physical units and thus one would expect at least some
complications in the nonperturbative sector. A straightforward fit
with three light and one heavy quark shows that the inclusion of the
charm quark reduces fit quality mainly in the $Q^2$ range below $1\
\gev^2$ (First two columns in Table~\ref{tab:charm_parameters}). This
is about as far as we can go in our analysis without introducing any
model elements that modify the low $Q^2$ behavior in some ad hoc
manner.
\begin{figure}[thb]
  \centering
   \includegraphics[width=.4\linewidth]{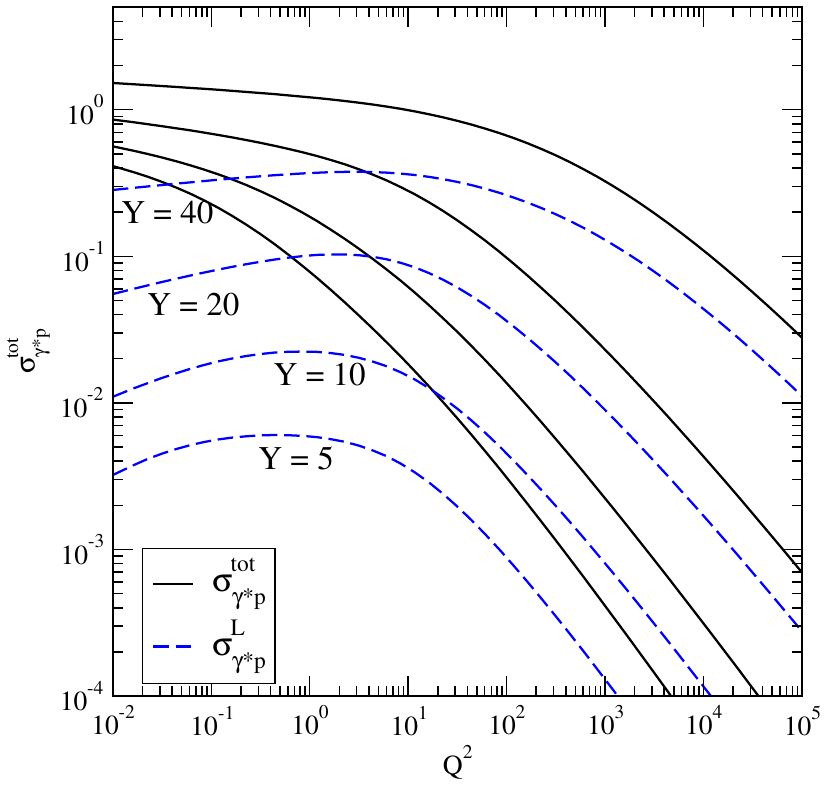}
 \hfill
 \includegraphics[width=.395\linewidth]{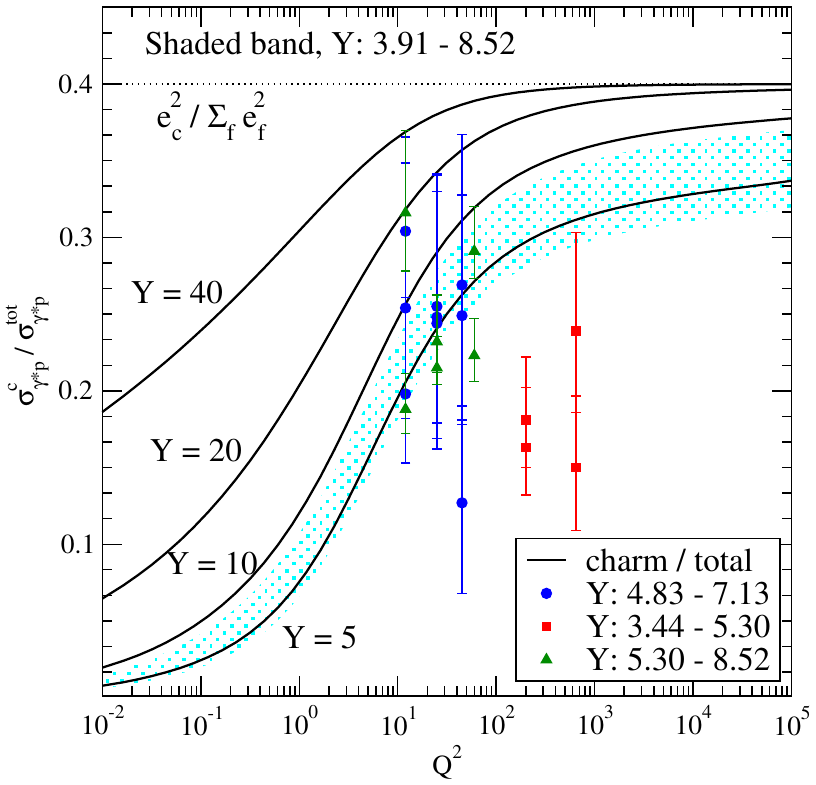}
 \caption{Cross sections    including charm quarks with $m_{u,d,s,c}=\{3,5,105,1270\}$ $\mev$
   for $Y=\{5,10,20,40\}$. The curves are calculated with the
   parameters corresponding to $x_{\text{eff}}=x(1+Q_s^2 /Q^2)$ (or
   equivalently $Y_{\text{eff}}=Y-\ln\left(1+Q_s^2 /Q^2\right)$) in
   Table~\ref{tab:charm_parameters}. \textbf{Left:}
   $\sigma_{\gasp}^{\text{tot}}$ and $\sigma_{\gasp}^{L}$ (blue
   dashed) plotted separately. Below the largest quark mass used,
   $\sigma^L$ starts to drop. This is incompatible with current
   conservation in the limit $Q^2\to 0$ and reduces the quality of the
   fit. \textbf{Right:} The charm fraction
   $\sigma_{\gasp}^{c}/\sigma_{\gasp}^{\text{tot}}$ as a function of
   $Q^2$. The horizontal line indicates the large $Q^2$ limit
   $e_c^2/\sum_f e_f^2=2/5$. The data are
   from~\cite{Adloff:1996xq,Aktas:2004az,Aktas:2005iw}. Agreement is
   clearly qualitative at best.}
  \label{fig:charm_demo}
\end{figure}

We should, however, at least qualitatively discuss a nonperturbative
modification introduced by Golec-Biernat and
Wüsthoff~\cite{Golec-Biernat:1998js}, a modeling device to accommodate
nonperturbative contributions at small $Q^2$. They have suggested to evaluate
the dipole cross section at
\begin{align}
  \label{eq:x_eff_GB_W}
  x_{\text{eff}}^{\text{GB-W}} 
  = x \frac{Q^2+4m_f^2}{Q^2} 
  = \frac{Q^2+4m_f^2}{Q^2+W^2}
\end{align}
instead of $x$ in order to guarantee $W$-independent cross sections at
small $Q^2$ as required by current conservation in the photo-production
limit at $Q^2=0$.  We have found that with the data set used
here,~(\ref{eq:x_eff_GB_W}) in fact improves the fit for $Q^2 \leq 45\
\gev^2$ in the GB-W model.  When using asymptotic solutions to the
evolution equation, however, fit quality goes down (taken over the
full $Q^2$ range) as shown in the third column of
Table~\ref{tab:charm_parameters}. In fact even the ``inverse''
modification $x_{\text{eff}}=x\frac{Q^2}{Q^2+4m_f^2}$ works better
from a $\chi^2$ perspective (fourth column), despite a clear lack of
supportive arguments.  One possible reason for the failure
of~\eqref{eq:x_eff_GB_W} is conceptual: it is not compatible with the
factorization into impact factors and Wilson line correlators that is
at the core of our renormalization group picture. Quark masses are
properties of projectile constituents and have every reason to show up
in impact factors. On the other hand, it is hard to imagine that they
should be resummed into a feature of the energy dependence of the
Wilson line correlators, which are purely determined by target
properties. But this is exactly what is done when
using~(\ref{eq:x_eff_GB_W}). From this perspective the only scale
available for use in a modification at small $Q^2$ is in fact
$Q_s^2(x)$. We tentatively suggest to replace~(\ref{eq:x_eff_GB_W})
with
\begin{align}
  \label{eq:x_eff_Qs}
  x_{\text{eff}} 
  = x \frac{Q^2+Q_s^2(x)}{Q^2} 
  = \frac{Q^2+Q_s^2(x)}{Q^2+W^2}
\ .
\end{align}
While this does not freeze the cross sections at fixed $W$, it does flatten
them noticeably at small $Q^2$ and leads to an improvement of the fit quality,
both with a schematic mass pattern of $m_{u,d,s,c}=\{5,5,5,1400\}\ \mev$ and
the current quark mass pattern $m_{u,d,s,c}=\{3,5,105,1270\}\ \mev$ as shown
in the last two columns of Table~\ref{tab:charm_parameters}. We emphasize
that~(\ref{eq:x_eff_Qs}) is only a conjecture to resum part of the
nonperturbative contributions below $Q_s$, one should not take too much
encouragement from the improvement of $\chi^2$ alone, see the ad hoc success
of the redefinition in column 4.
\begin{table}[H]
 \centering
 \begin{tabular}{lrrrrrr}   $x_{\text{eff}}
   $ & $x$ & $x$ & $x\frac{Q^2+4m_f^2}{Q^2}$  &  $x\frac{Q^2}{Q^2+4m_f^2}$  
   &  $x\frac{Q^2+Q_s^2 }{Q^2}$  &  $x\frac{Q^2+Q_s^2}{Q^2}$    \\
   & & {\tiny ($Q^2 > 1\ \gev^2$)} & & {\tiny ad hoc mod.} & & {\tiny (phys. masses)}  
   \\ [2mm] 
   $Y_{\text{off}}$  & $-5.86$  & $-6.87$  & $-5.63$ & $-5.99$ & $-6.50$ & $-6.73$  \\ 
   $\Lambda$ $[\mev]$  & 88.1  & 63.5 & 89.7 & 90.5  & 70.3 & 66.1  \\ 
   $\sigma_0$ $\gev^{-2}$ & 57.48 & 73.04  & 61.20 & 51.69 & 68.94 & 70.52  \\
   $\chi^2/\text{dof}$ & 1.41 & 1.05 & 1.59 & 1.23 & 1.13 & 1.14 \\ 
 \end{tabular}
 \caption{
   Including a charm quark; nonperturbative modifications. For the
   second column data with $Q^2 \leq 1\ \gev^2$ are removed, reducing 
   the dof from 295 to 221. Quark masses are $m_{u,d,s,c} = \{5,5,5,1400\}\ 
   \mev$ in all but the rightmost column  where
   $m_{u,d,s,c}=\{3,5,105,1270\}\ \mev$.
 }
\label{tab:charm_parameters}
\end{table}

\section{Lessons from the diffractive data}
\label{sec:lessons-from-diffr-data}

\subsection{The need for NLO contributions to the impact factors}
\label{sec:need-NLO-impact-factors}

Diffractive HERA data extend down to $\beta \sim .04$. Thus overall
$Y=\ln(1/x)$ and $Y_{\text{gap}}=\ln(1/x_{\mathbbm{P}})$ remain
comparable while $Y_{\text{frag}} = \ln(1/\beta)$ remains too small
for multiple gluon emission to build up within the projectile
fragmentation region -- contributions from the $q\Bar q g$-component
of the impact factor (which has its first contribution at NLO),
however start to play a role even at such moderate $\beta$ values as
already observed in the pioneering papers of~\cite{Wusthoff:1997fz,
  GolecBiernat:1998js} and reiterated
in~\cite{GolecBiernat:2008gk}. The main reason for that is that the
$q\Bar q$ contributions given in Eq.~(\ref{eq:dsigmadbeta}) strictly
vanish at $\beta\to 0$, even with NLO effects to the evolution of the
Wilson line correlators taken into account (see
Figs.~\ref{fig:diff_components}). Any gluon component in the impact
factor, on the other hand, will generate a nonvanishing cross section
in this region of phase space: the NLO contributions to the impact
factors are the \emph{leading} contribution at small $\beta$.

Unfortunately, no full expression for the $q\Bar q g$-component is
available, only the large $Q^2$ and small $\beta$ limits (without
$\beta$-evolution) are known exactly. However, an interpolating form
has been suggested in~\cite{Marquet:2007nf}. Below we find that
nonperturbative contributions related to the target profiles by far
dominate the uncertainties, and, for simplicity, we content ourselves
with the large $Q^2$ expressions of ~\cite{Wusthoff:1997fz,
  GolecBiernat:1998js} to estimate the contributions.

The starting point then is
\begin{align}
  \label{eq:dsigmadbeta-three-term-sum}
 \frac{d\sigma^{\gamma^* A\to X p}}{d\beta} =
 \frac{d\sigma^{\gamma^* A\to X p}_{q\Bar q,T}}{d\beta}
   +\frac{d\sigma^{\gamma^* A\to X p}_{q\Bar q,L}}{d\beta}
     +\frac{d\sigma^{\gamma^* A\to X p}_{q\Bar q g,T}}{d\beta}
\bigg\vert_{\text{LL}(Q^2)}
\end{align}
where the corresponding structure functions $x_{\mathbb{P}}
F^{D(3)}_{q\Bar q,T}$, $x_{\mathbb{P}} F^{D(3)}_{q\Bar q,L}$, and
$x_{\mathbb{P}} F^{D(3)}_{g g,T}$ [all functions of $(x_{\mathbb{P}}
,Q^2,\beta)$] are obtained by dividing out a factor
$\frac{4\pi^2\alpha_{\text{em}}}{Q^2\beta}$.  The first two terms
in~\eqref{eq:dsigmadbeta-three-term-sum} are given in
Eq.~(\ref{eq:dsigmadbeta}), they contain the LO impact factors with
only a $q\Bar q$ Fock component in the final state. The last term is
the large $Q^2$ part of the contribution of the NLO impact factor:
only the transverse part contributes, the longitudinal part being of
higher twist. At large $Q^2 $ the $q\Bar q g$-Fock state appears in a
configuration in which the inter-quark-distance is tiny compared to
$1/Q_s$ and one may take the corresponding coincidence limit. The
$q\Bar q$ part is then indistinguishable from a gluon and the analytic
expression may be cast in terms of gluon dipole amplitudes $N^{g
  g}_Y(\bm r,\bm b)$. The corresponding expression was first given by
Golec-Biernat and Wüsthoff~\cite{Golec-Biernat:1999qd}. In it, masses
are set to zero, in the spirit of a large $Q^2$ expansion.  As with
the first two terms of Eq.~\eqref{eq:dsigmadbeta-three-term-sum},
which were already given in Eq.~\eqref{eq:dsigmadbeta}, we present
this result in a notation inspired by~\cite{Marquet:2007nf}, but with
the $b$-integral not yet performed. This allows us to assess the
quantitative impact of our lack of precise knowledge of the
$b$-dependence in all three contributions. We retain the assumption
that the dipole amplitudes only depend on $|\bm r|$ and are
independent of the orientation of the dipole. Then
\begin{subequations}
  \label{eq:qqg-leading-Q^2}
\begin{align}
  \label{eq:qqg-leading-Q^2-expl}
  \frac{d\sigma^{\gamma^* A\to X p}_{q\Bar q
      g,T}}{d\beta}\bigg\vert_{\text{LL}(Q^2)} = &
  \frac{\alpha_{\text{em}} \alpha_s C_f N_c}{8\pi^2 Q^2} \int_\beta^1
  \frac{dz}{(1-z)^3}\Bigl[\Bigl(1-\tfrac{\beta}{z}\Bigr)^2
  +\Bigl(\tfrac{\beta}{z}\Bigr)^2\Bigr]
  \int\limits_0^{(1-z)Q^2}\hspace{-1em} d{\bm k}^2
  \ln\left(\tfrac{(1-z)Q^2}{{\bm k}^2}\right) \notag \\ &\times
  \int\limits_0^\infty d\bm r^2\; d{\bm r'}^2 \phi_{gg}(z,|\bm k|,|\bm
  r|,|\bm r'|) \int d^2\bm b\ N^{g g}_Y(\bm r,\bm b)\; N^{g g}_Y({\bm
    r}',\bm b)
\end{align}
where
\begin{align}
\label{eq:phiggdef}
\phi_{gg}(z,|\bm k|,|\bm r|,|\bm r'|) := & {\bm k}^4 J_2( |\bm k||\bm
r| ) K_2\left(\sqrt{\frac{z}{1-z}{\bm k}^2 \bm{r}^2}\right)
K_2\left(\sqrt{\frac{z}{1-z}{\bm k}^2 \bm{r}'^2}\right) J_2( |\bm k||\bm
r'| )\ .
\end{align}
\end{subequations}
Any inclusion of quark masses (or a dependence of the dipole
orientation) in such an expression would amount to a resummation of
subleading effects with little control over their relevance.
Similarly, we have no reliable argument to set the scale in the strong
coupling $\alpha_s$ that appears in the prefactor of this
expression. Both these assessments are reinforced once one starts
analyzing the nonperturbative uncertainties in
Eq.~(\ref{eq:dsigmadbeta-three-term-sum}), even after it has been
updated on the perturbative level with full NLO ingredients.  To be
consistent in our treatment below we will therefore also only consider
the massless limit for the $q\Bar q$ contributions.

In fact, the largest uncertainty in~\eqref{eq:qqg-leading-Q^2} and in
the corresponding expressions for the quarks,
Eq.~\eqref{eq:dsigmadbeta}, is related to the impact parameter
integral which is not under perturbative control. Even after using
data to set the overall normalization of the total cross section (the
main nonperturbative parameter entering the LO total cross section),
already the leading order diffractive contributions (the $q\Bar
q$-terms in~(\ref{eq:dsigmadbeta-three-term-sum})) require additional
nonperturbative input. This only gets more pronounced at NLO: higher
order Fock components in the projectile wave function couple to higher
n-point functions of Wilson lines, each of which is affected in its
own way by non-perturbative effects. This affects the relative
normalizations of the terms in
Eq.~(\ref{eq:dsigmadbeta-three-term-sum}) as well as the relative
normalization of total and diffractive cross sections. In practice,
this manifests itself in a strong model dependence of the
normalization of individual cross sections. The main issues here are
the relative normalization of total and diffractive cross sections on
the one hand and the weight of individual Fock components such as the
$q\Bar q$ and $q\Bar q g$ contributions in
Eq.~(\ref{eq:dsigmadbeta-three-term-sum}) on the other.

\subsection{Nonperturbative aspects of  \texorpdfstring{$\bm
    b$}{b}-dependence and profile functions}
\label{sec:nonp-infl-b-dep}

The general attitude for the total cross section -- to assume a fixed,
$x$-independent target size, which at LO implies taking the $q\Bar
q$-dipole amplitude as the product of a profile function $T(\bm b)$
and $r$-dependent remainder $N^{q\bar q}_Y(\bm r)$ (normalized to one
at $r\to\infty$) leaves us with only a single fit parameter, the area
resulting from the $b$-integration. In the diffractive case, the
choice of profile strongly affects overall and relative normalization
of
\begin{align}
  \label{eq:box-profile-Nqqbar2}
  \int d^2{\bm b} \ N^{q\bar q}_Y(\bm r,\bm b)N^{q\bar q}_Y(\bm r',\bm b)
  \intertext{and}
  \label{eq:box-profile-Ngg2}
  \int d^2{\bm b} \ N^{g g}_Y(\bm r,\bm b)N^{g g}_Y(\bm r',\bm b)
\end{align}
featuring in the formulae above as well as the more general correlators in
their full NLO generalizations.

The simplest treatment would associate a factorized profile with each of the
amplitudes, \emph{identical} for both quarks and gluons according to (again
with $\cal R$ labeling the representation)
\begin{align}
  \label{eq:globally-factorized}
  N^{\cal R}_Y(\bm r,\bm b) \to T(\bm b) N^{\cal R}_Y(\bm r)
\end{align}
and already in this case, the relative weight of diffractive and total cross
sections are highly model dependent: A box profile $T_{\text{box}}(\bm b)$ of
height one, normalized in width to produce a factor
\begin{align}
  \label{eq:S-norm-total}
\sigma_0
= 2\int d^2{\bm b}\  T(\bm b)
\end{align}
for the total cross section, results in a factor $\int d^2{\bm b}\,
T^2_{\text{box}}(\bm b) = \sigma_0/2$ in the diffractive case. A
Gaussian profile, which has some phenomenological justification at
large $|\bm b|$,\footnote{This is based on successful parametrizations
  of meson production data via $e^{-B_d |t|}
  \left.\frac{d\sigma}{dt}\right\vert_{t=0}$ and constrains the shape
  for $|\bm b| > .3\,\text{fm}$~\cite{Munier:2001nr}.} produces an
additional factor $\frac12$ in the diffractive case compared to the
box profile: the area under $T_{\text{Gauss}}^2(\bm b)$ is half the
area under $T_{\text{Gauss}}(\bm b)$.  Clearly, the relative
normalization of the total and the diffractive contributions is
strongly dependent on the shape of the profile: Arbitrary factors of
this sort can already be obtained by varying the width and the height
of the box profile while keeping~(\ref{eq:S-norm-total}) fixed. While
one may dismiss box profiles as unphysical, the issue remains: there
is by no means a canonically prescribed physical profile that would
outright eliminate such modeling choices.

Also a Gaussian profile, justified as it may be in some $b$-ranges for
quarks, leaves intrinsic uncertainties: Not only is the overall
normalization an issue, but also the relative normalization of parton
species (here quarks vs gluons) is affected. To explore this in more
detail, we will invoke Casimir scaling, which we expect to hold at
least at small $b$, as a guiding principle. Assuming the same profile
factor in front of both quark and gluon amplitudes breaks Casimir
scaling for the full $b$- and $r$-dependent amplitude $S^{\cal
  R}_Y(\bm r,\bm b) = 1-N^{\cal R}_Y(\bm r,\bm b) $ for effectively
all $\bm b^2 >0$ and hence, is not compatible with the notion that the
energy dependence is dominated by perturbative gluon emission as
encoded in the evolution equations and their Gaussian truncation. To
guarantee this then requires that Casimir scaling is at best weakly
broken for central collisions.  Models with Casimir scaling restored
completely are readily constructed: One may, for example, exponentiate
the profile in the spirit of the IPSat and bCGC models according to
\begin{align}
  \label{eq:bCGCdef}
  N^{\cal R}_Y(\bm r,\bm b) \to 1 -e^{-C_{\cal R} T(\bm b) {\cal
      G}_Y(\bm r) }\ .
\end{align}
Alternatively, one might start with a factorized (Gaussian) profile for quarks
to define ${\cal G}_Y(\bm r,\bm b)$ for general use in the Gaussian truncation
via $T(\bm b)N^{q\Bar q}_Y(\bm r) =: 1-e^{-C_f {\cal G}_Y(\bm r,\bm
  b)}$. This allows to calculate general $n$-point functions in the Gaussian
truncation with $b$-dependence and yields Casimir scaling for dipoles according
to
\begin{align}
  \label{eq:new-b-ansatz}
  N^{\cal R}_Y(\bm r,\bm b) \to 
  1-\bigl[1-T(\bm b) N^{q\bar q}_Y(\bm r)\bigr]^{\frac{C_{\cal R}}{C_f}}
\ .
\end{align}
Taking Casimir scaling as a guiding principle modifies both $\bm b$
and $\bm r$ dependence in a $C_{\cal R}$ dependent manner. In
particular it leads to sizable changes in relative normalization of
quark and gluon contributions as compared to the height-one box
profile, for which all of these definitions are equivalent.

Plain exponentiation of the profile as in~(\ref{eq:bCGCdef}) enhances
the influence from nonperturbative regions of phase space as $Y$
increases: with increasing $Y$ the large $|\bm r|$-growth of ${\cal
  G}_Y(\bm r)$ will progressively lift up any nonvanishing large $|\bm
b|$ tails of the profile function if they exist at all. This in turn
leads to $Y$-dependent growth of the overall normalization of the
dipole cross section (after the $b$-integration is done). This results
in an inconsistent, unphysical interplay of perturbative gluon
emission with non-perturbative long range physics. One might attempt
to regulate the large $r$-behavior of ${\cal G}_Y(\bm r)$ to preclude
that, but to do this in a defensible way would clearly require
non-perturbative input completely outside the scope of JIMWLK
evolution.

The ansatz~(\ref{eq:new-b-ansatz}) on the other hand does not require
any such additional input: large $|\bm r|$ and $|\bm b|$ behavior
decouple. Therefore we use this model below to estimate the impact on
relative normalizations of cross sections.

With an eye to the total cross section, we first note that the model
leads to representation dependent normalization for the $b$-integrated
dipole cross section. Assuming a Gaussian profile, we find that the
$b$-integrated profiles acquire a representation-dependent $\bm
r$-dependence we denote $N'^{\cal R}_Y(\bm r)$ as well as a nontrivial
large $|\bm r|$-normalization which we choose to display explicitly
\begin{align}
  \label{eq:b-int-N-new-norm}
  \int d^2{\bm b}\ N^{\cal R}_Y(\bm r,\bm b) := 2\pi B_d\
  H({\frac{C_{{\cal R}}}{C_f}}) N'^{\cal R}_Y(\bm r) \xrightarrow{|\bm
    r|\to\infty} 2\pi B_d H({\frac{C_{{\cal R}}}{C_f}}) \approx 2\pi
  B_d
  \begin{cases}
    1 & \text{for quarks} \\
1.6   & \text{for gluons}
  \end{cases}
\end{align}
where $H(z) := \psi_0(1-z) +\gamma_E$ is the harmonic number of $z$
(expressed via the digamma function $\psi_0$).  Diffractive
normalizations differ from the normalizations in the total cross
section, and in addition, the $\bm r$- and $\bm r'$-dependence after
$b$-integration for a generic representation only factorizes
approximately\footnote{For~(\ref{eq:bCGCdef}) factorization of $\bm
  r$-dependence is generically a bad approximation and the result
  depends strongly on the IR regularization at large $r$.  }
\begin{subequations}
  \label{eq:calR-approx-fact}
\begin{align}
  \label{eq:calR-approx-factd-expr}
  \int d^2{\bm b}\ N^{\cal R}_Y (\bm r,\bm b) N^{\cal R}_Y ({\bm
    r}',\bm b) \approx 2\pi B_d\ \Bigl( 2H\big(\frac{C_{\cal
      R}}{C_f}\bigr) -H\big(\frac{2C_{\cal R}}{C_f}\bigr) \Bigr)\
  \Tilde N^{\cal R}_Y (\bm r) \Tilde N^{\cal R}_Y (\bm r')
\end{align}
with
\begin{align}
  \label{eq:tildeN-def}
  \Tilde N^{\cal R}_Y (\bm r) := \left[ \frac{\int d^2{\bm b}\ [
      N^{\cal R}_Y(\bm r,\bm b) ]^2 }{ 2\pi B_d\ \Bigl(
      2H\big(\frac{C_{\cal R}}{C_f}\bigr) -H\big(\frac{2C_{\cal
          R}}{C_f}\bigr)\Bigr)} \right]^{\frac12}\ .
\end{align}
\end{subequations}
Unsurprisingly, normalizations of quark and gluon contributions differ
again: the relative diffractive normalizations for quarks and gluons
are reliably assessed along the diagonal $\bm r=\bm r'$, where one
finds
\begin{align}
  \label{eq:b-int-N2-new-norm}
 \int d^2{\bm b}\ [ N^{\cal R}_Y(\bm r,\bm b)]^2 
  = 2\pi B_d\ \bigl(2H({\frac{C_{{\cal R}}}{C_f}})-H({\frac{2C_{{\cal
          R}}}{C_f}})\bigr) 
\xrightarrow{|\bm r|\to\infty}
 2\pi B_d
  \begin{cases}
    \tfrac12 \ \ \text{for quarks} \\
  1.  \ \ \text{for gluons}
  \end{cases}
\ ,
\end{align}
i.e. the gluon contribution is enhanced by a factor of two compared
the quarks. This is not the case for the
ansatz~\eqref{eq:globally-factorized} or the use of a box profile of
height one.  We take this as a practical indication that the
nonperturbative uncertainties are indeed large and that a precision
fit at NLO requires refined nonperturbative input or independent
phenomenological constraints on the individual normalization of each
Fock component.

On the practical side one finds that factorization is exact for quarks
and a good approximation for gluons.
For quarks $\Tilde N^{q\Bar q}_Y(\bm r) = N^{q\Bar q}_Y(\bm r)$ and
$(2H({\frac{C_f}{C_f}}) - H({\frac{2C_{f}}{C_f}})) = \frac12$.  For
gluons one may, for simplicity, even use $N'^{g g}_Y(\bm r)$ as
inspired by~\eqref{eq:b-int-N-new-norm} to approximate
\begin{align}
  \label{eq:gluon-b-int-simple-approx}
  & \int d^2{\bm b}\ N^{g g}_Y(\bm r,\bm b) N^{g g}_Y({\bm r}',\bm b)
  \approx 2\pi B_d\ C\ N'^{g g}_Y(\bm r)N'^{g g}_Y(\bm r')
\end{align}
where $C = 1$, in line with~\eqref{eq:b-int-N2-new-norm}.

These may be used in the momentum space formulae originally derived
with height one box profiles (which factorize trivially) after
properly including the normalization factors -- the factorization
error is much smaller than any nonperturbative uncertainty inherent in
the choice of the profile model.

\subsection{Diffractive fits to HERA data}
\label{sec:diffr-fit-results}

Given the uncertainties arising both from incomplete NLO impact
factors (Sec.~\ref{sec:need-NLO-impact-factors}) and nonperturbative
aspects of the impact parameter dependence
(Sec.~\ref{sec:nonp-infl-b-dep}) precision fit to diffractive data are
out of the question.

The fits presented below are done with this in mind -- mainly to
assess if one can get a qualitative agreement with data, based on the
fit parameters $Y_{\text{off}}$ and $\Lambda$ obtained from the total
cross section. To optimize the diffractive fits, an independent
normalization $B_d$ is allowed and its relation to
$B_d^{\text{tot}}=\sigma_0/4\pi$ is ignored in the fit. How closely
the fit results match is treated a consistency check. As already
indicated, the scale of $\alpha_s$ in the $q\Bar q g$ component of the
diffractive cross section in Eq.~\eqref{eq:qqg-leading-Q^2-expl} is
undetermined. This factor of $\alpha_s$ will therefore be treated as a
further fit parameter. While $B_d$ and $\alpha_s$ must be meaningful
in the context of HERA physics, we need a means to accommodate the
uncertainties exposed above. In this sense, we take the precise values
of these parameters as scenario dependent.

\begin{figure}[htb]
 \centering
 \includegraphics[width=.39\linewidth]{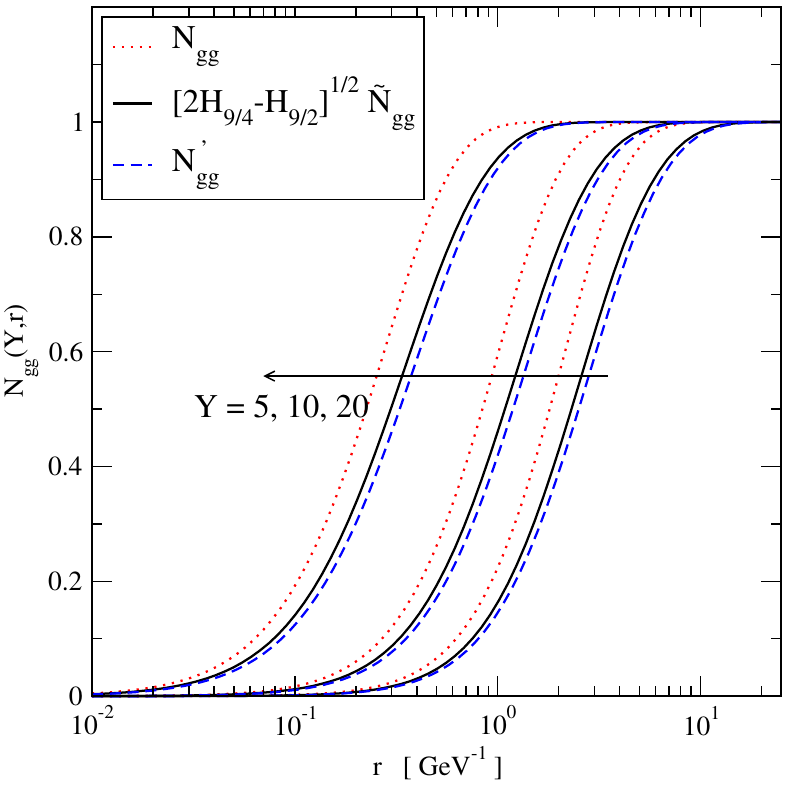}
 \hfill
 \includegraphics[width=.4\linewidth]{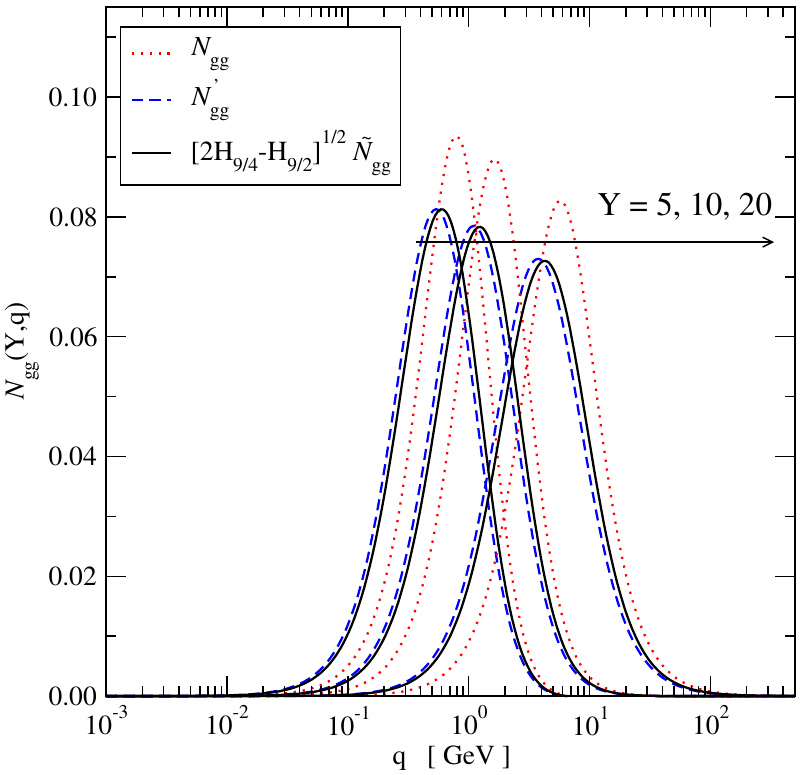}
 \caption{Three gluon dipole inputs: globally factorized $N^{gg}_Y(\bm
   r)$ Eq.~\eqref{eq:globally-factorized}, Casimir scaling based
   $b$-integrated $(2H_{9/4} - H_{9/2})^{\frac12}\Tilde N^{gg}_Y(\bm
   r)$ Eq.~\eqref{eq:calR-approx-fact} and $N'^{gg}_Y(\bm r)$ in
   coordinate space (\textbf{left panel}) and momentum space (see
   App.~\ref{sec:transformations} for precise definitions)
   (\textbf{right panel}).}
 \label{fig:gluon_inputs}
\end{figure}

In the fits to the total cross section, the differences between the
different scenarios using the energy conservation correction turned
out to be small. For this reason, it is sufficient to consider only GT
results from our simulations. We retain the GB-W model results for
comparison. In what follows, the overall normalization of the $\qqp$
components is the same in all cases since for the quarks the profile
function factorizes trivially. For the gluon component, the relative
normalization is scenario dependent and we explicitly consider the
following two cases:
\begin{itemize}
\item \textbf{[fact]:} $b$-dependence completely factorized: the
  simplest case given in~\eqref{eq:globally-factorized}. The $\qqp$
  and $\qqp g$ components are equally weighted.

\item \textbf{[sc]:} $b$-dependence based on Casimir scaling after
  using the approximation of Eq.~\eqref{eq:calR-approx-fact}. The
  $\qqp g$ component is enhanced by a factor of two relative to the
  ansatz~\eqref{eq:globally-factorized}, its shape is slightly
  modified. (The $q\Bar q$ components remain completely factorized.)
\end{itemize}
$N_{Y}^{gg}(\bm{r})$ for the two schemes and the corresponding
momentum space amplitudes ${\cal N}_{Y}^{gg}(\bm{q})$ (see
App.~\ref{sec:diffr-cross-sect-mom-space} for definitions), are
illustrated in Fig.~\ref{fig:gluon_inputs}. 
The difference between the factorized and Casimir scaling schemes are
pronounced: The main effect of [sc] is to increase the effective $R_s$
after $b$-integration is done compared to the factorized scheme
[fact].\footnote{We note in passing that the gluon amplitudes
  $N'^{gg}_{Y}(\bm{r})$ (dashed blue) and $\tilde{N}^{gg}_{Y}(\bm{r})$
  (solid black) are almost the same: a tiny rescaling in $r$ is
  required in order to (perfectly) match the curves.}  As a result the
[sc]-gluon stays closer to the quark result than the [fact]-gluon, see
Fig.~\ref{fig:compare_quark_gluon} in
App.~\ref{sec:tools-diffr-cross}.

\begin{figure}[bht]
 \centering
 \includegraphics[width=.33\linewidth]{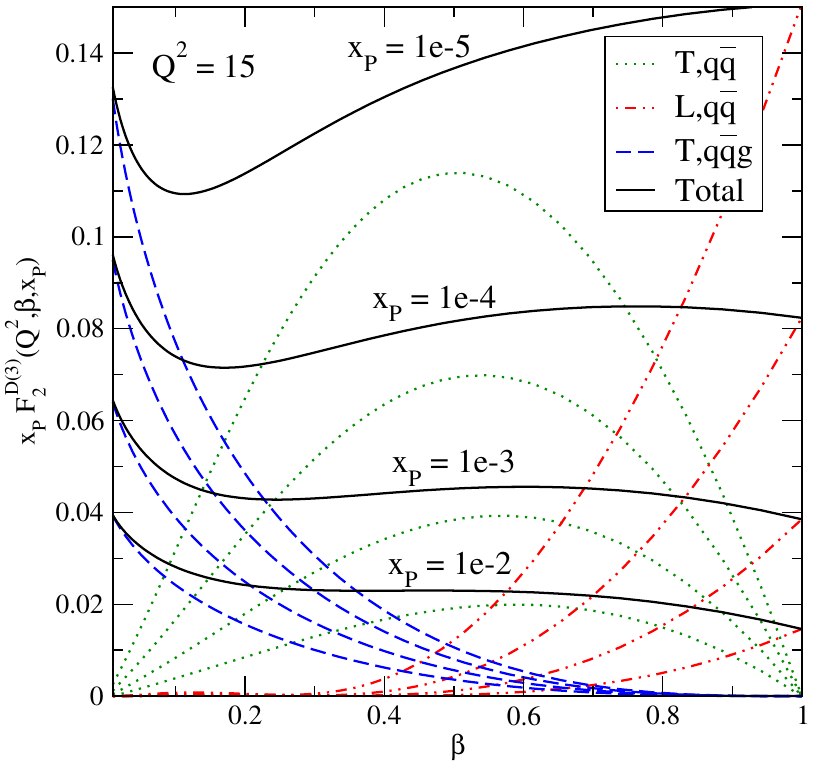}
 \hfill
 \includegraphics[width=.3101\linewidth]{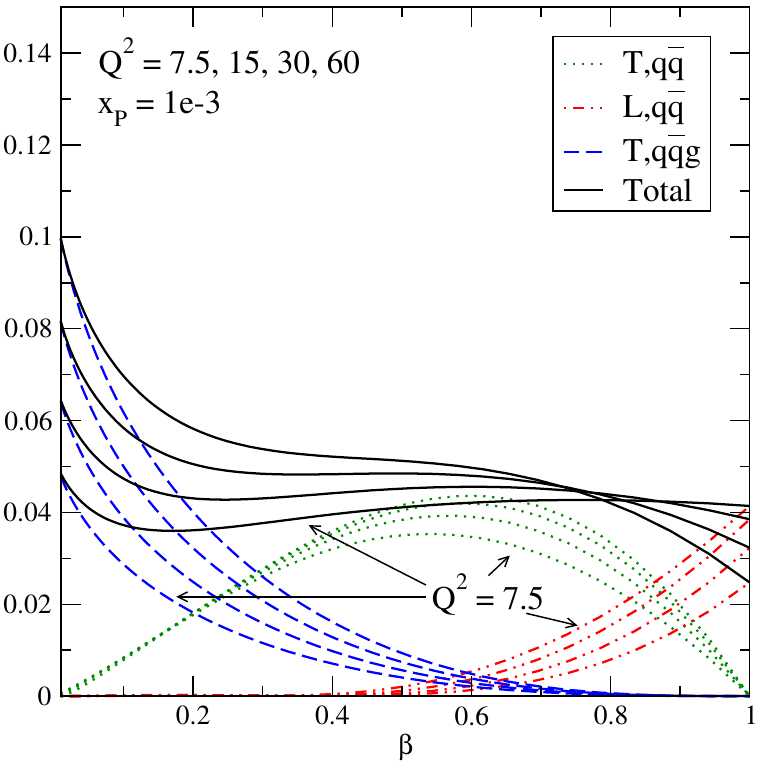}
\hfill
\includegraphics[width=.3101\linewidth]{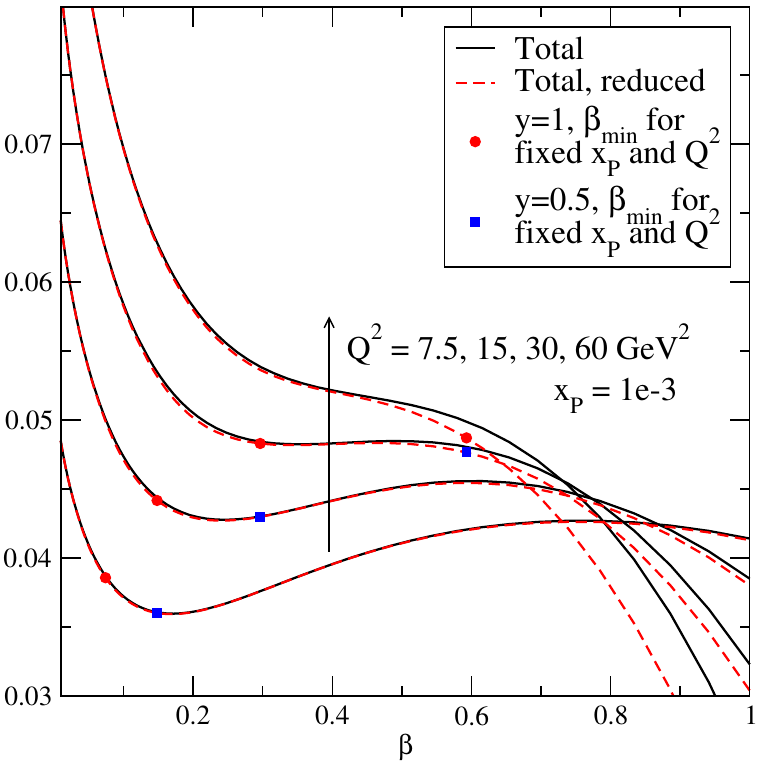}
\caption{Behavior of the contributions given in
  Eq.~\eqref{eq:mom-space-diffr}. \textbf{Left:} Change in
  $x_{\mathbbm{P}} F_2^{D(3)}$ as $x_{\mathbbm{P}}$ is decreased with
  fixed $Q^2$. \textbf{Middle:} $Q^2$ is increased with fixed
  $x_{\mathbbm{P}}$. \textbf{Right:} A comparison of $\xp\sigma_r^{D(3)}$
  and $\xp F_2^{D(3)}$ according to
  Eq.~\ref{eq:diff_cross_reduced}. See the text for more details.}
 \label{fig:diff_components}
\end{figure}

Since the literature lists data directly for structure functions $\xp
F_{2}^{D(3)}=\frac{Q^2\beta}{(2\pi)^2\alpha_{\text{em}}}\frac{d\sigma^{\gamma^*
    A\to X p}}{d\beta}$, the actual task is to compute each component
of the cross section Eq.~\eqref{eq:dsigmadbeta-three-term-sum} with
the parameters obtained from the total cross section and then to
optimize $B_d$ and $\alpha_s$ in the following expression,
\begin{align}
 \label{eq:linear_combo}
 \xp\ F_2^{D(3)}& = \bm{B_d} \left[
   \xp\ F_{\qqp,T}^{D(3)}\big\arrowvert_{B_d=1}
   +
   \xp\ F_{\qqp,L}^{D(3)}\big\arrowvert_{B_d=1}
   +
   \bm{\alpha_s}\xp\
   F_{\qqp g,T}^{D(3)}\big\arrowvert_{B_d=1,\alpha_s=1}
 \right]\ .
\end{align}
The numerical burden of this procedure can be vastly reduced by
replacing the coordinate space expressions for these contributions
shown in the text by their momentum space counterparts, see
App.~\ref{sec:tools-diffr-cross}.  $\xp F_{2}^{D(3)}$ and its
components are visualized in Fig.~\ref{fig:diff_components}: the
overall behavior of $\xp F_2^{D(3)}$ is analogous in all cases,
including the GB-W model.

\begin{table}[!thb]
  \centering
\subfloat{\begin{tabular}{lll}
    $N_{Y}^{{\cal R}}(\bm{r})$:~\eqref{eq:globally-factorized}  & GB-W & GT  \\ [2mm]
    $B_d$  & 2.85  & 4.70 \\    
    $\alpha_s$ & 0.54  & 0.44 \\
    $\chi^{2}/552$ & 1.44  & 1.44  \\ 
  \end{tabular}
}
\hspace{3cm}
\subfloat{\begin{tabular}{ll} 
    $\tilde{N}_{Y}^{{\cal R}}(\bm{r})$:~\eqref{eq:calR-approx-fact} & GT  \\ [2mm]
    $B_d$  & 4.70 \\    
    $\alpha_s$ & 0.31 \\
    $\chi^{2}/552$ & 1.47  \\
  \end{tabular}
}
\caption{The results of the fits to the diffractive
  data~\cite{Adloff:1997sc,Derrick:1995wv,Breitweg:1997aa,
    Breitweg:1998gc,Chekanov:2008cw}. \textbf{Left:} [fact] Fully
  factorized $b$-dependence of Eq.~\eqref{eq:globally-factorized}.
  \textbf{Right:} [sc] $b$-dependence based on Casimir scaling
  according to
  Eqs.~\eqref{eq:calR-approx-fact},~\eqref{eq:b-int-N2-new-norm}
  and~\eqref{eq:gluon-b-int-simple-approx}.}
\label{tab:enter_diffractive}
\end{table}
\begin{figure}[p]
 \centering
 \includegraphics[width=.939\linewidth]{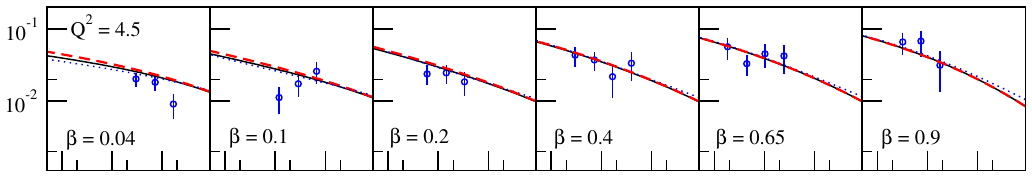}
 \vspace{-0.22cm}
\\
 \includegraphics[width=.939\linewidth]{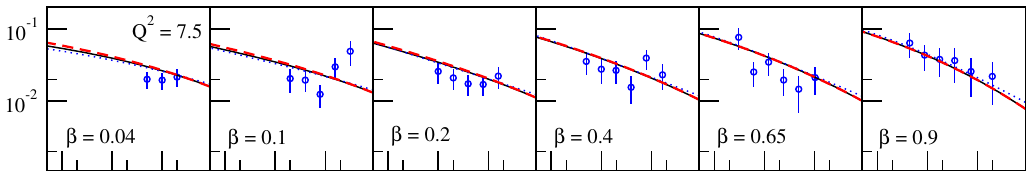}
 \vspace{-0.22cm}
\\
 \includegraphics[width=.939\linewidth]{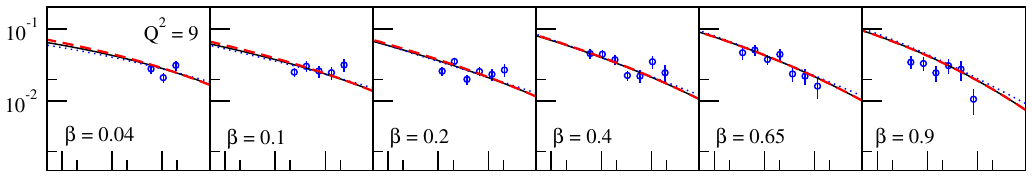}
 \vspace{-0.22cm}
\\
 \includegraphics[width=.939\linewidth]{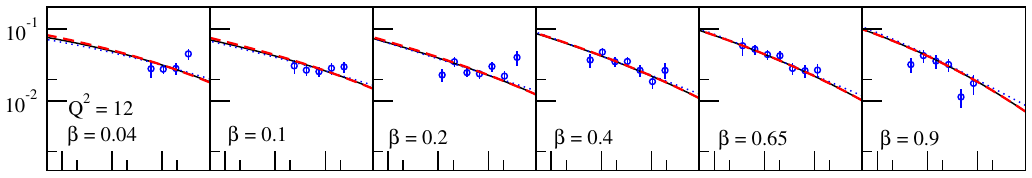}
 \vspace{-0.22cm}
\\
 \includegraphics[width=.939\linewidth]{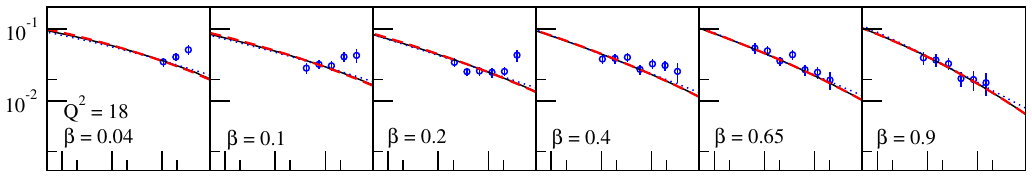}
 \vspace{-0.22cm}
\\
 \includegraphics[width=.939\linewidth]{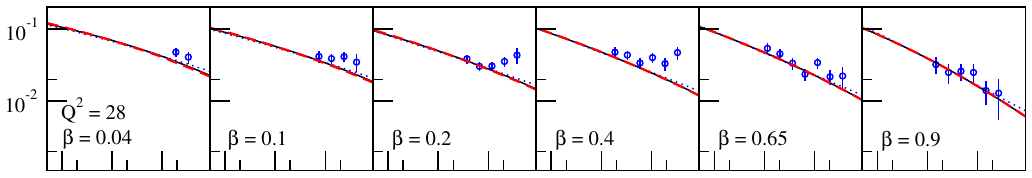}
 \vspace{-0.22cm}  
\\
 \includegraphics[width=.939\linewidth]{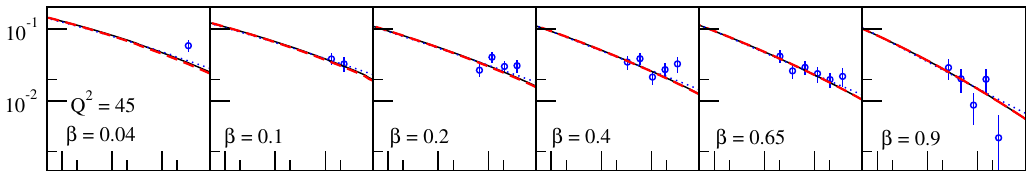}
 \vspace{-0.22cm}
\\
\includegraphics[width=.939\linewidth]{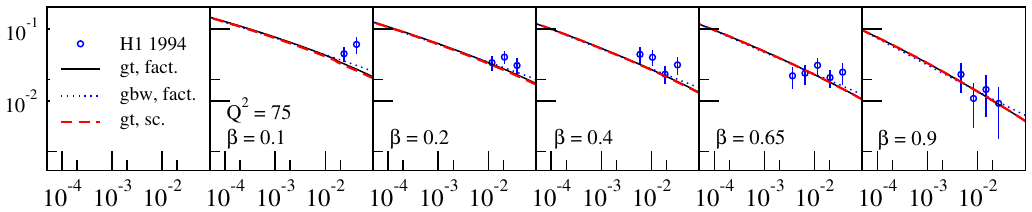}
\vspace{-0.15cm}
 \vspace{+0.15cm}
 \caption{$x_{\mathbbm{P}}
   F_{2}^{\text{D}(3)}(x_{\mathbbm{P}}Q^2,\beta)$ as a function of
   $x_{\mathbbm{P}}$. The data are from H1~\cite{Adloff:1997sc}.}
 \label{fig:diff_f2_h1}
\end{figure}

Fig.~\ref{fig:diff_f2_h1} presents a subset\footnote{Although not
  shown, the data from
  ZEUS~\cite{Breitweg:1998gc,Derrick:1995wv,Breitweg:1997aa,Chekanov:2008cw}
  were also included in the fits.} of $x_{\mathbbm{P}}
F_{2}^{\text{D}(3)}(Q^2,\beta,\xp)$ data from H1~\cite{Adloff:1997sc}
together with the theoretical predictions. We show three practically
indistinguishable scenarios:
\begin{itemize}
\item {}[fact]: GT with the factorized
ansatz~\eqref{eq:globally-factorized} (denoted ``gt, fact.'')]
\item {}[sc]: GT based on the Casimir scaling ansatz
  of~\eqref{eq:calR-approx-fact} (denoted ``gt, sc.'').
\item GB-W with the factorized ansatz~\eqref{eq:globally-factorized}
  (denoted ``gbw, fact.'')
\end{itemize}
The corresponding fit parameters are presented in
Tables~\ref{tab:enter_diffractive}.

The parameter $\alpha_s$ turns out to be strongly scenario dependent,
reflecting the differences in the shapes, scales and overall
normalizations of the gluon amplitudes shown in
Fig.~\ref{fig:gluon_inputs}. For GT, the common normalization $B_d$ as
well as the fit quality $\chi^2/\text{dof}$ are of the same order in
all cases. The values of $B_d$ are systematically smaller than the
experimental value $B_d=7.1\ \gev^{-2}$ reported
in~\cite{Breitweg:1997aa} but, they \emph{are} consistent with
$B_d^{\text{tot}}=\sigma_0/4\pi$ obtained from the total cross
section.\footnote{Note that the use of $T_{\text{box}}({\bm b})$ of
  height one induces an overall factor of two in \emph{all} components
  relative to $T_{\text{Gauss}}({\bm b})$ in the framework
  of~\eqref{eq:globally-factorized}. Since it can only be absorbed by
  a redefinition of $B_d$ in the diffractive fits, this would lead to
  major inconsistencies. } BK evolution (not shown) results in fits of
the same systematic behavior and fit quality as the GT fits presented
here.

Fit quality varies noticeably across phase space: splitting the data
into two subsets $\beta \leq 0.5$ and $\beta > 0.5$ or equivalently
$Q^2 \leq M_X^2$ and $Q^2 > M_X^2$,\footnote{As with the data for the
  total cross section, there is kinematical correlation in the data
  range: $\beta$ is small if $Q^2 \ll M_X^2$. In this case,
  $\xbj=\xp\beta$ is also small and thus $\sgp\approx Q^2/\xbj$ is
  large. Since $\sgp\ll \sep$ in the experiments is finite, small
  values of $\beta$ are more likely paired with low $Q^2$.} we
observe that the best match with the data is obtained at large $\beta$
where the $\qqp$ components dominate, see
Table~\ref{tab:splitted_in_q2}.
\begin{table}
 \centering
 \begin{tabular}{lll} $\text{GT}$  & $\beta \leq 0.5$ & $\beta > 0.5$ 
   \\ [2mm]
   $N_{Y}^{{\cal R}}(\bm{r})$:~\eqref{eq:globally-factorized}  & $\chi^2/264=1.58$ & $\chi^2/288=1.30$  \\ [1mm]
   $\tilde{N}_{Y}^{{\cal R}}(\bm{r})$:~\eqref{eq:calR-approx-fact}  & $\chi^2/264=1.64$ & $\chi^2/288=1.31$ \\
\end{tabular}
\caption{The diffractive data split into two subsets $\beta \leq 0.5$
  and $\beta > 0.5$. The data with $\beta>0.5$ fit somewhat better (a
  region where $q\bar{q}$ components dominate).}
\label{tab:splitted_in_q2} 
\end{table}
This reinforces our statements that a better treatment of the $q\Bar q
g$ amplitude is required. The presently implemented improvements in
our treatment of the $b$-dependence only lead to a tiny improvements
in fit quality: the main differences between the gluon amplitudes
resulting from the two schemes dubbed [fact] and [sc] are effectively
absorbed into the normalization of the $\qqp g$ component via
parameter $\alpha_s$.

As a check of consistency, the fits are also performed by using the
reduced cross section (as was done in~\cite{Marquet:2007nf})
\begin{align}\label{eq:diff_cross_reduced}
  \sigma_r^{D(3)} =F_2^{D(3)}\left(1 -
    \frac{y^2}{1+(1-y)^2}\frac{F_{\qqp,L}^{D(3)}}{F_2^{D(3)}}\right)\quad\quad
  ;\quad\quad y \approx\frac{Q^2}{\sep \xbj}\ ,
\end{align}
where $\sep= 318^2\ \gev^2$ at HERA. We note that for the present
diffractive data $y<0.5$ so that $\sigma_r^{D(3)}$ hardly deviates
from $F_2^{D(3)}$ over the bulk of the data range\footnote{For 322
  points out of 512 $y < 1/4$ implying that the factor in front of
  $F_{\qqp,L}^{D(3)}$ in~\eqref{eq:diff_cross_reduced} is $<0.04$. The
  data with $1/4 \leq y \leq 1/2$ are most likely associated with
  large $Q^2$.}. As a consequence, one would at best expect a tiny
improvement in the fit quality in any approach once the reduced cross
section is used. We find indeed that $\chi^2$ values improve almost
imperceptibly to $\chi^2/552=1.39$ and $\chi^2/552=1.42$,
respectively. Consistent with this, the parameters remain practically
unchanged: $B_d=4.75\ \gev^{-2}$ and $\alpha_s=0.43$
for~\eqref{eq:globally-factorized} and $B_d=4.75\ \gev^{-2}$ and
$\alpha_s=0.31$ for~\eqref{eq:calR-approx-fact}.

To pinpoint precisely where the differences arise consider the
rightmost panel in Fig.~\ref{fig:diff_components}: this compares
$\sigma_r^{D(3)}$ with $F_2^{D(3)}$ for fixed $\xp$ and $Q^2$ as a
function of $\beta$. It should be noted that if the diffractive phase
space is presented in this way, one must take into account that the
kinematic limit $y\leq 1$ is violated at small $\beta$. Two conditions
$y=1$ and $y=0.5$ marked by the red circles and blue squares overlaid
on each curve are indicating the smallest possible $\beta$ for fixed
$\xp=10^{-3}$, $Q^2=\{7.5,15,30,60\}\ \gev^2$ and $\sep=318^2\ \gev^2$
(set by $\beta_{\text{min}}=Q^2/(\xp\sep y)$). The former is the
absolute kinematic limit of HERA whereas the latter approximately
marks the lowest values of $\beta$ of the existing data.

\subsection{Ratios with the total cross section}
\label{sec:ratios}

The approximate constant ratio of the diffractive to inclusive cross
sections was originally observed at HERA by ZEUS
collaboration~\cite{Breitweg:1998gc}. To compute this observable, we
use the relation (see ~\cite{Breitweg:1998gc,Chekanov:2008cw})
\begin{align} \label{eq:connect_sigmad_f2d} \frac{1}{2M_X}\frac{d
    \sigma_{\gamma^{*}p\rightarrow X p}^{\text{diff}}(Q^2,M_X,\xp)}{d
    M_X}\approx\frac{(2\pi)^2\ \alpha_{\text{em}}}{Q^2(Q^2 +M_X^2)}\
  \xp\ F_{2}^{D(3)}(Q^2,M_X,\xp)
\end{align}
for the diffractive cross section. The ratio of the diffractive cross
section to the total cross section can then be computed by
\begin{align}
 \label{eq:rdiff_tot}
 R_{\text{tot}}^{\text{diff}}(\sgp) = \frac{\int_{M_a}^{M_b}d M_X\
   d\sigma_{\gamma_{*}p\rightarrow
     Xp}^{\text{diff}}(Q^2,M_X,\xp)\big/ d
   M_X}{\sigma_{\gamma^{*}p}^{\text{tot}}(\xbj,Q^2)}\quad ;\quad Q^2
 \ll \sgp \ll \sep\ ,
\end{align}
where the integration boundaries, i.e. the bins for the diffractive
mass $M_X$, are determined by the experimental setup (see
Table.~\ref{tab:ratio_results}).

In the calculation of $R_{\text{tot}}^{\text{diff}}$ given above, no
new free parameters are introduced. Basically,
$R_{\text{tot}}^{\text{diff}}$ is parameter free since the
normalizations cancel trivially. However, as mentioned earlier, we
ignored the value of $B_d^{\text{tot}}=\sigma_0/4\pi$ from the total
cross section and allowed a distinct normalization in the diffractive
fits. For instance, in the case of GT the ratio of the optimal
normalizations is found to be $B_d/B_d^{\text{tot}}\approx 1.07$,
i.e. fairly close to one. In practice this means that \emph{all} HERA
data can be resolved by the same normalization
$B_d^{\text{tot}}=\sigma_0/4\pi\approx 4.40\ \gev^{-2}$: a single
parameter fit to the diffractive cross section then results in
$\alpha_s=0.34$ with a reasonable $\chi^2/552=1.52$ for the
scenario~\eqref{eq:calR-approx-fact}. The region $\beta\leq 0.5$ is
responsible of the reduction of the overall $\chi^2/\text{dof}$: at
$\beta> 0.5$ the fit quality is actually slightly improved.
\begin{figure}[p]
 \centering
 \includegraphics[width=1.0\linewidth]{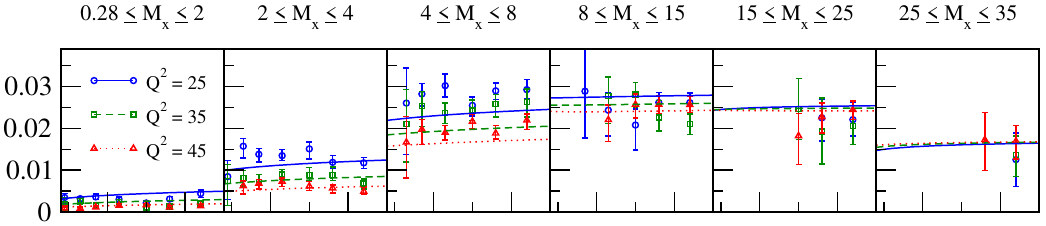}
 \vspace{-0.75cm}
\\
 \includegraphics[width=1.0\linewidth]{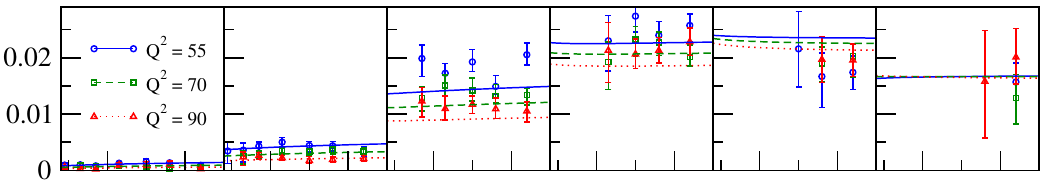}
 \vspace{-0.75cm}
\\
\includegraphics[width=1.0\linewidth]{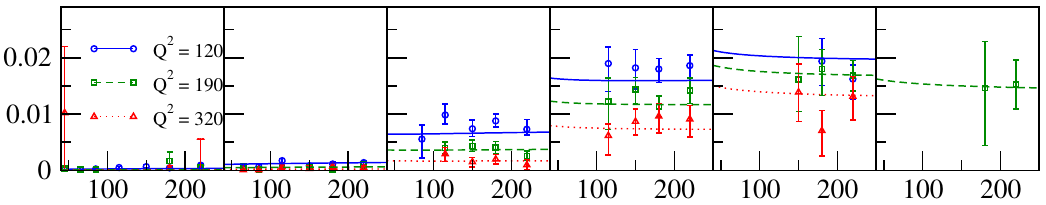}
\caption{The ratio of the diffractive versus the inclusive cross
  sections according to Eq.~\ref{eq:rdiff_tot} as a function of
  $\sqrt{\sgp}$ for different values of $Q^2$ and bins of diffractive
  mass $M_X$. The smallest values of $\beta$ are located in the top
  right corner: from there, following the panels to the left or down
  increases $\beta$. The ratios are based on the parameters for GT in
  Tables~\ref{tab:inclusive_results} $(\sigma_{\text{tot}})$
  and~\ref{tab:enter_diffractive} ($\sigma_{\text{diff}}$,
  Eq.~\eqref{eq:calR-approx-fact}). The data are
  from~\cite{Chekanov:2008cw}.}
 \label{fig:diff_tot_ratio}
\end{figure}
\begin{table}[p]
 \centering
 \begin{tabular}{lrlllllll}  & $M_X$ $[\gev]$  & 0.28-2 & 2-4 &
   4-8 & 8-15 & 15-25 & 25-35 & $\sum$ \\ [2mm]
Treatment   & dof & 54 & 57 & 46 & 37 & 23 & 10 & 227 \\
   \multicolumn{2}{l}{$F_2^{D(3)}$ ; $N_{Y}^{{\cal R}}(\bm{r})$:~\eqref{eq:globally-factorized}} & 2.05 & 1.22 & 2.27 & 0.77 & 1.16 &
   0.29 & 1.51  \\ [1mm]
   \multicolumn{2}{l}{$F_2^{D(3)}$ ; $\tilde{N}_{Y}^{{\cal R}}(\bm{r})$:~\eqref{eq:calR-approx-fact}} & 2.05 & 1.22 & 2.30 & 0.82 & 0.88 & 0.24 & 1.49 \\ [2mm]
   \multicolumn{2}{l}{$\sigma_r^{D(3)}$ ; $N_{Y}^{{\cal R}}(\bm{r})$:~\eqref{eq:globally-factorized}} & 1.78 & 1.16 & 2.30 & 0.77 & 1.16 &
   0.29 & 1.44  \\ [1mm]
   \multicolumn{2}{l}{$\sigma_r^{D(3)}$ ; $\tilde{N}_{Y}^{{\cal R}}(\bm{r})$:~\eqref{eq:calR-approx-fact}} & 1.78 & 1.16 & 2.33 & 0.82 & 0.88 &  0.23 &  1.42 \\
 \end{tabular}
 \caption{The fit results of $R_{\text{tot}}^{\text{diff}}$ for
   each $M_x$ bin (GT, diffractive cross section vs. reduced cross section). The data are
   from~\cite{Chekanov:2008cw}.}
\label{tab:ratio_results}
\end{table}
\begin{figure}
  \centering
 \includegraphics[width=0.52\linewidth]{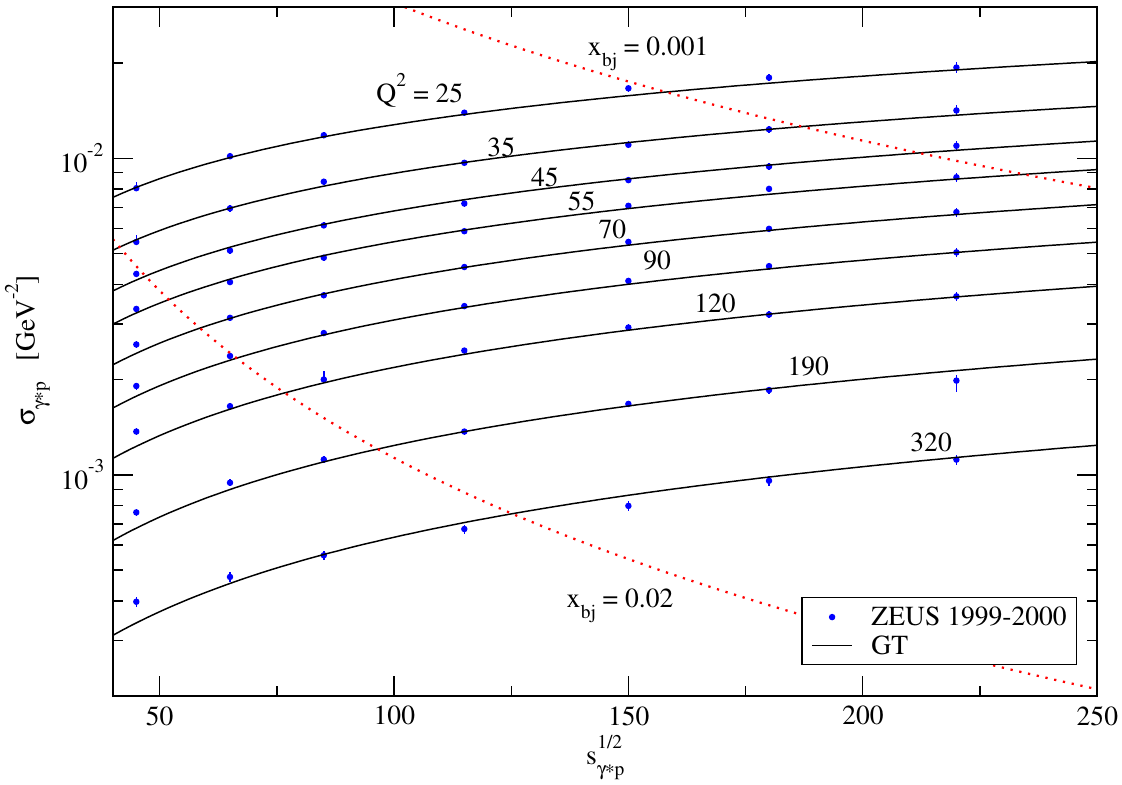}
\hspace{0.0cm}
\caption{The $\gasp$ cross section as a function of
  $\sqrt{\sgp}=\sqrt{Q^2(1/\xbj -1)}$ for different values of
  $Q^2$. The data are from~\cite{Chekanov:2008cw}, i.e. it is the
  total cross section part of $R_{\text{tot}}^{\text{diff}}$ data
  shown in Fig.~\ref{fig:diff_tot_ratio}. As long as $\xbj =
  Q^2/(Q^2+\sgp)\lesssim 0.02$ the theoretical predictions match
  excellently with the experiments. The breakdown at large $\xbj$
  becomes evident as one approaches the lower left corner with fixed
  $\sqrt{\sgp}\approx 45-65$ $\gev$.}
 \label{fig:xsect_only}
\end{figure}

In Fig.~\ref{fig:diff_tot_ratio} are shown the latest
$R_{\text{tot}}^{\text{diff}}$ data from~\cite{Chekanov:2008cw}
together with the theoretical predictions based on the fits to the
inclusive and diffractive cross sections presented
earlier\footnote{More $R_{\text{tot}}^{\text{diff}}$ data with $Q^2\in
  [2.7,55]$ $\gev^2$ can be found in~\cite{Chekanov:2005vv}. The fits
  to this data turned out to be slightly worse.}. As seen, the overall
behavior of $R_{\text{tot}}^{\text{diff}}$ is well produced with
fairly good fit qualities (see Table~\ref{tab:ratio_results}). In the
case of the reduced cross section $\sigma_r^{D(3)}$ only the bins
$M_X\in [0.28,2]$ and $[2,4]$ $\gev$ (the largest $\beta$) are
affected, resulting in a slightly better $\chi^2/\text{dof}$. In
Fig.~\ref{fig:xsect_only} we show the corresponding total cross
section alone: the match with the data is strikingly good indicating
that the source of the uncertainties in $R_{\text{tot}}^{\text{diff}}$
is the incomplete description for the diffractive cross section.

\section{Asymptotic versus pre-asymptotic fits, a comparison}
\label{sec:pre-as-examples}

As we have discussed at length for the total cross section in
Sec.~\ref{sec:lessons-from-total-cross-section}, an asymptotic fit
without the energy conservation included is not feasible. This is
caused by a strong fit tension arising between too large an evolution
speed at small $Y$ and too shallow a dipole correlator shape at large
$Y$. Once energy conservation is included, both features improve
towards what is needed to match the data: evolution speed is uniformly
lowered and correlators become steeper overall. The upshot is a fit
whose main ingredients are determined perturbatively -- both shape and
evolution speed in the asymptotic region are
predominantly\footnote{Aside from regulator effects on the coupling
  which are visible due to the $R_s$ values inherent to the kinematic
  properties of the HERA experiments.} determined by the nonlinear
structure of the evolution equation and its kernel. As reported, it
works flawlessly up to $Q^2$ of $1200\ \gev^2$, i.e. the largest $Q^2$
values available at HERA for $x\le 0.02$.

The fits performed in~\cite{Albacete:2009fh} by contrast include
running coupling effects but no energy conservation correction. They
obtain a fit that is almost as good as the GT fits described in this
paper. To achieve this, they \emph{must} move away from the asymptotic
pseudo-scaling region and use pre-asymptotic features of an evolution
trajectory. This allows them to simultaneously satisfy the speed and
steepness requirements of the dipole correlators and improve the fit
quality over what is possible in the asymptotic (pseudo-scaling)
domain without the energy conservation correction included. The
features of the correlators along the part of the evolution trajectory
used in the fit differ drastically from those in the asymptotic case
and a detailed comparison is in order. We will perform this comparison
for both the total and diffractive cross sections
(which~\cite{Albacete:2009fh} does \emph{not} consider), to illustrate
once more that, with present theoretical limitations, there is no hope
to use the more differential diffractive cross section to
differentiate between theoretical approaches.

In doing so we have to deal with secondary differences of the fit
procedures in our case and in~\cite{Albacete:2009fh}. Where we have
argued for the use of current quark masses and have included a charm
quark contribution successfully using (optionally) an $m_f$-independent
remapping of $x \mapsto x_{\text{eff}}=x\frac{Q^2+Q_s^2(x)}{Q^2}
$, \cite{Albacete:2009fh} uses three quarks with $m_f=140\ \mev$ and
an $m_f$-dependent remapping of $x \mapsto
x_{\text{eff}}^{\text{GB-W}}=x\frac{Q^2+4m_f^2}{Q^2} $, which, if used
with our solutions degrades the $\chi^2$ considerably. The fit and
parameters of~\cite{Albacete:2009fh} were done for $x\le 0.01$ instead
of the $x\le 0.02$ we have used. For comparison we have to restrict
ourselves to the same range. This reduces the $Q^2$ range of available
data from $Q^2 \le 1200\ \gev^2$ to $Q^2 \le 650\ \gev^2$. Fortunately
the separation scheme used to define the running coupling contribution
is the same in both treatments. 

For the pre-asymptotic fit scenario, inclusion of the charm quark is
virtually impossible without reworking most of the ingredients. To
simplify the comparison we have also left out the charm quark in the
asymptotic fits we compare to explicitly -- for its inclusion see
Sec.~\ref{sec:quark-masses}.

\begin{figure}[htb]
\centering
\includegraphics[width=.323\linewidth]{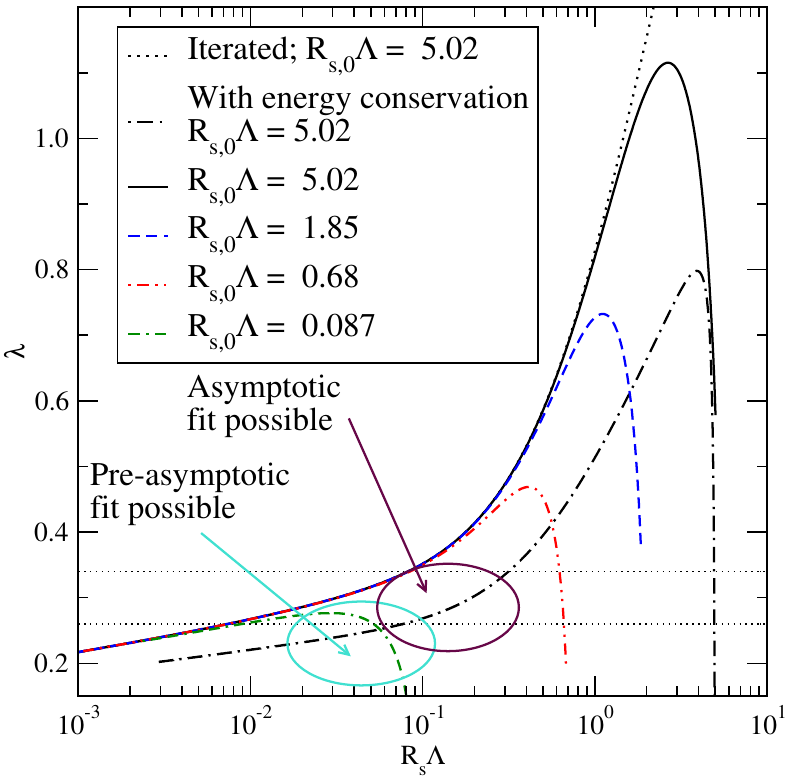}
\hfill
\includegraphics[width=.318\linewidth]{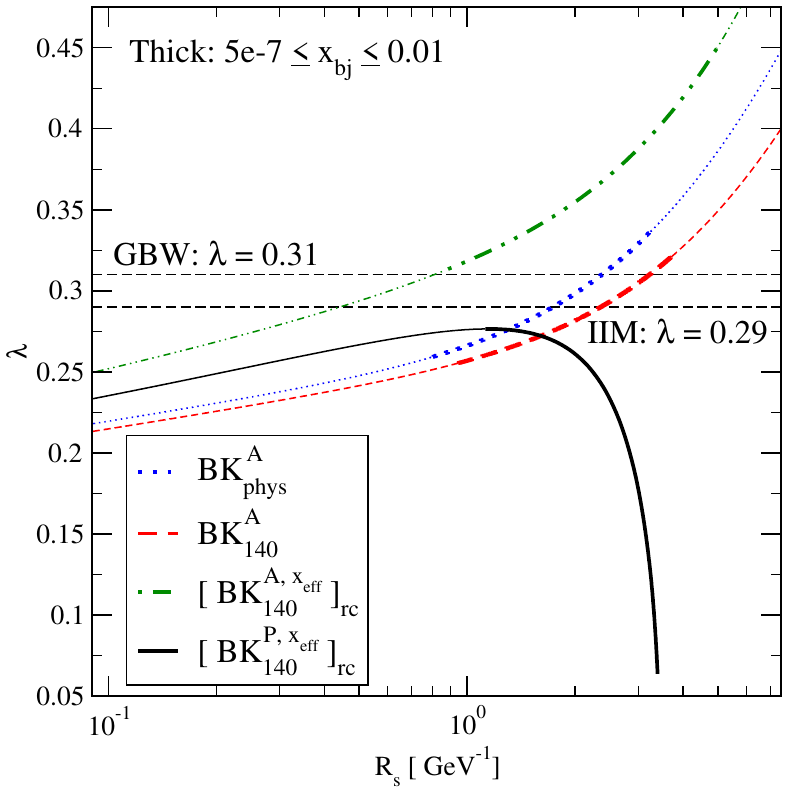}
\hfill
\includegraphics[width=.33\linewidth]{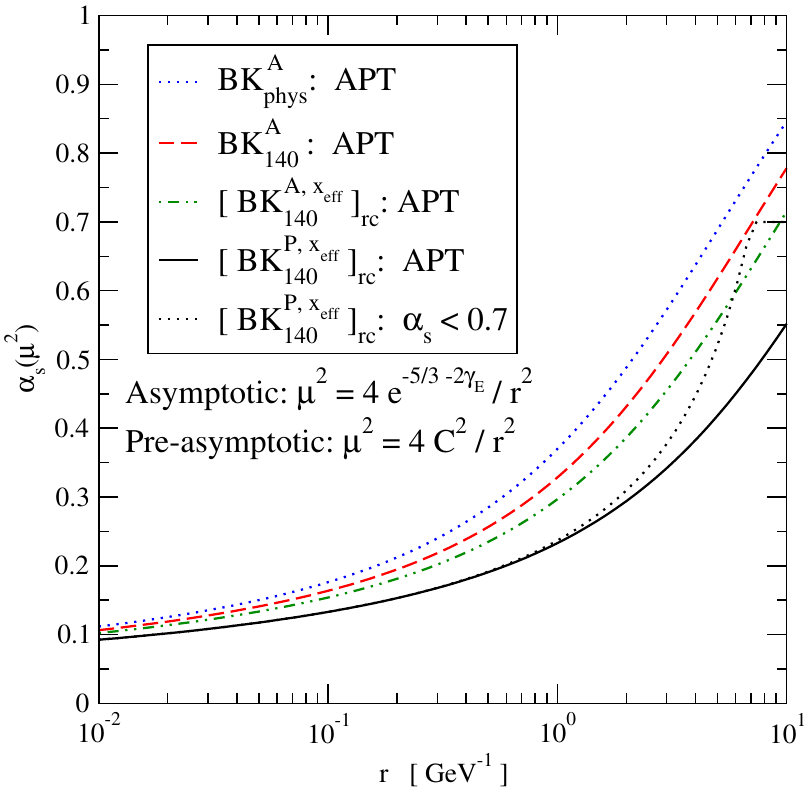}
\caption{\textbf{Left:} Comparison of evolution speeds of
  pre-asymptotic fits without energy conservation
  correction~\cite{Albacete:2009fh} with the asymptotic fits including
  the energy conservation correction. The marked areas indicate the
  ranges of $\lambda(R_s(Y)\Lambda)$ where the different fit
  strategies match to data. The curve starting from
  $R_{s,0}\Lambda=0.087$ corresponds to the pre-asymptotic
  fit. \textbf{Middle:} The situation after the ratio ${\cal
    C}/\Lambda$ is known from the data fit. Due to the unique energy
  dependence of each scenario, the fit range $5\times
  10^{-7}\leq\xbj\leq0.01$ is emphasized by thicker line width for
  each curve. \textbf{Right} $\alpha_s(\mu)$ for each scenario after
  the ratio ${\cal C}/\Lambda$ is determined by the data fit.  ${\cal
    C}/\Lambda \approx 5.2$, $7.1$ and $19.1$ for the scenarios
  $\text{BK}_{\text{phys}}^{\text{A}}$,
  $\text{BK}_{\text{140}}^{\text{A}}$ and
  $[\text{BK}_{\text{140}}^{\text{P},x_{\text{eff}}}]_{\text{rc}}$,
  respectively. This gives an impression of the size of the slowdown
  effect originating from the energy conservation correction. The
  black dashed line shows coupling used in~\cite{Albacete:2009fh},
  regulated in the IR by freezing it at maximum value of $.7$.}

 \label{fig:two_scenarios}
\end{figure}

We have attempted to expose the effects of the secondary differences
in the fit procedures by comparing a set of fit scenarios that permute
some of these ingredients. The fits below are tagged as
$\text{BK}_{\text{phys}}^{\text{A}}$, $\text{BK}_{140}^{\text{A}}$,
$\text{BK}_{140}^{\text{A},x_{\text{eff}}}$, 
 $[\text{BK}_{140}^{\text{A},x_{\text{eff}}}]_{\text{rc}}$,
and
$[\text{BK}_{140}^{\text{P},x_{\text{eff}}}]_{\text{rc}}$ and defined
as follows
\begin{itemize}
\item $\text{BK}_{\text{phys}}^{\text{A}}$: asymptotic
  (pseudo-scaling), energy conservation correction included, three
  quarks, mass pattern $m_f = \{3,5,105\}\ \mev$. An $x$ remapping
  induces no discernible differences.
\item $\text{BK}_{140}^{\text{A}}$: asymptotic (pseudo-scaling),
  energy conservation correction included, three quarks, mass pattern
  $m_f = \{140, 140, 140\}\ \mev$.
\item $\text{BK}_{140}^{\text{A},x_{\text{eff}}}$: asymptotic
  (pseudo-scaling), energy conservation correction included, three
  quarks, mass pattern $m_f = \{140, 140, 140\}\ \mev$, remapping of
  $x$ according to $x \mapsto
  x_{\text{eff}}^{\text{GB-W}}=x\frac{Q^2+4m_f^2}{Q^2} $.
\item $[\text{BK}_{140}^{\text{A},x_{\text{eff}}}]_{\text{rc}}$: same as previous,
  but with the energy conservation correction omitted. This attempts a
  fit on the asymptotic line the fit of~\cite{Albacete:2009fh}
  eventually merges onto.
\item $[\text{BK}_{140}^{\text{P},x_{\text{eff}}}]_{\text{rc}}$, the
  fit developed in~\cite{Albacete:2009fh}: pre-asymptotic, NLO
  corrections restricted to running coupling corrections (energy
  conservation correction excluded), three quarks, mass pattern $m_f =
  \{140, 140, 140\}\ \mev$, remapping of $x$ according to $x \mapsto
  x_{\text{eff}}^{\text{GB-W}}=x\frac{Q^2+4m_f^2}{Q^2}$. Evolution
  starts from a GB-W like initial state $S_{Y_0=\ln(1/0.01)}^{q\Bar
    q}(r)=\exp\left[-(r Q_{s,0}/2)^2\right]$ with $Q_{s,0}^2=0.241\
  \gev^2$. The scale choice for the running coupling in the fit is
  parametrized differently from our treatment: In the argument of a
  running coupling we use an $r$ dependent scale in the form
  $\frac{\mu^2(r)}{\Lambda^2} = \frac{{\cal C}^2}{r^2 \Lambda^2}$ with
  ${\cal C}^2 = 4 e^{-2\gamma_E-\frac53}$ and vary $\Lambda$.  They
  use $\frac{\mu^2(r)}{\Lambda^2} = \frac{4 C^2}{r^2 \Lambda^2}$, with
  $\Lambda$ set to $.241\ \gev$ a priori, and obtain $C^2=5.3$ from
  the fit, see Fig.~\ref{fig:two_scenarios}, right panel.
\end{itemize}
In all treatments including
$[\text{BK}_{140}^{\text{P},x_{\text{eff}}}]_{\text{rc}}$ an APT
regulator is used while~\cite{Albacete:2009fh} regulate the coupling
by freezing it at $\alpha_s^\text{max} = .7$, see
Fig.~\ref{fig:two_scenarios}, right panel. Note that the choice of
initial condition in
$[\text{BK}_{140}^{\text{P},x_{\text{eff}}}]_{\text{rc}}$ is quite
restrictive and no effort is made to vary the shape of the correlator
other than allowing for some offset rapidity before matching evolution
results to data. A systematic study of the impact of varying the
shape, even a theoretical exploration of which shape features might be
responsible for what kind of physics property of the cross section is
still outstanding. Here we only attempt to contrast asymptotic fits
with \emph{one} example of a preasymptotic one.

\begin{figure}[!thb]
 \centering
 \includegraphics[width=.320\linewidth]{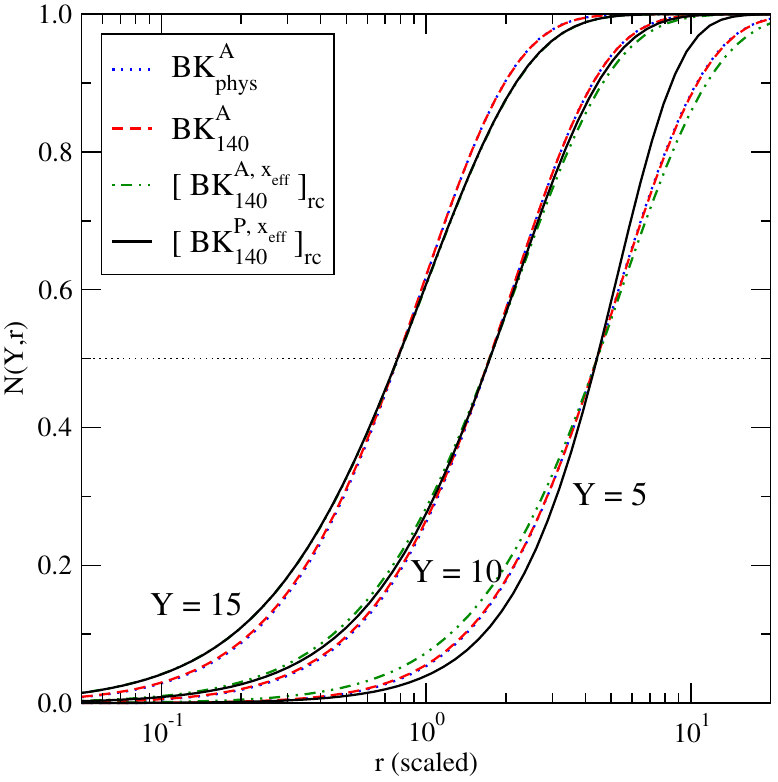}
 \hfill
 \includegraphics[width=.324\linewidth]{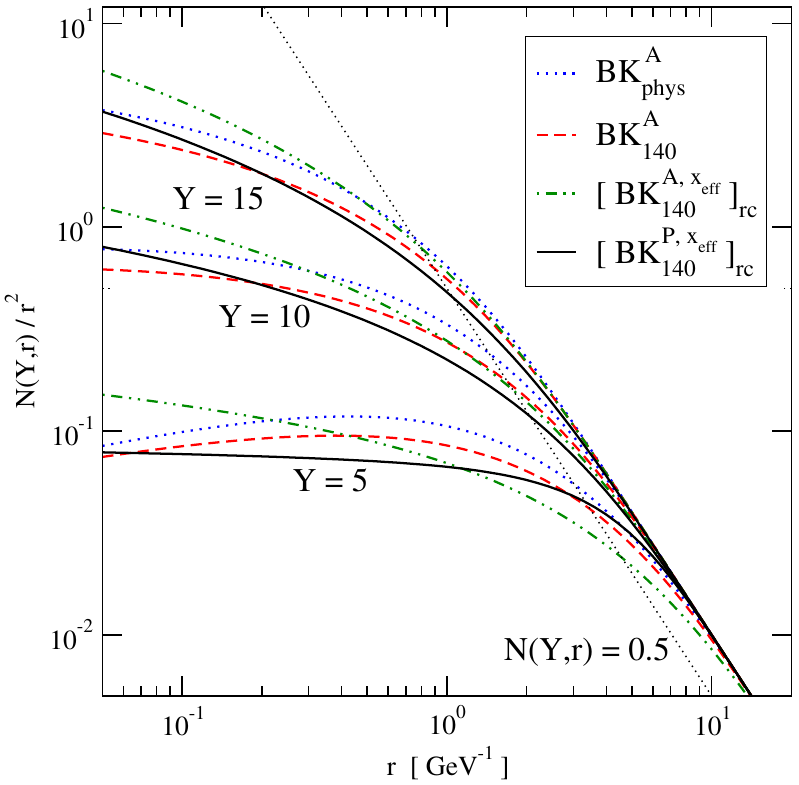}
\hfill
\includegraphics[width=.33\linewidth]{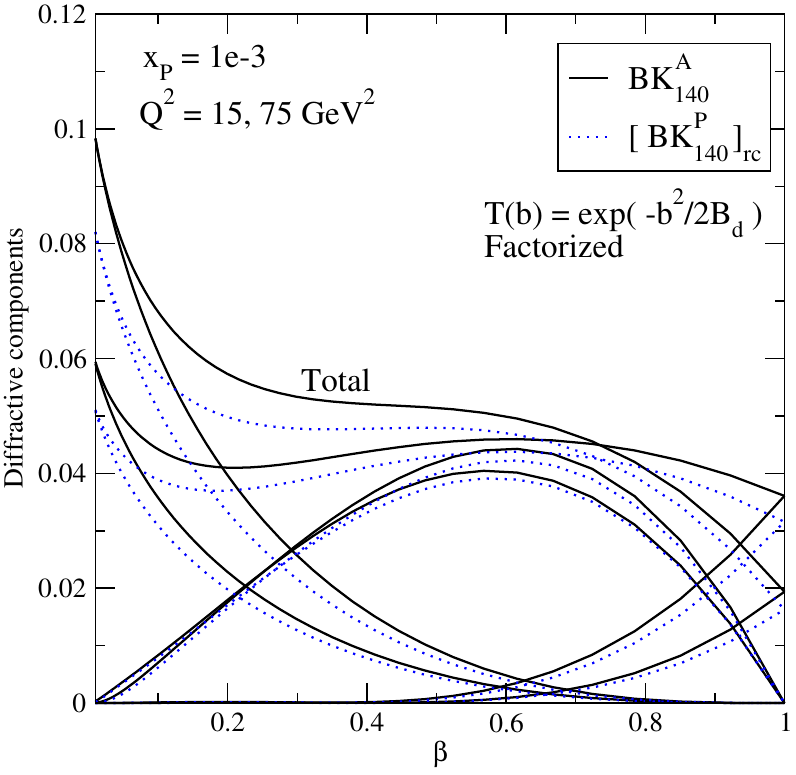}
\caption{Correlator shapes obtained from the fits to the total cross
  section (Table~\ref{tab:pre-asymp-total}). \textbf{Left:} A
  comparison of the shapes of $N^{q\Bar q}_Y(r)$. The curves are
  shifted to match at $N^{q\Bar q}_Y(r)=0.5$ for each value of
  $Y$. \textbf{Middle:} $N^{q\Bar q}_Y(r)/r^2$: the behavior at small
  $r$. \textbf{Right:} Contributions to the diffractive cross section
  for $\xp= 10^{-3}$ using a factorized $b$-profile.}
 \label{fig:fit_comparison}
\end{figure}

We intentionally only show BK based fits to allow for a direct
comparison, despite the fact that GT fits have better $\chi^2$ and
note that the modifications from $\text{BK}_{\text{phys}}^{\text{A}}$
through $\text{BK}_{140}^{\text{A}}$ to
$[\text{BK}_{140}^{\text{A},x_{\text{eff}}}]_{\text{rc}}$
incrementally reduce fit quality. Fit quality recovers only for
$[\text{BK}_{140}^{\text{P},x_{\text{eff}}}]_{\text{rc}}$ and relies
on the freedom gained once one allows for correlator shapes away from
the pseudo-scaling behavior. See Table~\ref{tab:pre-asymp-total}.
\begin{table}
 \centering
 \begin{tabular}{lllllll} $\xbj\leq 0.01$ \T & & $\text{BK}_{\text{phys}}^{\text{A}}$ & $\text{BK}_{\text{140}}^{\text{A}}$ &  $\text{BK}_{\text{140}}^{\text{A},x_{\text{eff}}}$
   & $[\text{BK}_{\text{140}}^{\text{A},x_{\text{eff}}}]_{\text{rc}}$  &  $[\text{BK}_{\text{140}}^{\text{P},x_{\text{eff}}}]_{\text{rc}}$ \\ [2mm]
   &  $\Lambda$ $[\mev]$ & 87.3 & 59.0  & 52.7 & 31.2 & 241   \\
   $Q^2 \leq 45$  &  $\chi^{2}/200$ & 0.82 & 0.95 & 1.01 & 2.23 & 0.97  \\
   $\gev^2$   & $\sigma_{0}$ $[\gev^{-2}]$ & 55.68  & 72.3 & 77.3 & 79.3 & 81.6 
   \\ [2mm]
   &  $\Lambda$ $[\mev]$ & 93.7  & 68.9  & 63.1 & 52.4  &  241   \\
   $Q^2 \leq 650$ \T   &  $\chi^{2}/230$ & 0.86 & 1.02 & 1.09 & 3.42 & 1.01 \\
   $\gev^2$  & $\sigma_{0}$ $[\gev^{-2}]$ & 55.91 & 69.9 & 74.2 & 71.4  & 81.7
   \\ 
  \end{tabular}
  \caption{A comparison of the fits to the total cross
section. Parameters are based on the full range $\xbj \le .01$,
$Q^2\le 650\ \mev$. The superscripts ``A'' and ``P'' refer to the
asymptotic and pre-asymptotic fits, respectively. The quark masses are
$m_{u,d,s}=$ $\{3,5,105\}$ $\mev$ (physical) or $m_{u,d,s}=$
$\{140,140,140\}$ $\mev$ (ad hoc).}
\label{tab:pre-asymp-total}
\end{table}
The first qualitative difference of the fit scenarios is captured
in a plot of evolution speeds (left panel in
Fig.~\ref{fig:two_scenarios}) which indicates the parts of
trajectories used in the fit of~\cite{Albacete:2009fh},
$[\text{BK}_{140}^{\text{P},x_{\text{eff}}}]_{\text{rc}}$, and our
favored asymptotic fits. The $R_s\Lambda$ ranges of the successful
pre-asymptotic and asymptotic fits are strongly shifted against each
other as required by the constraints on evolution speeds.
Note that the fit interval matched onto data on the pre-asymptotic fit
trajectory ends before the asymptotic shape is reached (and the
correlator shape would have become too shallow): the fit strategy
applied in the pre-asymptotic fit
$[\text{BK}_{140}^{\text{P},x_{\text{eff}}}]_{\text{rc}}$~\cite{Albacete:2009fh}
relies on features of the initial condition
\emph{throughout}. Strikingly, the evolution speed increases over the
whole fit interval of the pre-asymptotic fit, while it monotonically
decreases in the asymptotic case. In addition, the dipole cross
section of $[\text{BK}_{140}^{\text{P},x_{\text{eff}}}]_{\text{rc}}$
interpolates between a steep, almost GB-W shape at $\xbj=.01$, and a
final shape at the small $\xbj$ end of the data range that remains
still steeper than the (too shallow) pseudo-scaling shape obtained
from running coupling BK evolution without the energy conservation
correction . (See Fig.~\ref{fig:fit_comparison}, left panel.)
Note that detailed short distance information in this fit is
predominantly imprinted through the choice of initial
condition. Evolution effects, which propagate into the UV just like in
the linear BFKL case, have not yet reached the short distance tail of
the correlators. This is different in our asymptotic fits, where the
shape is determined entirely by the structure of r.h.s. of the
evolution equation, which in its entirely is based on a (highly
resummed) perturbative calculation.\footnote{Operationally this is
  imprinted on the solution by having the solution evolve into the
  asymptotic region over a long $Y$ interval before the fit range is
  reached and the correlators have reached pseudo-scaling
  shapes. Practically this will only affect the UV within the BFKL
  diffusion radius, but this easily covers the range relevant to HERA
  fits.} To highlight the actual differences at short distances we
show part of the short distance asymptotics of the fits by plotting
$\frac{N_Y^{q\Bar q}(r)}{r^2}$ in Fig.~\ref{fig:fit_comparison}, right
panel.\footnote{For a direct comparison in momentum space, consult
  App.~\ref{sec:transformations}, in particular
  Fig.~\ref{fig:compare_quark_gluon}, both the middle and right
  panels.}

We conclude our comparison with a look at the diffractive cross
sections, see Table~\ref{tab:pre-asympt-diffr}. 
\begin{table}[htb]
  \centering
  \resizebox{\textwidth}{!}{
  \begin{tabular}{lllllllll}
    $\xp\leq 0.01$ & $B_d$ & $\alpha_s$ & $\beta\leq 0.5$ & $\beta>0.5$ & Total & $Q^2\! \leq\! 10$ & $10\! <\! Q^2\! \leq\! 45$ & $Q^2\! >\! 45$ \B  \\ 
    $\chi^2 / \text{dof}$    & &  &  $\chi^2 /153$ & $\chi^2 / 215$ & $ \chi^2/368 $ & $\chi^2 /97$ & $\chi^2 /174$ & $\chi^2 /97$   \\ [2mm]
    $\text{BK}_{\text{140}}^{\text{A}}$, fact. & 6.30  & 0.40 & 1.39  & 1.37 & 1.38 & 1.53  & 1.35  & 1.48 \\ [1mm]
    $[\text{BK}_{\text{140}}^{\text{P}}]_{\text{rc}}$ fact. & 7.47 & 0.34  & 1.29 & 1.22 & 1.25 & 1.50  & 1.17  & 1.15  \\ 
 \end{tabular}
}
\caption{The fits to the diffractive data with the parameters shown in
  Table~\ref{tab:pre-asymp-total}. Tag ``fact.'' refers to the globally factorized
  case~\eqref{eq:globally-factorized}, however, with the large-$N_c$ replacement $C_f\rightarrow N_c/2$.} 
\label{tab:pre-asympt-diffr}
\end{table}
The pattern of fit quality shown in the table repeats what we have
already seen for the total cross section: 
\begin{itemize}
\item The asymptotic fits shown in
  Sec.~\ref{sec:lessons-from-total-cross-section}
  and~\ref{sec:lessons-from-diffr-data} show a good fit quality, also
  when restricted to $\xbj\leq 0.01$ and $Q^2\leq 650$ -- i.e. the
  good fit quality shown earlier is \emph{not} driven by the large $x$
  and $Q^2$ part of phase space.  The choice $m_{u,d,s}=140\ \mev$ for
  the light flavors, although popular in the literature, is not
  optimal and the redefinition of the Bjorken variable $x_{\text{eff}}
  = \xbj(1+4m_f^2/Q^2)$ actually reduces the fit quality even
  further. Therefore, in Figs.~\ref{fig:fit_comparison} the curves
  corresponding to the modified BK evolution are based on
  $\text{BK}^{\text{A}}_{\text{phys}}$ in which
  $x_{\text{eff}}=\xbj$. The fit to the diffractive data is fairly
  good.
\item Asymptotic fits without the energy conservation correction
  ($[\text{BK}]_{\text{rc}}$ of
  Secs.~\ref{sec:lessons-from-total-cross-section}
  and~\ref{sec:lessons-from-diffr-data} or
  $[\text{BK}_{\text{140}}^\text{A}]_{\text{rc}}$ introduced here)
  result in a poor fit even within the range $\xbj \leq 0.01$ and $Q^2
  \leq 45$ $\gev^2$. The fit within the wider range $Q^2\leq 650\
  \gev^2$ is a failure. The modification $m_f = 140\rightarrow 5$
  $\mev$ improves the fit slightly, giving $\chi^2/200 = 1.65$ with
  $\Lambda = 52.7$ $\mev$ and $\sigma_0 = 56.1$ $\gev^{-2}$. The
  diffractive data are not considered.
\item For the pre-asymptotic procedure
  $[\text{BK}_{140}^{\text{P},x_{\text{eff}}}]_{\text{rc}}$
  of~\cite{Albacete:2009fh}, a good fit is obtained for $\xbj\leq
  0.01$ and $Q^2\leq 45\ \gev^2$. The fit quality remains good when
  the results are extrapolated to $Q^2=650\ \gev^2$. The fit to the
  diffractive data turns out to be excellent. The ratio
  $\sigma_{\gasp}^{\text{diff}}/\sigma_{\gasp}^{\text{tot}}$ is not
  investigated. It is clear, however, that if both the diffractive and
  the inclusive cross section data are well resolved, then the same
  holds also for their ratio.
\end{itemize}

Close inspection of Table~\ref{tab:pre-asympt-diffr} reveals that
$[\text{BK}_{140}^{\text{P},x_{\text{eff}}}]_{\text{rc}}$ shows the
best $\chi^2$-value for diffractive data presented in this paper. The
origin for this behavior can be discerned from the right panel of
Fig.~\ref{fig:fit_comparison}, which presents the diffractive
structure functions obtained with the asymptotic and pre-asymptotic
approaches. The overall shape of each component is the same in both
approaches but the magnitudes are systematically smaller for
pre-asymptotic case. This is especially pronounced at small $\beta$
and explains the somewhat better match with the data. Unfortunately,
it is the small $\beta$ region where our present theoretical input, in
particular the expression used to approximate the $\qqp g$
contribution is deficient. Thus one cannot make any drastic
conclusions based on this region: further corrections may change the
results fundamentally. Our present theoretical setup is not firm
enough to make use of the more detail information inherent in
diffractive data to further constrain our analysis.

\section{Conclusions}
\label{sec:final}

The main message from this study is that JIMWLK evolution at NLO
allows an asymptotic fit to all HERA data below $\xbj \le .02$, to
both total and diffractive cross sections. This becomes possible only
after \emph{all} NLO corrections -- including the energy conservation
correction -- are included. The energy conservation correction is the
decisive ingredient in this argument: an asymptotic fit without it
barely works for $Q^2 \le 45\ \gev^2$ and fails entirely
beyond. Contrary to that, with the energy conservation correction
included, the fit obtained for $Q^2 \le 45\ \gev^2$ simply
\emph{extrapolates} to a successful fit all the way up to $Q^2= 1200\
\gev^2$, i.e. the low $Q^2$ range determines the result fully, over a
range much larger that expected from naive BFKL based momentum space
diffusion arguments. While our study does not exclude that a
pre-asymptotic component is compatible with data also if one includes
the energy conservation correction, such a feature is not required
with present accuracy.

This observation is quite striking in that such an asymptotic fit is
highly constrained and largely determined by perturbation theory: the
shape of the asymptotic, pseudo-scaling correlators themselves is
fully determined by the structure of the r.h.s. of the evolution
equation, which is the result of a purely perturbative calculation.

The pre-asymptotic fits of~\cite{Albacete:2009fh, Albacete:2009ps},
which omit the energy conservation corrections, on the other hand,
rely strongly on the fact that the evolution trajectory matched onto
data has not yet reached the asymptotic domain. Both evolution speed
and correlator shapes can only be matched onto data as long as many of
the features of the initial conditions are not yet erased. This
remains true even though the \emph{precise} details of these initial
conditions are not too strongly constrained as soon as one allows
evolution trajectories that start with shapes not related to the GB-W
model by evolution.

This puts an additional emphasis on the role of NLO corrections:
without at least the inclusion of running coupling corrections a fit
to HERA data from evolution equations is virtually impossible -- it
becomes \emph{very} hard to find an initial condition that would allow
us to match the full HERA range, mostly due to too fast evolution. NLO
corrections bring both qualitatively new features (such as scale
breaking via the running of the coupling) as well as quantitative
modifications (slowdown via the energy conservation correction) that
make a data fit successively easier, until, with both included, even
the UV details are naturally imprinted by the evolution equation
itself.

All other results are secondary to these observations, but they round
out the picture and point us to what main theoretical improvements are
needed to step beyond what can be done presently.

The total cross section fits work best if one assumes massless quarks
-- a fit with physical quark masses starts to deviate from at $Q^2$
values below mass of the heaviest quark included in the
fit. Phenomenologically only u,d,s, and c quarks need to be considered
and only the charm quark with a mass of the order of $Q_s$ induces a
strong modification. Since all these modifications are concentrated in
the infrared, one should either drop this region from consideration
altogether and accept the loss of fit quality in this region, or adopt
non-perturbative arguments to improve the fit. One such strategy
improves fit quality in the infrared by distorting how one maps the
evolution trajectories onto phase space by replacing $\xbj$ by some
$Q^2$-dependent $x_{\text{eff}}$. Even with rough models for
$x_{\text{eff}}$, plausible charm fractions are achieved. \emph{Any}
refinement that aims at quantitative rather than qualitative
improvements, however, must incorporate nonperturbative information
from outside the scope of JIMWLK evolution.

The asymptotic fits to diffractive data use the parameters already
determined via the total cross section. The formalism clearly requires
contributions from NLO impact factors to fill in phase space at
$\beta\to 0$. Without them a fit to data is \emph{not} possible. While
this was known empirically already from fits within the GB-W model,
from an evolution perspective this becomes a consistency requirement.
As soon as more than one Fock-component comes into play the details of
how one models the impact parameter dependence of their scattering
begins to affect fit quality: the choice of model for this
non-perturbative aspect starts to affect the relative weight of these
Fock-components in the associated cross section. One of the few direct
experimental constraints in this respect comes from the ratio of total
to diffractive cross sections, and can be accommodated easily with
Gaussian profiles no matter if one uses a globally factorized form or
enforces exact Casimir scaling on the other extreme.

At present, two issues hamper precision fits that involve differential
cross sections with exclusive final states: the lack of complete NLO
impact factors, and a consistent treatment of the impact-parameter
dependence based on non-perturbative input. Only the first of these
issues is sure to be resolved in the near future, but it requires
considerable effort: The full NLO corrections to JIMWLK evolution
and the NLO impact factors have to be calculated. Beyond that a
treatment of resummed collinear corrections needs to be formulated
that eliminates any double counting issues that affect our current
inclusion of the energy conservation corrections. All of these can be
addressed within perturbation theory. Any progress on the impact
parameter dependence requires is a different matter altogether.

\section*{Acknowledgments}

J.K. and K.R. acknowledge the support from the Academy of Finland
grant number 1134018. J.K. has also been supported by Jenny and Antti
Wihuri Foundation and Oulu University Scholarship Foundation.

\appendix

\section{Kinematics and common approximations}
\label{sec:kin-approx}

For completeness we include a brief description of the kinematical variables
involved in the inclusive and diffractive deep inelastic scattering
processes.

Only two independent variables are required for the process
$\gamma^{*}(q)\ p(P)\rightarrow \text{anything}$, the photon
virtuality $Q^2$ and the Bjorken variable $x$:
\begin{subequations}
\label{eq:dis_invariants}
\begin{align}
Q^2 & := -q^2  := -(k-k')^2  \\
\xbj & :=  \frac{Q^2}{2P\cdot q}
\end{align}
\end{subequations}
where, at leading twist, $\xbj$ carries the interpretation of the
momentum fraction carried by the quark inside the target that is
struck by the virtual photon. The reduction to two kinematic variables
holds, wherever
\begin{align}
\label{eq:s-simplify}
P^2/Q^2 = m_p^2 /Q^2 \ll 1 \hspace{1cm}\text{and}\hspace{1cm} 1/\xbj \gg 1
\end{align}
to guarantee that $\sgp$, the total energy squared of the $\gamma^{*}(q)\
p(P)$-subprocess (frequently denoted $W^2$) can be expressed in terms of $Q^2$ and $\xbj$ only:
\begin{align}
\sgp &= (P+q)^2 =P^2 +2P\cdot q-Q^2 \approx Q^2 \left(\frac{1}{\xbj}-1\right)
\ . 
\end{align}
The conditions~\eqref{eq:s-simplify} are satisfied for the majority of
the experimental data.

In addition to the variables in Eqs.~\eqref{eq:dis_invariants}, two
additional kinematic variables are required to describe the kinematics
of the diffractive process $\gamma^{*}(q)\ p(P)\rightarrow X(M_{X})\
p(P') $:
\begin{subequations}
\begin{align} 
\label{eq:diff_variables} 
 x_{\mathbbm{P}} &= \frac{(P-P')\cdot
    q}{P\cdot q}= \frac{Q^2 + M_X^2 - t}{Q^2 + \sgp -m_p^2} \approx
  \frac{Q^2 + M_X^2}{Q^2 + \sgp}\ ,
  \\
  \beta &= \frac{-q^2}{2\ (P-P')\cdot q} = \frac{Q^2}{Q^2 + M_X^2 -t}
  \approx \frac{Q^2}{Q^2 + M_X^2} = \frac{\xbj}{x_{\mathbbm{P}}}\ ,
\end{align}  
\end{subequations}
where $x_{\mathbbm{P}}$ is the fraction of the proton four momentum
carried by the colorless diffractive exchange called the Pomeron. The
variable $\beta$ is an analogue of the $\xbj$ for the diffractive
system: it is the momentum fraction of the Pomeron carried by the
interacting parton inside the Pomeron. The variable $M_X$ is the
invariant mass of the diffractive hadronic final state denoted by $X$
and $t=(P-P')^2 \leq 0$ is the squared four momentum transfer.
Experimentally $\vert t\vert \ll Q^2,\ M_X^2$, thus $t$ is set to zero
in the above equations in addition to the proton mass $m_p$.

\section{Parameter optimization}\label{sec:definitions}

\begin{SCfigure}[50]
 \includegraphics[width=.40\linewidth]{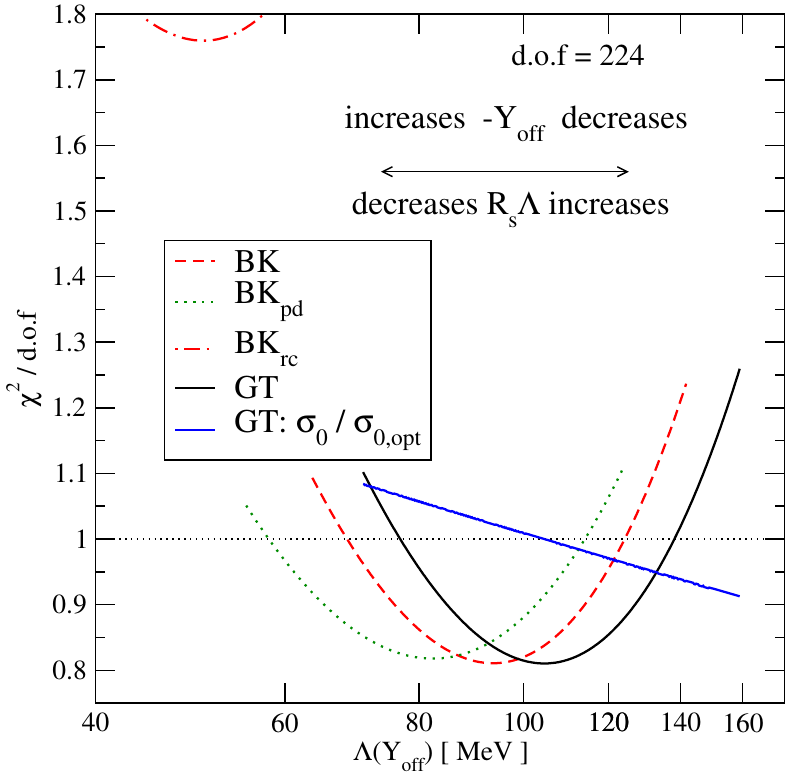}
 \caption{$\chi^{2}/\text{dof}$ as a function of
   $\Lambda(Y_{\text{off}})$.  Each evaluated $\chi^{2}/\text{dof}$
   corresponds to a different choice of $N(Y=0,r)$ with the optimal
   $\Lambda$ and $\sigma_0$. The line at $\chi^{2}/\text{dof} = 1$
   shows the extraction point of the error bars for $\Lambda$ (see
   Figs.~\ref{fig:rs_vs_lambda}). In the asymptotic approach the
   offsets $Y_{\text{off}}$ are actually negative meaning that one
   discards a certain number of configurations from the beginning of
   the evolution (typically $Y=0$ is in the pseudo-scaling stage). The
   horizontal crossing line (blue) indicates the normalization
   dependence via $\sigma_0/\sigma_{0,\text{opt}}$ where
   $\sigma_{0,\text{opt}}=55.33$ $\gev^{-2}$ is the normalization of
   the best overall $\chi^2/\text{dof}$. This log-linear dependence of
   the normalization is common to all descriptions. Note the red
   dot-dashed curve in the top left corner that corresponds to the
   data fit without the energy conservation
   correction. \vspace{.45cm}}
 \label{fig:LAMBDA_vs_chi}
\end{SCfigure}

The total error for the inclusive data is obtained by adding the
systematic error quadratically to the statistical error as follows,
\begin{align}\label{eq:tot_error}
  \epsilon_{\text{tot}} = \pm
  \sqrt{(\text{stat})^{2}+\text{max}[(\text{sys}_{\pm})^{2}]}\quad
  ;\quad\chi^{2}/\text{dof} =
  \sum_{i}\frac{(F_{2,\text{exper}.}^{i}-F_{2,\text{theor}.}^{i})^{2}}{
    (\epsilon_{\text{tot}}^{i})^{2}}
  \Big/ \text{dof}\ ,
\end{align}
which results in symmetric error bars. For the inclusive data this is
fine since only a small number of data points have asymmetric
systematic errors. However, in the case of the diffractive data the
asymmetric error bars are kept due to the large differences between
$|\text{sys}_{+}|$ and $|\text{sys}_{-}|$: a bias of this size in the
error propagation cannot be ignored. A common $\chi^2$-test for the
goodness of fit is applied, shown on the right
in~\eqref{eq:tot_error}. The parameter optimizing procedure is
illustrated in Fig.~\ref{fig:LAMBDA_vs_chi} where $\chi^2 /\text{dof}$
is plotted as a function of the (correlated) parameters
$\Lambda(Y_{\text{off}})$.

In some figures and tables we show the error bars for the BK based
descriptions. The origin of these is the data fits and they are based
on the arbitrary choice made by us. Even though there are three
parameters to be fitted, it turned out that the procedure of seeking
the minima of $\chi^2$ is effectively one parameter fit: two of the
parameters are correlated and the third one is just a trivial
normalization. We illustrate this in Fig.~\ref{fig:LAMBDA_vs_chi}
where $\chi^2 /\text{dof}$ is plotted as a function of the correlated
parameters $\Lambda(Y_{\text{off}})$. The horizontal dotted line
indicates the point from where the error bars are extracted: it is
\emph{chosen to be} at $\chi^2 /\text{dof}=1$.

\section{Numerical implementation of the energy conservation
  correction}\label{sec:energy-conservation}

The numerical treatment of the BK- and Gaussian truncations of JIMWLK
evolution is identical, an appropriate invertible map relates the two
truncations for $q \Bar q$-dipoles~\cite{Kovchegov:2008mk}, despite
any differences encountered in the application to more general
correlators. This statement in fact extends to the inclusion of the
energy conservation corrections as shown in
Eqs.~(\ref{eq:bks_modified}) and (\ref{eq:GT_modified}). Since the
actual numerical implementation used is done in BK-form, we phrase the
discussion here in BK form only. With the exception of the energy
conservation correction, the discretization and the numerical
evaluation of the evolution equation
follow~\cite{Kovchegov:2008mk}. Since the computational cost of a
single right-hand side evaluation scales like $N_{\text{parent}}
\times N_{\text{daughter}}\times N_{\text{angle}}$ (where $N_i$ denote
the number of grid points in polar coordinates), optimization of the
number of discretization points is highly recommended when solving the
modified BK equation\footnote{High computational cost of the R.H.S
  evaluation in the iterative procedure together with the requirement
  a small rapidity step $\delta Y$ makes the overall numerical cost of
  a single evolution trajectory far too large.}.

The $Y$-derivative appearing on the r.h.s. of
Eq.~(\ref{eq:bks_modified}) can not be evaluated directly, to cope
with it we have implemented an iterative procedure. We first write the
evolution equation in finite difference form
\begin{align}\label{eq:iterate}
  \frac{S_{i+1}^{k}-S_i}{\delta Y} =
  f[S_i]-\left[\frac{f[S_{i+1}^{k-1}]-f[S_i]}{\delta
      Y}\right]\equiv F[S_{i+1}^{k-1}]\quad ;\quad f[S]=\text{r.h.s.
    of BK}\ ,
\end{align}
where the second term on the r.h.s. represents the energy conservation
correction.

In~(\ref{eq:iterate}), $S$ is a function of dipole size and $i$ labels
discrete rapidity steps. $k$ is used to label iteration steps at fixed
rapidity used to deform the solution of the equation without energy
conservation correction, i.e. $S_{i+1}^0$, defined via
\begin{align}
  \label{eq:initialization}
  \frac{S_{i+1}^{0}-S_i}{\delta Y} =
  f[S_i]
\end{align}
into a solution of the full equation~(\ref{eq:iterate}) at $S_{i+1} :=
S_{i+1}^{k\to\infty}$, where the limit $k\to\infty$ assumes convergence of the
procedure.  Leaving convergence issues aside for the moment, the iteration
with~\eqref{eq:iterate} proceeds as follows:
\begin{align}\label{eq:steps}
  \nonumber &1)\ \text{evaluate}\ f[S_{i+1}^{k-1}]\quad 2)\
  \text{calculate}\ F[S_{i+1}^{k-1}]\quad 3)\ \text{evolve}\
  S_{i+1}^{k}=S_i^0 +\delta Y F[S_{i+1}^{k-1}]
  \\
  &4)\ \text{return to 1) with new configuration from step 3).}
\end{align}
The iterative solution is accepted as the solution $S_{i+1}$ for the next time
step at some finite $k$, once left and right hand sides of the discretized
equation~(\ref{eq:iterate}) agree to some desired accuracy
$\epsilon$
\begin{align}\label{eq:ending}
  \frac{S_{i+1}^{k}-S_i^0}{\delta Y} -
  F[S_{i+1}^{k}]  < \epsilon 
\ ,
\end{align}
or equivalently $F[S_{i+1}^{k}]- F[S_{i+1}^{k-1}] < \epsilon$.

Note that the definition of evolution speed $\lambda$ in
Eq.~(\ref{eq:lambda-def}) can be used to directly translate the accuracy
criterion of Eq.~(\ref{eq:ending}) into the error implied for $\lambda$:
\begin{align}\label{eq:eestimate}
  \lambda_\epsilon := 2\int\frac{d r}{r}[
  \text{l.h.s(\ref{eq:iterate})}-\text{r.h.s(\ref{eq:iterate})}]=
  2\int\frac{d r}{r}\left[ \frac{S_{i+1}^{k}-S_i^0}{\delta
      Y}-F[S_{i+1}^{k}]\right]\ .
\end{align}
The main issue with this procedure is that it is not stable for
arbitrary shapes of the dipole function $S$: convergence
towards~\eqref{eq:ending} occurs only for a small number of iteration
steps before the iterations start to diverge in the sense of an
asymptotic series. Convergence can be improved by reducing $\delta Y$
(to, in general, impractical values) with the pseudo-scaling region
showing the best convergence properties.

A numerical exploration reveals that it is the nonlinear region $\bm
r^2 > R_s^2(Y)$, where $N_Y(\bm r) = 1- S_Y(\bm r)$ approaches one,
that is least stable. The source of the instability is the
nonlinearity, but the convergence properties of the asymptotic series
and the obtainable accuracy can be improved by reducing the step size
$\delta Y$.

The manner in which $\delta Y$ affects convergence of the iteration
procedure is slightly peculiar since the energy conservation term [in
the update step (Eq.~(\ref{eq:steps}), step 3))] carries no explicit
overall power of $\delta Y$. However, its initial size (in the first
iteration step at $k=1$) is determined by the difference between $S_i$
and $S_{i+1}^0$ as induced by~(\ref{eq:initialization}). This
difference \emph{is} proportional to $\delta Y$ and thus step size
imprints itself on the whole subsequent iteration procedure:
convergence can be improved by reducing $\delta Y$.

The main criterion for convergence therefore is the difference between
$S_i$ and $S_{i+1}^0$, and this not only depends on $\delta Y$ itself,
but also on the shape of $S_i$: a solution near the pseudo-scaling
regime generically leads to a smaller energy conservation correction
and better stability than a solution that is far from pseudo-scaling,
such as the Gaussian shape used by GB-W.

Since stability is most precarious at $\bm r^2> R_s^2(Y)$, there is an
interplay between IR regulators applied to tame the Landau pole and
the convergence of the iteration procedure: Irrespective of the shape
of the solutions, stability is improved whenever $\delta Y K_{\bm{x z
    y}} R^{\text{eff}}_{\bm{x z y}}$ is small. $R^{\text{eff}}_{\bm{x
    z y}}$ plays the role of an effective coupling and we find that
\begin{itemize}
\item the linear region $r\lesssim R_s(Y)$ is stable even with the
  relatively large fixed coupling $\alpha_s=0.4$. A fast convergence occurs
  in all cases.
\item the non-linear region $r>R_s(Y)$ is unstable even with the
  relatively small fixed coupling $\alpha_s=0.2$. The case
  $\alpha_s=0.4$ is already challenging and requires an impractically
  small step size $\delta Y$. The convergence is generally slow.
\end{itemize}
For the realistic case with the full running coupling kernel (whose
size is essentially determined by the size of the parent dipole $r$)
this implies that the problems caused by the presence of the
nonlinearities are exacerbated by the running of the coupling. Thus,
the iteration is quick and stable at $r\lesssim R_s(Y)$ thanks to both
the smallness of the kernel and absence of nonlinearities. In contrast
to that, the region $r>R_s(Y)$ is difficult (especially at low
rapidities) since the kernel is large and the equation is dominated by
non-linear effects.

In practical terms, the convergence properties of the iteration
procedure preclude the use of~(\ref{eq:steps}) away from the
asymptotic line. We have applied~(\ref{eq:steps}) in the
pseudo-scaling region, iteratively reducing $\delta Y$ to push the
step of minimal error $k_{\text{min}}$ to larger $k$ and minimize the
error in a brute force approach to set a baseline. As is typical with
iteration procedures, convergence properties can be strongly affected
by a modification of the iteration procedure. As a compromise between
speed and accuracy near the pseudo-scaling region we have amended
steps $1)-2)$ of~\eqref{eq:steps} by a set of re-weighting steps:
\begin{align}\label{eq:mix_it}
  \nonumber &3)\ S_{i+1}^{k}=S_i^0 +\delta Y
  F[S_{i+1}^{k-1}]\quad\quad;\quad\quad
  G_{i+1}^{k}=2S_{i+1}^{k-1}-S_{i+1}^{k}
  \\
  \nonumber &4)\ \text{evaluate}\ f[S_{i+1}^{k}]\ \text{and}\
  \text{calculate}\ F[S_{i+1}^{k}]
  \\
  \nonumber &5)\ S_{i+1}^{k+1}=S_i^0 +\delta Y
  F[S_{i+1}^{k}]\quad\quad;\quad\quad
  G_{i+1}^{k+1}=2S_{i+1}^{k}-S_{i+1}^{k+1}
  \\
  \nonumber &6)\ \text{calculate}\ {\cal
    W}_{i+1}^{k}=S_{i+1}^{k}-\Delta_s \left( G_{i+1}^{k}- S_{i+1}^{k}
  \right)/\left(\Delta_g-\Delta_s\right)
  \\
  \nonumber &\text{where}\ \Delta_s=\sum_r (S_{i+1}^{k+1}-
  S_{i+1}^{k})\quad ;\quad\Delta_g=\sum_r (G_{i+1}^{k+1}- G_{i+1}^{k})
  \\
  &7)\ \text{return to 1) with weighted}\ {\cal W}_{i+1}^{k}\ \text{of 6)
    taking the place of $S_{i+1}^k$}
\end{align}
The procedure terminates when$||\Delta_s/\Delta_g| -
1|<\epsilon$. Then further iterations of ${\cal W}_{i+1}^{k}$ do not
lead to any improvement\footnote{The condition $\Delta_s \approx
  -\Delta_g$ yields ${\cal W}_{i+1}^{k}\approx
  (S_{i+1}^{k}+G_{i+1}^{k})/2 $ regardless of $|\Delta_{s,g}|\ll 1$
  which obviously is a bad solution for any $r$.} and one proceeds
with $S_{i+1}^{k}$ in step 4). Initially (typically) $|\Delta_g| \gg
|\Delta_s|$ and so $G_{i+1}^{k}$ gets weighted less than $S_{i+1}^{k}$
in ${\cal W}_{i+1}^{k}$.

\begin{figure}
 \centering
 \includegraphics[width=.328\linewidth]{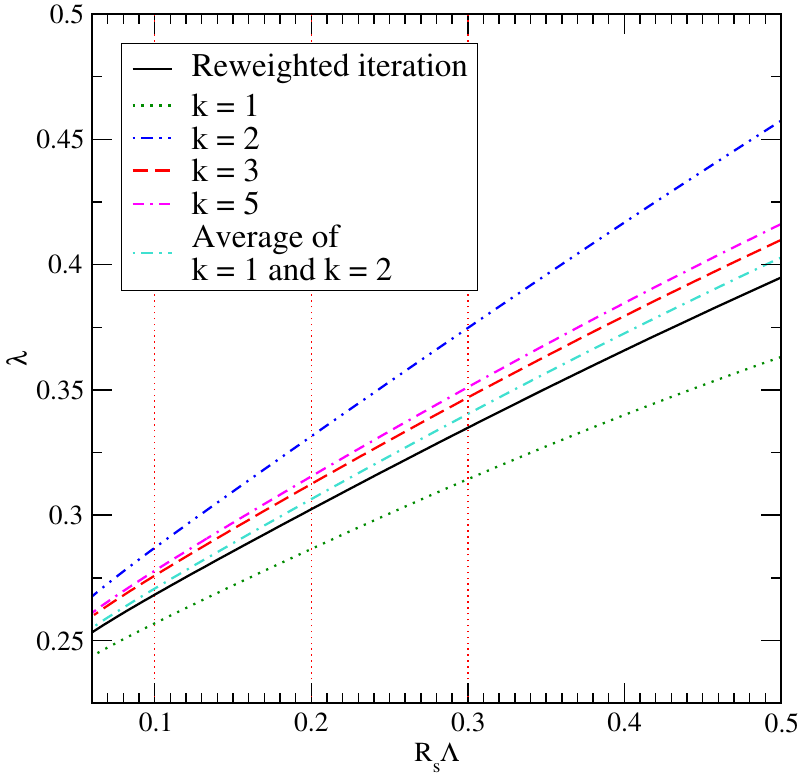}
 \hfill
 \includegraphics[width=.328\linewidth]{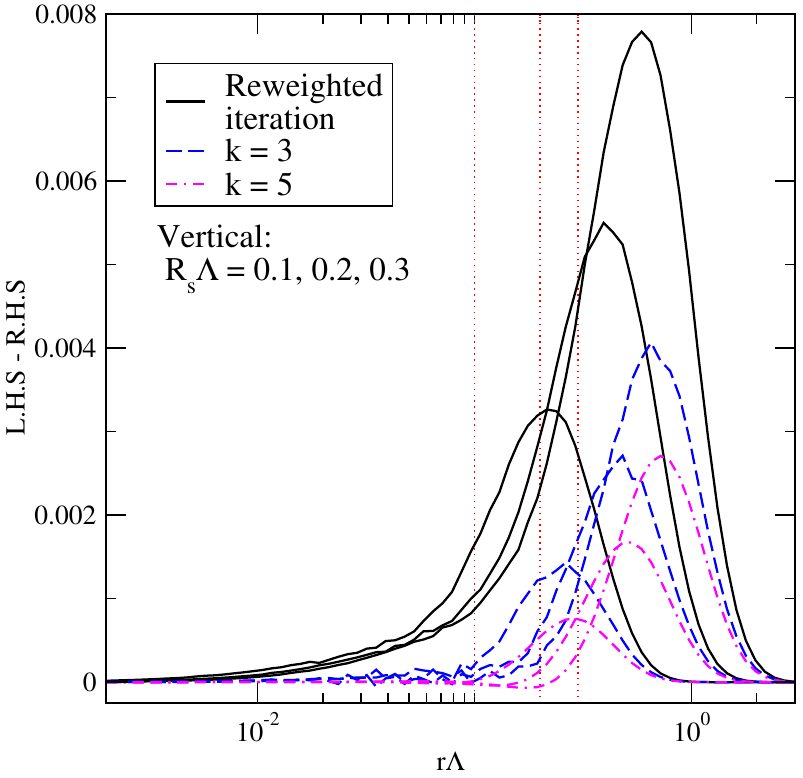}
\hfill
 \includegraphics[width=.328\linewidth]{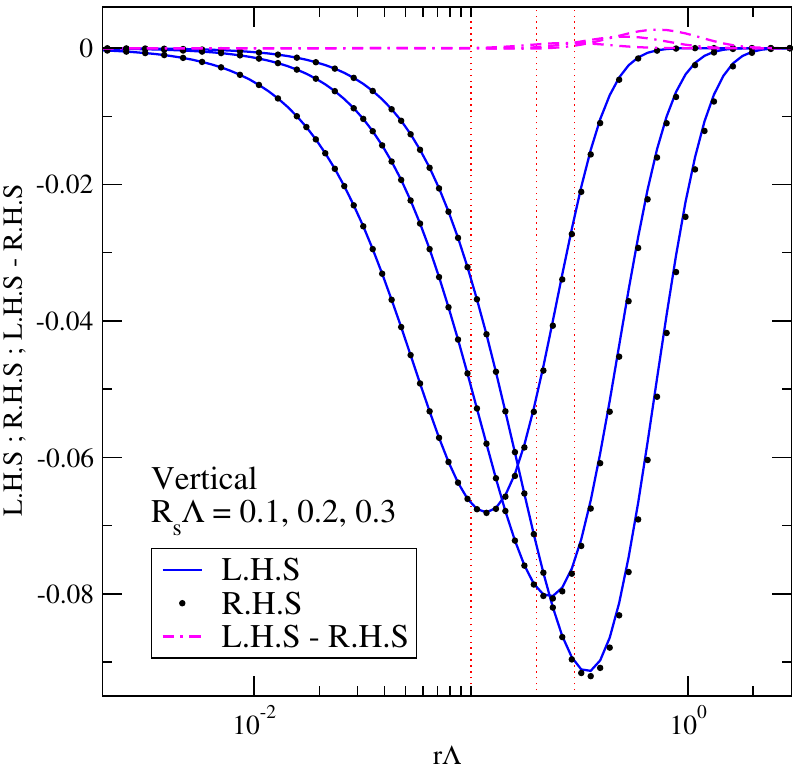}
 \caption{\textbf{Left:} $\lambda$: a comparison of iterations $k=1,2,3,5$
   of~\eqref{eq:steps}, the average of $k=1,2$ and the
   procedure~\eqref{eq:mix_it}. \textbf{Middle:} Accuracy of the solution
   from~\eqref{eq:ending} as a function of $r$ for $R_s\Lambda=$
   $\{0.1,0.2,0.3\}$ (corresponds $R_s\approx\{1,2,3\}$ $\gev^{-1}$). {\bf
     Right:} The solution of the case $k=5$ is, for all practical purposes,
   accurate but the requirement of computing time is huge. L.H.S. and
   R.H.S. refer to the discretized evolution equation~\eqref{eq:iterate}}
 \label{fig:error_estimate}
\end{figure}

\begin{SCfigure}[50]
 \centering
 \includegraphics[width=.40\linewidth]{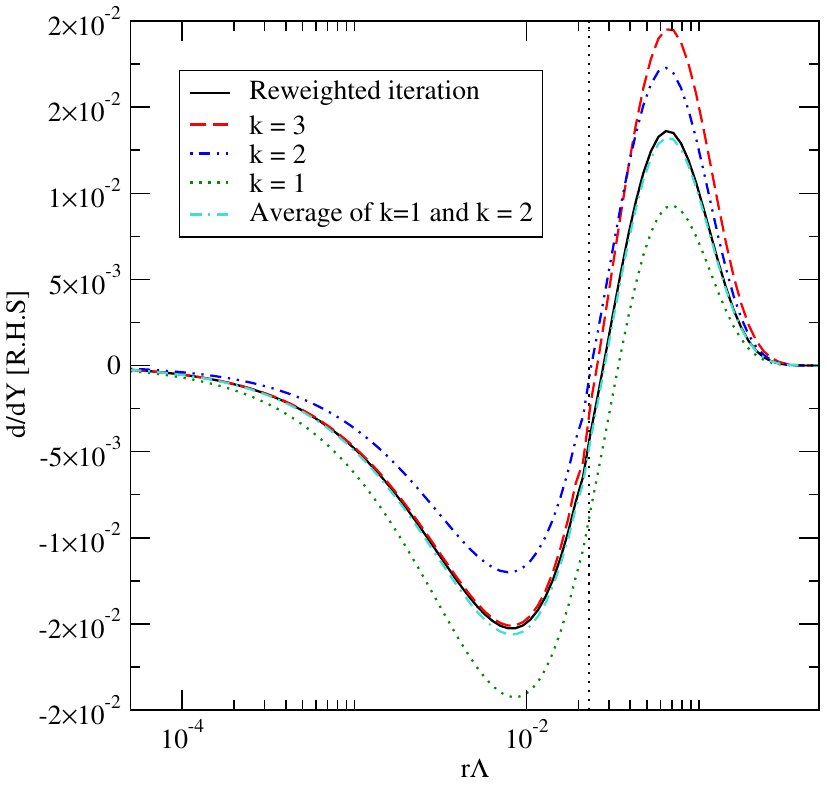}
\hspace{0.0cm}
\caption{The energy conservation correction 
  $\big(f[S_{i+1}^{k-1}]-f[S_i^{0}]\big)/\delta Y$
  for the cases $k=1,2,3$ of~\eqref{eq:steps}, an average of $k=1,2$
  and the procedure~\eqref{eq:mix_it}. The latter approximately
  coincides with the average of $k=1,2$ for all $r$ and thus the
  easiest way to get a fairly good solution is taking the average of
  $k=1,2$. A sharp slowly converging peak at large $r$ is a
  consequence of the non-linearity since it appears even with a
  relatively small fixed coupling $\alpha_s=0.2$. The running
  coupling, being large at these scales, gives rise to the additional
  stability problems.}
 \label{fig:partial_rhs}
\end{SCfigure}

The upshot is high precision at small $r$ after few iterations already at
$\delta Y = 0.0025$, at the price of comparatively limited precision in the
large $r$ part and with very few iteration steps. The procedure always
terminates but can, without additional modifications, not exceed the precision
indicated in Figs.~\ref{fig:error_estimate} and~\ref{fig:partial_rhs}.

We observe that the unmodified method~(\ref{eq:steps}) at $k=1$, i.e. the
crudest approximation, greatly underestimates $\lambda$ whereas the first
iteration $k=2$ overestimates it, as seen in the left panel of
Fig.~\ref{fig:error_estimate}.

From $k=3$, the brute force solution to~(\ref{eq:steps}) is more
accurate at large $r$ than the approximative
solution~\eqref{eq:mix_it}, however, this solution is increasingly
difficult to obtain: in order to keep the iteration stable one has to
reduce the change in shape of $S$ with $Y$ by working near the
pseudo-scaling region,\footnote{This requires some numerical
  optimization of its own.} and \emph{in addition} needs to employ a
very small rapidity step $\delta Y\lesssim 10^{-3}$. To
stabilize~(\ref{eq:steps}) also for $k=4,5$ requires $\delta Y\lesssim
10^{-4},10^{-5}$ respectively which increases the numerical effort
prohibitively. This is illustrated in the right panel of
Fig.~\ref{fig:error_estimate} which compares the three step iteration
of~(\ref{eq:steps}) with the result of the re-weighted result
from~(\ref{eq:mix_it}) obtained with almost an order of magnitude
smaller CPU time. Fig.~\ref{fig:partial_rhs} shows the size of the
energy conservation correction in the different cases. For both
Figs.~\ref{fig:error_estimate} and~\ref{fig:partial_rhs}, the
configurations are selected from the $Y$- or $R_s\Lambda$-range
typically used in the data fits\footnote{In the data fits
  $\Lambda\approx 0.1\ \gev$ and so
  $R_s\Lambda=\{0.1,0.2,0.3\}\rightarrow R_s\approx\{1,2,3\}\
  \gev^{-1}$.}.

The induced error of evolution speed according to~\eqref{eq:eestimate} is
shown in Table~\ref{tab:errors}
\begin{table}[!thb]
  \centering
  \begin{tabular}{llll}
    & $\delta Y$  & $\lambda_\epsilon$ & $\vert \lambda_\epsilon(r\leq R_s)\vert/\lambda_\epsilon$  \\ [1mm] 
    reweighted        & $0.0025$              & $\{0.025,0.018,0.011\}$  & $\{0.256,0.250,0.245\}$        \\
    brute force $k=3$ & $10^{-3}$               & $\{0.010,0.007,0.003\}$  & $\{0.106,0.092,0.079\}$      \\
    brute force $k=5$ & $10^{-5}$    & $\{5.824,3.541,1.561\}\times 10^{-3}$ & $\{6.437,0.448,4.590\}\times 10^{-3}$  \\
 \end{tabular}
 \caption{The error estimates for $\lambda$ based
on~\eqref{eq:eestimate} for different iteration procedures (first
column). The numerical cost is measured by $\delta Y$. Most of the
error comes from $|\bm r|> R_s(Y)$ (last column). The values
correspond to correlation lengths $R_s\Lambda=\{0.3,0.2,0.1\}$.} 
\label{tab:errors}
\end{table} 
for three samples corresponding to $R_s\Lambda=\{0.3,0.2,0.1\}$. The
region $r\leq R_s$ is fairly accurate in all cases and especially the
case $k=3$ is close to the accurate solution even at $r>R_s$. The peak
values (Figs.~\ref{fig:error_estimate}, middle) correspond to
$S_{Y,\bm{xy}}\approx 0.2$. As seen, the iterations $k=4,5$ do not
bring any \emph{essential} improvement and for $k>5$ the iteration
procedure turned out to be highly unstable\footnote{The iteration
  seems to be approaching (alternatingly) the fixed point but the
  requirement of a very small $\delta Y$ makes the brute force method
  impractical.} at $r>R_s$ for any relevant $\delta Y$. 

It should denoted that a convergent solution in the offset
region\footnote{The region of large running coupling that is cut off
  by the parameter $Y_{\text{off}}$.} is easier to obtain with some
other, say stronger, regulator but, however, inside the actual fit
range this kind of modification does not bring any improvement.

\section{Tools to efficiently address diffractive cross sections}
\label{sec:tools-diffr-cross}

A lot of simplifications used in earlier treatments with simple height
one box profiles can be carried over to more general profile models if
the $\bm r$ and $\bm r'$ dependence in the $\bm b$-integrated product
of dipole amplitudes factorizes according
to\footnote{In~\eqref{eq:b-int-prod-of-dip-amp} the constants
  $\sigma_{\cal R}^{(2)}$ are adjusted such that $N_{{\cal
      R},Y}^{(2)}(\bm r) \xrightarrow{\bm r\to \infty} 1$ for
  convenience.}
\begin{align}
  \label{eq:b-int-prod-of-dip-amp}
  \int d^2{\bm b}\ N_{{\cal R},Y}^*(\bm r,\bm b)N_{{\cal R},Y}(\bm
  r',\bm b) = \sigma_{\cal R}^{(2)}\ [N_{{\cal R},Y}^{(2)}(\bm r)]^*
  N_{{\cal R},Y}^{(2)}(\bm r') \ ,
\end{align}
or if such factorized behavior is a good approximation to the full
result c.f.~\eqref{eq:calR-approx-fact}. 

Wherever~\eqref{eq:b-int-prod-of-dip-amp} holds, a momentum space
variant of the cross sections of~\eqref{eq:dsigmadbeta-three-term-sum}
offers an efficient route to perform the data fits. It allows to
pre-calculate most of the numerical integrals in the final expressions
for the diffractive structure function and speeds up the parameter
seeking process tremendously. The tools to do so are collected in the
remainder of this section.
 
\subsection{Diffractive cross sections in momentum space}
\label{sec:diffr-cross-sect-mom-space}

The momentum space expressions for the three terms in
Eq.~\eqref{eq:dsigmadbeta-three-term-sum} read
\begin{subequations}
  \label{eq:mom-space-diffr}
\begin{align}
& \frac{d\sigma^{\gamma^* A\to X p}_{q\Bar q,T}}{d\beta}
= 
\frac{(2\pi)^2\alpha_{\text{em}}}{Q^2\beta}
\sum\limits_f e_f^2
\int\limits_{z^-_f}^{z^+_f}  dz\ 
\frac{ Q^2(z^2+\Bar z^2) N_c
}{32\pi \Bigl(\Bar\beta
-\frac{m_f^2}{\frac{z\Bar z Q^2}\beta}\Bigr)}
\notag \\ & \hspace{.2cm} \times
\int\!\! d^2\bm b\  
\Biggl[
 \int 
  \frac{dq^2}{q^2}{\cal N}^{q\Bar q}_Y(q,b) 
\
\Biggl(
1-2\beta
- \frac{2 m_f^2}{\frac{z\Bar z Q^2}{\beta}} 
-
\frac{(1-2\beta)\, \frac{z\Bar z Q^2}{\beta}-(q^2+2m_f^2)
}{
\sqrt{(\frac{z\Bar z Q^2}{\beta}+q^2)^2
-4(\Bar\beta\, \frac{z\Bar z Q^2}{\beta}-m_f^2)\, q^2}}
\Biggr)
\Biggr]^2
\notag \\ & \hspace{2cm} + 
\frac{(2\pi)^2\alpha_{\text{em}}}{Q^2\beta}
\sum\limits_f e_f^2
\int\limits_{z^-_f}^{z^+_f}  dz\
\frac{\beta m_f^2 N_c}{8\pi z\Bar z}
\notag \\ & \hspace{.2cm}\times
\int d^2\bm b\ \Biggl[
 \int 
  \frac{dq^2}{q^2}{\cal N}^{q\Bar q}_Y(q,b) 
\Biggl(
1
- \frac{\frac{z\Bar z Q^2}{\beta}}{
\sqrt{(\frac{z\Bar z Q^2}{\beta}+q^2)^2
-4(\Bar\beta\, \frac{z\Bar z Q^2}{\beta}-m_f^2)\, q^2}}
\Biggr)
\Biggr]^2  
\\
& \frac{d\sigma^{\gamma^* A\to X p}_{q\Bar q,L}}{d\beta} = 
\frac{(2\pi)^2\alpha_{\text{em}}}{Q^2\beta}
\sum\limits_f e_f^2
\int\limits_{z^-_f}^{z^+_f}  dz\
\frac{\beta 4Q^2(z\Bar z)^2 N_c
}{8\pi z\Bar z}
\notag \\ & \hspace{.2cm}\times
\int d^2\bm b\ \Biggl[
 \int 
  \frac{dq^2}{q^2}{\cal N}^{q\Bar q}_Y(q,b) 
\Biggl(
1
- \frac{\frac{z\Bar z Q^2}{\beta}}{
\sqrt{(\frac{z\Bar z Q^2}{\beta}+q^2)^2
-4(\Bar\beta\, \frac{z\Bar z Q^2}{\beta}-m_f^2)\, q^2}}
\Biggr)
\Biggr]^2 
\\ &
 \frac{d\sigma^{\gamma^* A\to X p}_{q\Bar q g,T}}{d\beta}\bigg\vert_{\text{LL}Q^2} 
= 
\frac{(2\pi)^2\alpha_{\text{em}}}{Q^2\beta}
\frac{\alpha_s \beta C_f N_c}{32\pi^2}
\sum\limits_f e_f^2 
\int_\beta^1 \frac{dz}{z^2\Bar z^2}\left[\left(1-\tfrac{\beta}{z}\right)^2
\!+\!
\left(\tfrac{\beta}{z}\right)^2\right]
\int\limits_0^{Q^2} dk^2\ln\left(\tfrac{Q^2}{k^2}\right)
\notag \\ & \times
\int\!\! d^2\bm b
\Biggl[\int \frac{dq^2}{{q}^2}  {\cal N}^{gg}_Y(q,b)
 \Biggl(z^2 +\Bar z^2
+\frac{q^2}{k^2}
-\frac{(q^2-k^2(1-2z))^2+2 z\Bar z k^4}{k^2\sqrt{(q^2+k^2)^2-4(1-z) q^2 k^2}}
\Biggr)
\Biggr]^2
\end{align}
\end{subequations}
Where we have used the shorthand expressions $\Bar z = 1-z$ and
$z_f^\pm = \frac12(1\pm\sqrt{1-4m_f^2/M_X^2})$. In the massless limit,
these expressions match up with their counterparts
in~\cite{Wusthoff:1997fz,Golec-Biernat:1999qd}\footnote{Use $\alpha_s
  {\cal F}(x,\bm q) = \frac{N_c\sigma_0}{4\pi} {\cal N}_{\Bar q
    q,\ln(1/x)}(\bm q)$ to translate the $q\Bar q$ expressions into
  those of~\cite{Wusthoff:1997fz,Golec-Biernat:1999qd}. The $q\Bar q
  g$-term in the original literature suffers from an additional
  incorrect rescaling by a factor of $(C_f/N_c)^2$ and otherwise
  substitute ${\cal N}_{q \Bar q}$ for ${\cal
    N}_{gg}$~\cite{Marquet:2007nf,GolecBiernat:2008gk}.} once the
corresponding $b$-profiles\footnote{Where ever these papers make use
  of the $d\sigma/dt|_{t=0}$ in their expressions one \emph{must} use
  factorized Gaussian profiles for this procedure, the corresponding
  expressions are valid only in this case.} have been inserted and the
impact parameter integral has been carried out.

The cumbersome expressions in brackets arise from integrations over
the orientation of $\bm q$ of rather simple Fourier expressions for
the McDonald $K_i$ appearing in the coordinate space variants.  The
solutions of the evolution equations enter via ${\cal N}^{\cal
  R}_Y(q,b)$, which are determined from the coordinate space dipole
amplitudes $N^{\cal R}_Y(r,b)$. These are \emph{not} related by a
direct Fourier transform, see Sec.~\ref{sec:transformations} for
definitions and properties.

The benefit of using the momentum space forms of the equations shows
up in the numerical implementation: the required rapidity range of
${\cal N}$ can be pre-calculated whereas the coordinate space
expressions have highly oscillatory integrands in which no part of the
nested integrals can be pre-calculated.

We have cross-checked our results using both coordinate and momentum
space variants numerically. A direct comparison of the quark
contributions at different masses gives an idea of how they affect the
cross sections: non-zero quark masses reduce $\xp
F_{T,q\bar{q}}^{D(3)}$ whereas $\xp F_{L,q\bar{q}}^{D(3)}$ is
practically unaffected. In any case, it is expected that considering
the non-zero quark masses would only lead to a small rescaling of the
normalizations in Eq.~\eqref{eq:dsigmadbeta-three-term-sum}
\begin{SCfigure}[50]
 \centering
 \includegraphics[width=0.52\linewidth]{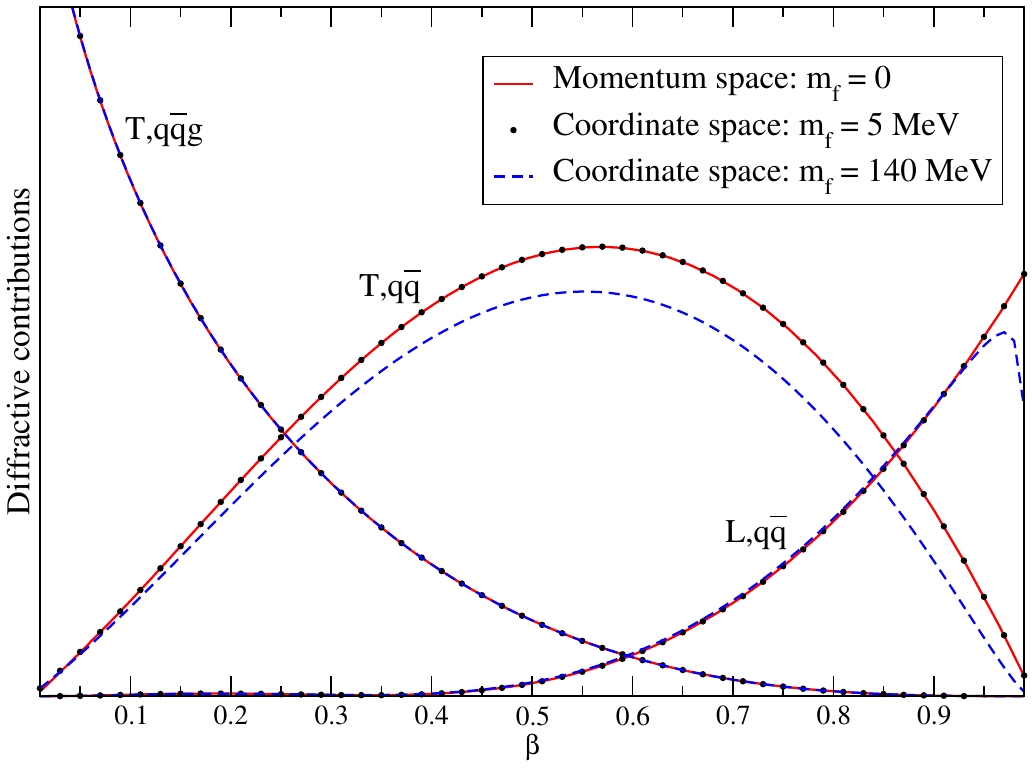}
 \caption{A comparison between the momentum and coordinate space
   formulae of the diffractive final states. Here, the same dipole
   input (a quark dipole) is used in all contributions and the
   parameters $B_d$ and $\alpha_s$ are set to one. In the case of the
   coordinate space equations, three light flavors with equal masses
   $m_{u,d,s} = 5,140$ $\mev$ are considered whereas for the momentum
   space equations $m_f=0$ (coincides perfectly with the case
   $m_{u,d,s}=5$ $\mev$). The curves are calculated with fixed
   $Q^2=15$ $\gev^2$ and $\xp=0.001$.}
\label{fig:mom_vs_co_masses}
\end{SCfigure}

\subsection{Integral transformations for dipole amplitudes}
\label{sec:transformations}

The key ingredient is the non-standard ``Fourier''-transform of the
dipole amplitude (here adapted to the azimuthally symmetric forward
case relevant in conjunction with~\eqref{eq:calR-approx-fact})
\begin{align}
  \label{eq:dipole-amp-via-mom}
  N^{\cal R}_Y(\bm r) = \int\frac{d^2\bm{q}}{\bm q^2} \big(1-e^{i\bm
    r\cdot\bm q}\big) {\cal N}^{\cal R}_Y(\bm q)
\end{align}
(${\cal R}=\Bar q q,gg$ etc. labels the representation;
see~\cite{Kovchegov:2006wf} for a broader exposition on the structure
of the exponentials) together with its inverse\footnote{This can be
  derived via an inverse Mellin
  transform~\cite{Braun:2000wr,Armesto:2001fa}
  (see~\cite{Kutak:2004ym} for a step by step exposition).}
\begin{align}
  \label{eq:dipole-mom-from-coord}
  {\cal N}^{\cal R}_Y(\bm q) = \Big(\frac{d}{d\ln(\bm q^2)}\Big)^2
  \phi_{{\cal R},Y}(\bm q) \hspace{.5cm}\text{where}\hspace{.5cm}
  \phi_{{\cal R},Y}(\bm q) :=\frac1{\pi^2}\int\frac{d^2\bm{r}}{\bm r^2}
  e^{-i\bm r\cdot\bm q} N^{\cal R}_Y(\bm r)\ .
\end{align}
Note that $\cal N$ as used here is dimensionless just as the dipole
amplitudes. This differs from the analogous quantities
in~\cite{Kovchegov:2006wf} or in the work of GB-W.  We note that
${\cal N}/\bm q^2$ is normalized to the saturation value of the dipole
amplitude:
\begin{align}
  \label{eq:n-mom-norm}
  \int\frac{d^2\bm{q}}{\bm q^2} {\cal N}^{\cal R}_Y(\bm q) 
    = N^{\cal R}_Y(|\bm r| \to \infty) \ ;
\end{align}
correspondingly, $S^{\cal R}_Y(\bm r) = \int\frac{d^2\bm{q}}{\bm q^2}
e^{i\bm r\cdot\bm q} {\cal N}^{\cal R}_Y(\bm q)$.
Eq.~(\ref{eq:n-mom-norm}) provides a stringent check for our numerical
tools, which also faithfully resolve the chain of transformations
$N^{\cal R}_Y(\bm r) \rightarrow \phi^{\cal R}_Y(\bm q)
\rightarrow {\cal N}^{\cal R}_Y(\bm q)$ for the GB-W model, where
all the steps can be determined analytically from
Eq.~(\ref{eq:dipole-mom-from-coord}).

\begin{figure}[H]
 \centering
 \includegraphics[width=.32\linewidth]{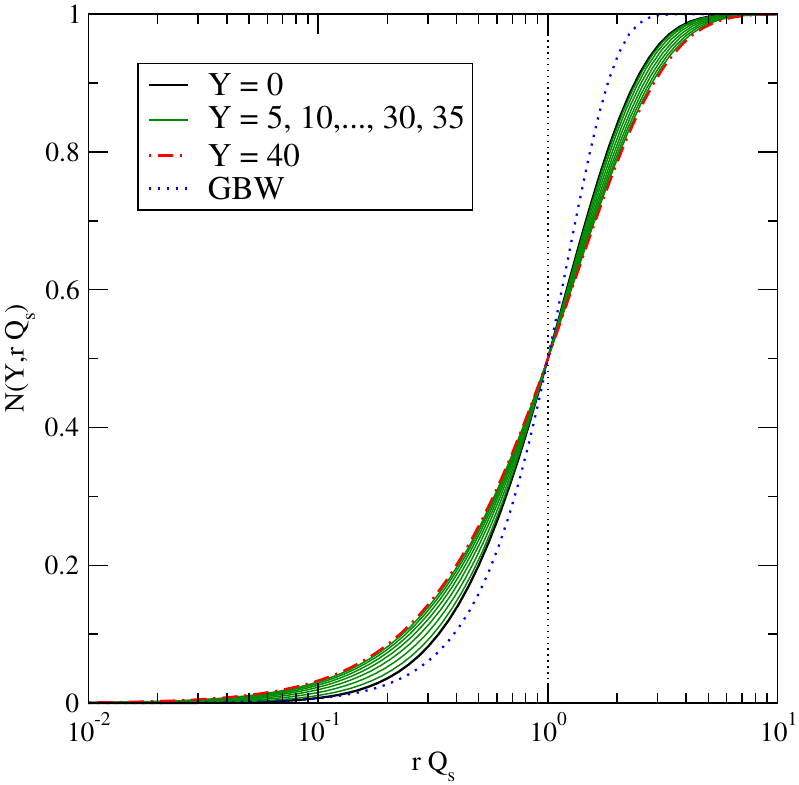}
 \hfill
 \includegraphics[width=.32\linewidth]{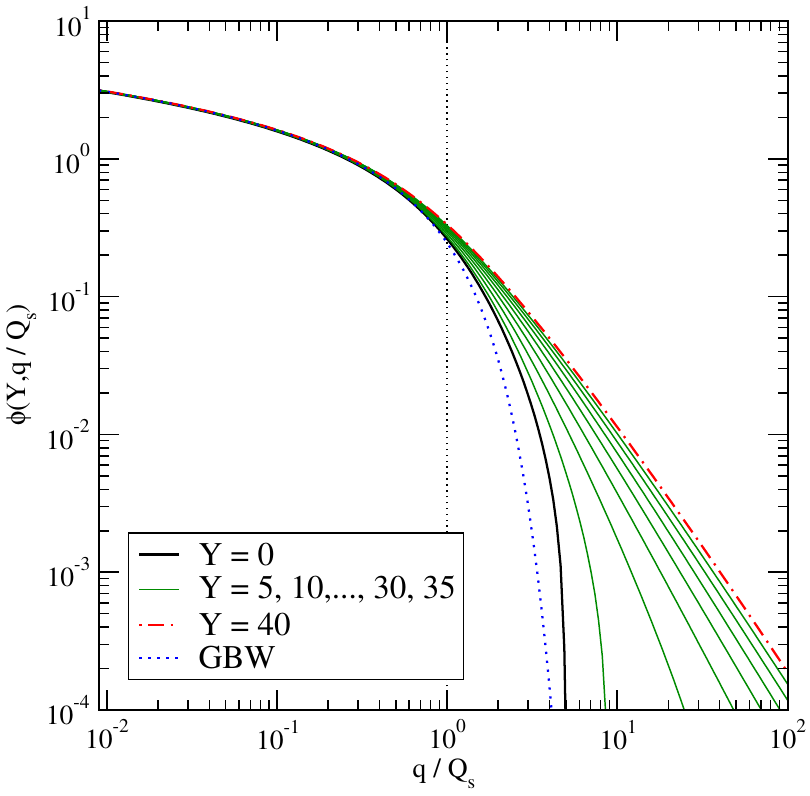}
 \hfill
 \includegraphics[width=.327\linewidth]{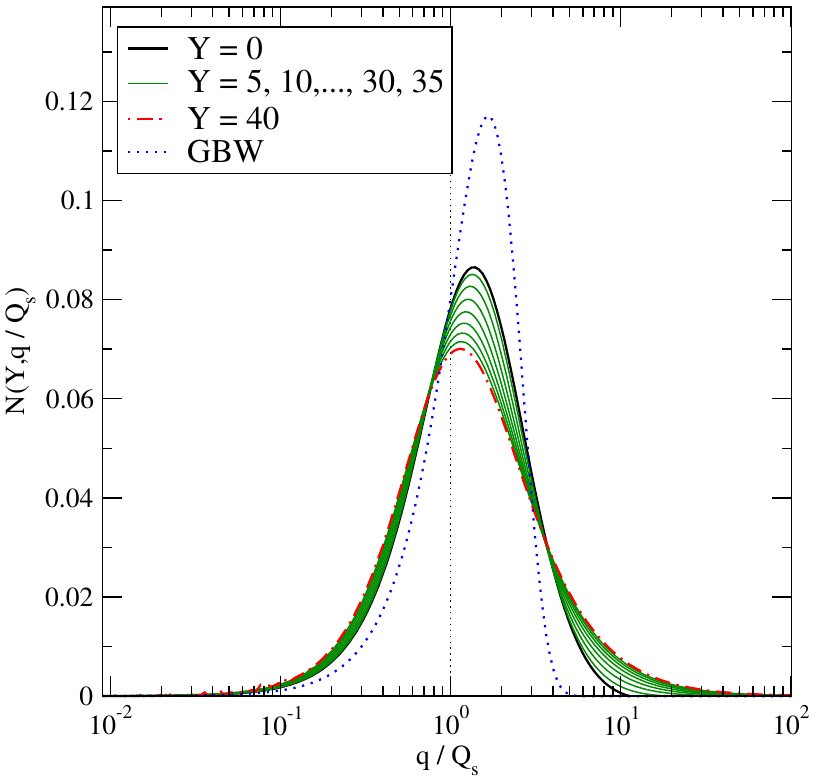}
 \caption{From left to right: $N^{q\Bar q}_Y(\bm r) \rightarrow
   \phi^{q\Bar q}_Y(\bm q) \rightarrow {\cal N}^{q\Bar q}_Y(\bm q)$
   transformations as a function of scaling variables $r Q_s$ and $l_t
   /Q_s$. The vertical lines indicate the position of the saturation
   scale $Q_s = 1/R_s$.  }
 \label{fig:transform_n_phi_caln}
\end{figure}

\begin{figure}[htb]
 \centering
 \includegraphics[width=.32\linewidth]{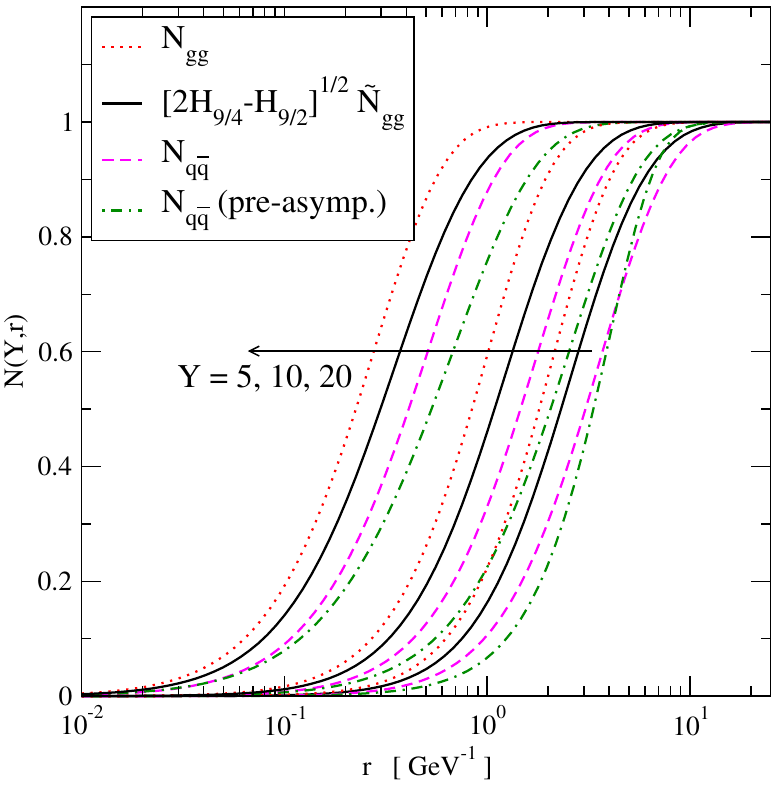}
 \hfill
 \includegraphics[width=.327\linewidth]{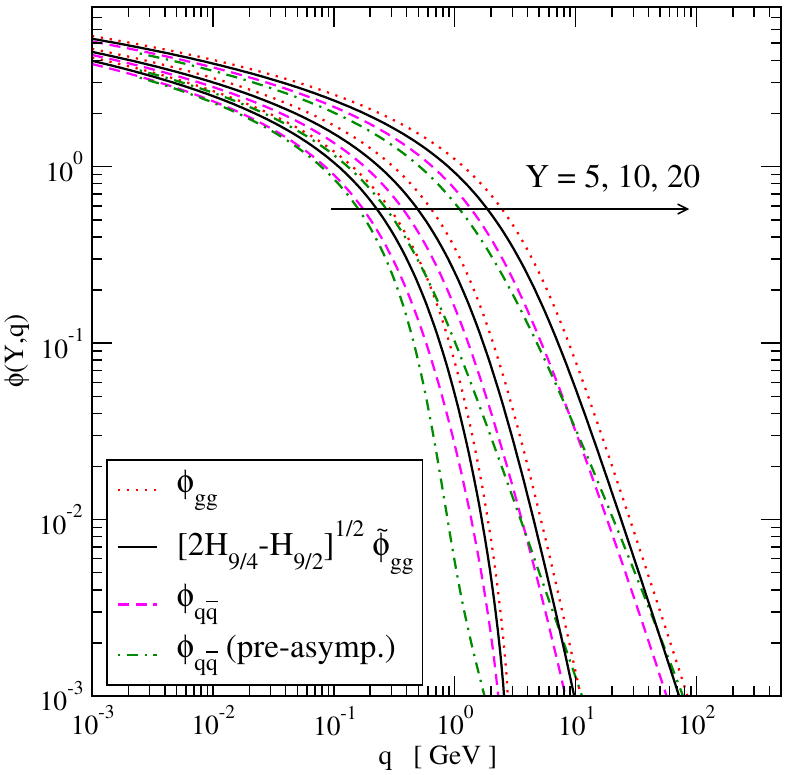}
 \hfill
 \includegraphics[width=.327\linewidth]{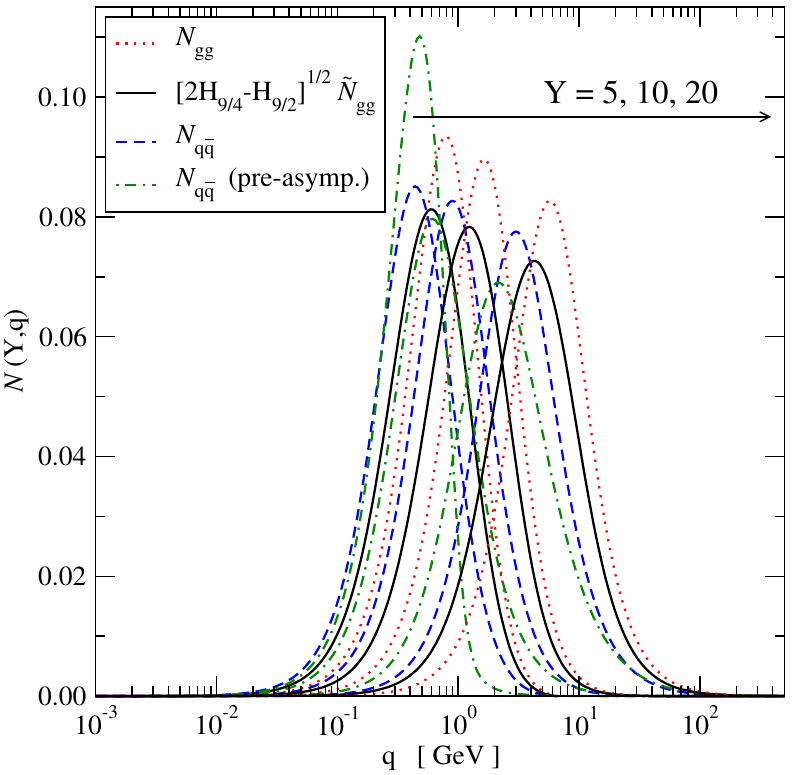}
 \caption{A comparison of the quark and gluon dipole inputs. From left
   to right: $N^{\cal R}_Y(\bm r) \rightarrow \phi^{\cal R}_Y(\bm q)
   \rightarrow {\cal N}^{\cal R}_Y(\bm q)$. Green dot-dashed curves
   are for the pre-asymptotic fit scenario studied in
   Sec.~\ref{sec:pre-as-examples} (starting from $Y=\ln(1/0.01)\approx
   4.61$).}
 \label{fig:compare_quark_gluon}
\end{figure}

The chain of transformations $N^{\cal R}_Y(\bm r) \rightarrow
\phi^{\cal R}_Y(\bm q) \rightarrow {\cal N}^{\cal R}_Y(\bm q)$ is
illustrated in Fig.~\ref{fig:transform_n_phi_caln}, quark and gluon
dipoles are compared in Fig.~\ref{fig:compare_quark_gluon}.

\section{Consistency checks}
\label{sec:con-checks}

With the importance of NLO contributions firmly established, one should at
least attempt to understand in which sense the results obtained
from~(\ref{eq:bks_modified}) or (\ref{eq:GT_modified}) are stable against
modifications. One obvious modification is a simple numerical check for
stability against higher order corrections such as higher order running
coupling contributions. While this has already been discussed in the quite
sophisticated framework of renormalon corrections in\cite{Gardi:2006rp} a
brief numerical check on the quantitative impact of such corrections on a data
fit gives an alternative ballpark impression of their impact. 
\begin{figure}[htb]
  \centering
   \includegraphics[width=.325\linewidth]{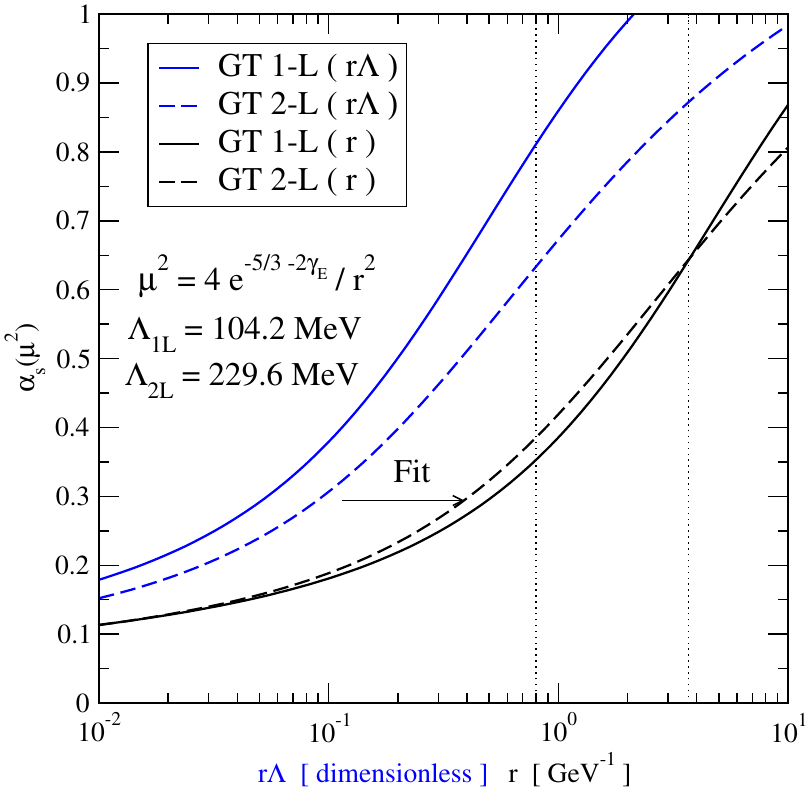}
 \hfill
 \includegraphics[width=.316\linewidth]{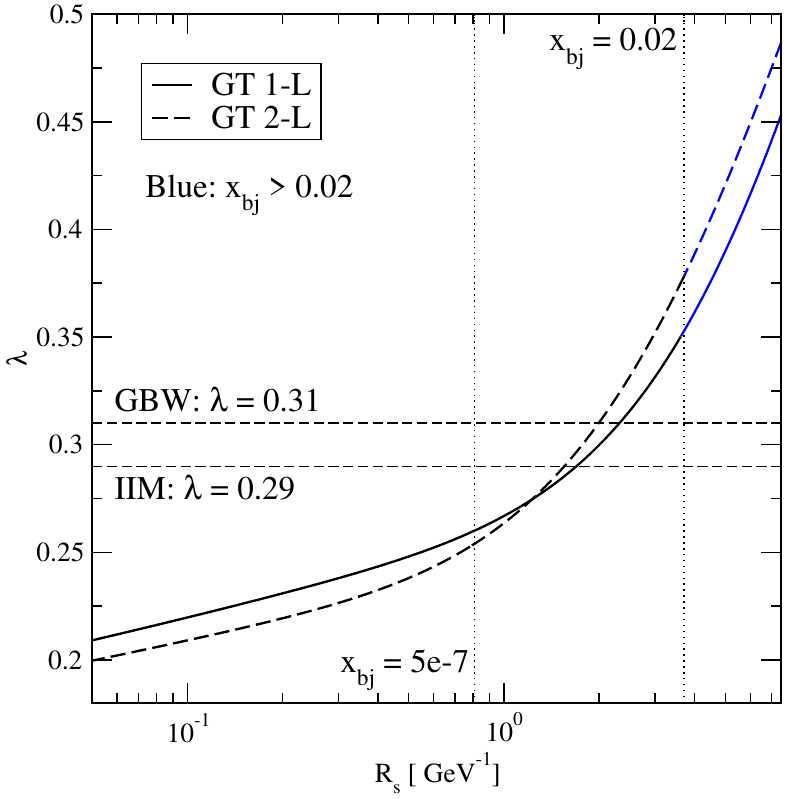}
 \hfill
 \includegraphics[width=.325\linewidth]{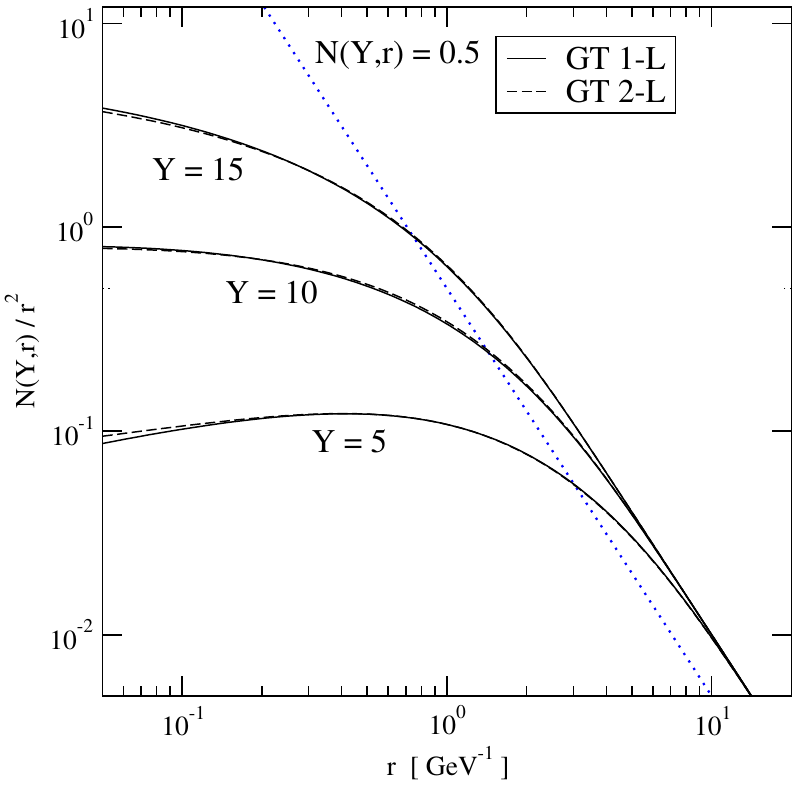}
 \caption[2-loop coupling]{The slowdown due to the running coupling is
   stable against higher order corrections. The vertical lines bracket
   the $R_s$-range of HERA. \textbf{Left:} One loop and two loop running
   couplings coincide after readjusting $\Lambda$. \textbf{Middle:}
   Evolution speeds differ only slightly in the HERA $x$-range. {\bf
     Right:} The dipole correlators (magnified by $r^{-2}$) agree
   after adjusting the fit parameters with only tiny differences
   remaining at short distances.}
  \label{fig:rc_slowdown}
\end{figure}

The fit is repeated by using the two loop running coupling
\begin{align}\label{eq:two_loop2}
  \alpha_s^{2L}(\mu)=\frac{4\pi}{\beta_0}\left[\frac{1}{\nat(\mu^2/\Lambda^2)}-\frac{\beta_1}{\beta_0^2}\frac{\nat\big(\nat(\mu^2
      /\Lambda^2)\big)}{\nat^2(\mu^2/\Lambda^2)}\right]\quad ;\
  \beta_0 = 9,\ \beta_1 = 64\ \ \text{for}\ \ N_f = 3\ ,
\end{align}
which was also regulated by the APT method\footnote{In practice,
  $\alpha_s^{2L}(\mu)$ must be evaluated by using a spectral integral
  representation, see~\cite{Shirkov:1997wi,Solovtsov:1999in}.}. Once
the optimal $\Lambda_{1L,2L}$ and the corresponding rapidity offsets
are known, the one and two loop couplings approximately coincide (see
Fig.\ref{fig:rc_slowdown}, left). Furthermore, the evolution speeds
(Fig.~\ref{fig:rc_slowdown}, middle) and the dipole correlators
(Fig.~\ref{fig:rc_slowdown}, right) coincide around $Y=10$, which is a
direct consequence of the fact that most data points are located
around this rapidity. The fit quality is not essentially affected, see
Table~\ref{tab:two_loop_results}.

\begin{table}[thb!]
  \centering
  \subfloat{\begin{tabular}{lllllll} GT ; $\xbj\leq 0.02$ & $Q^2\leq$ & dof & $\chi^2
        /\text{dof}$ & $\Lambda_{1L,2L}$ $[\mev]$ & $\sigma_0$ $[\gev^{-2}]$ & $\Lambda_{2L}/\Lambda_{1L}$ \\ [2mm]
        One loop &  $45$ $\gev^2$   & 224 & 0.81 & 104.7 & 55.33 & \\
        One loop  &  $1200$ $\gev^2$  & 295 & 0.80 & 104.2  & 55.24 & \\ [2mm]
        Two loop &  $45$ $\gev^2$  & 224 & 0.81 & 213.8 & 55.41 & 2.04  \\
        Two loop &  $1200$ $\gev^2$  & 295 & 0.83 & 229.6 & 54.59 & 2.20 \\
 \end{tabular}}
\caption{Fit results for GT
  with two loop running coupling. The quark masses are $m_{u,d,s}=5$ $\mev$.}
\label{tab:two_loop_results}
\vfill
\subfloat{\begin{tabular}{llllll} BK ; $\xbj\leq 0.02$  & $Q^2\leq$  & dof & $\chi^2/\text{dof}$  & $\Lambda$ $[\mev]$ & $\sigma_0$ $[\gev^{-2}]$ \\ [2mm]
      regulator: &  $45$ $\gev^2$  &  224 & 0.81 & 93.7 & 55.05 \\
      APT  &  $1200$ $\gev^2$  &   295 & 0.80 & 97.7  & 54.50 \\ [2mm]
      regulator:  &  $45$ $\gev^2$   &  224 & 0.96 & 111.7 & 56.57 \\
      alt.  &  $1200$ $\gev^2$  &   295 & 1.01 & 102.9  & 57.24 \\
 \end{tabular}}
\caption{Fit results corresponding the regulators in Eq.~\eqref{eq:alt_reg}. The case of the alternative regulator corresponds to the choice $\# = \exp(1)$ with the limit $\alpha_s^{alt}(0)=4\pi/\beta_0$.}
\label{tab:alternative_regulator}
\end{table}

\begin{SCfigure}[50]
\centering
 \includegraphics[width=.40\linewidth]{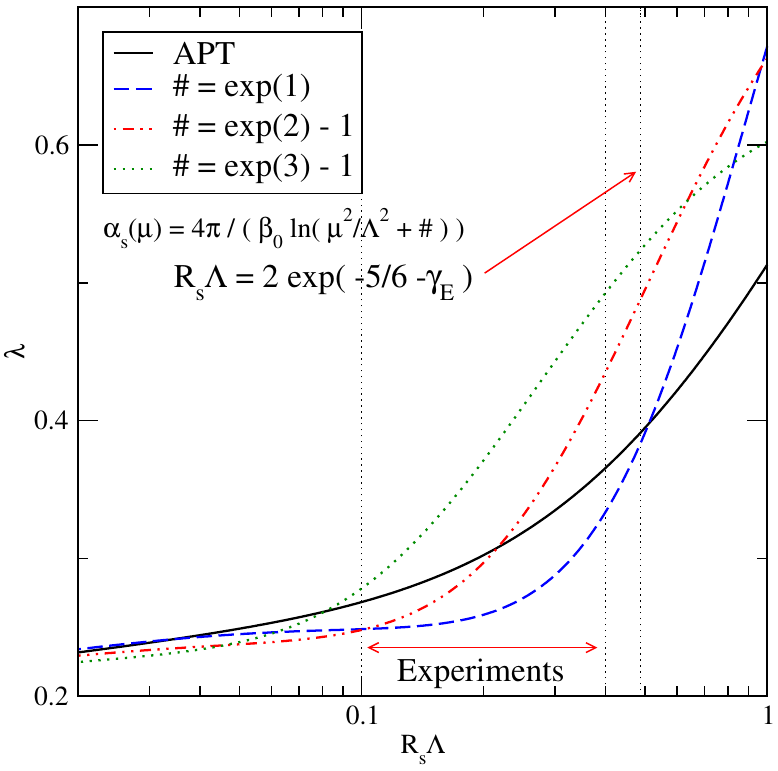}
 \caption{A comparison of the alternative regulators and APT regulator
   presented in Eq.~\eqref{eq:alt_reg}. The effect on the evolution
   speed is not what one would naively expect it to be: on the one
   hand, a weaker regulator with a larger effective running coupling
   $R_{\text{eff}}$ produces a faster evolution speed at the
   pre-asymptotic stage than the APT (as it should) but suddenly drops
   below it inside the fit range $R_s\Lambda \approx 0.1-0.4$. On the
   other hand, a stronger regulator, and hence a smaller
   $R_{\text{eff}}$, extends the pre-asymptotic stage remarkably
   which eventually leads to a faster evolution speed inside the fit
   range. The evolution with the APT regulator matches the best with
   the data. The length scale $R_s\Lambda=2e^{-5/6-\gamma_E}$
   indicates the location of the Landau pole.}
 \label{fig:lambda_regulators}
\end{SCfigure}
The sensitivity to the infrared regulator is investigated. Whereas the
APT regulator (\eqref{eq:alt_reg}, right) offers a smooth crossing
over the Landau pole with the limit $\alpha_s^{APT}(0)=4\pi/\beta_0$,
for instance, adding of a constant inside the logarithm as follows,
\begin{align}\label{eq:alt_reg}
  \alpha_s^{alt}(\mu)=\frac{4\pi}{\beta_0}\frac{1}{\nat(\mu^2/\Lambda^2
    +\#)}\quad\quad;\quad\quad
  \alpha_s^{APT}(\mu)=\frac{4\pi}{\beta_0}\left(\frac{1}{\nat(\mu^2/\Lambda^2)}-\frac{1}{\mu^2/\Lambda^2
      -1}\right)\ ,
\end{align}
results in a steeply increasing (for not too large $\#$) coupling near
the Landau pole. As seen in Fig.~\ref{fig:lambda_regulators}, the
effect induced by this type of modification extends to the actual fit
range. Despite the deviating evolution speeds, all cases shown can
resolve the data with a good $\chi^2/\text{dof}\lesssim 1$. However,
as seen in Table~\ref{tab:alternative_regulator}, the evolution with
the APT regulator yields the best fit to the data.

\begin{SCtable}
 \centering
 \begin{tabular}{llll} BK ; $\xbj\leq 0.02$ &
   $\chi^2/295$ & $\Lambda$ $[\mev]$  & $\sigma_0$ $[\gev^{-2}]$ \\ [2mm]
   reweighted, Eq.~\eqref{eq:mix_it}  & 0.80 & 97.7   & 55.50 \\
   $k=3$, Eq.~\eqref{eq:steps}  &   0.86  & 98.2  & 53.89 \\
   $k=5$, Eq.~\eqref{eq:steps}  &  0.95  & 95.0  & 54.18  \\
 \end{tabular}
 \caption{Fit results corresponding to the different iterative
   solutions presented in App.~\ref{sec:energy-conservation}. In all
   cases, a wider data range $Q^2\leq 1200$ $\gev^2$ is considered.}
\label{tab:iter_comparison}
\end{SCtable}
The last check concerns the solution of the modified BK equation
introduced in App.~\ref{sec:energy-conservation}. The fit with
$m_{u,d,s}=5\ \mev$ is repeated by using more accurate solutions for
the modified BK equation, i.e. the cases $k=3,5$ of the
procedure~\eqref{eq:steps}. The results of these fits are presented in
Table~\ref{tab:iter_comparison}. In both cases $\chi^2/\text{dof}$ is
increased if compared with the reweighted case shown in the first row
but, however, remains below one. The parameters $\Lambda$ and
$\sigma_0$ are altered as well but are still in the same ballpark as
the ones shown in the first row. The experimental data seem to favor
slower evolution speed obtained by the approximate solution.
\begin{figure}[!thb]
\centering
 \includegraphics[width=.33\linewidth]{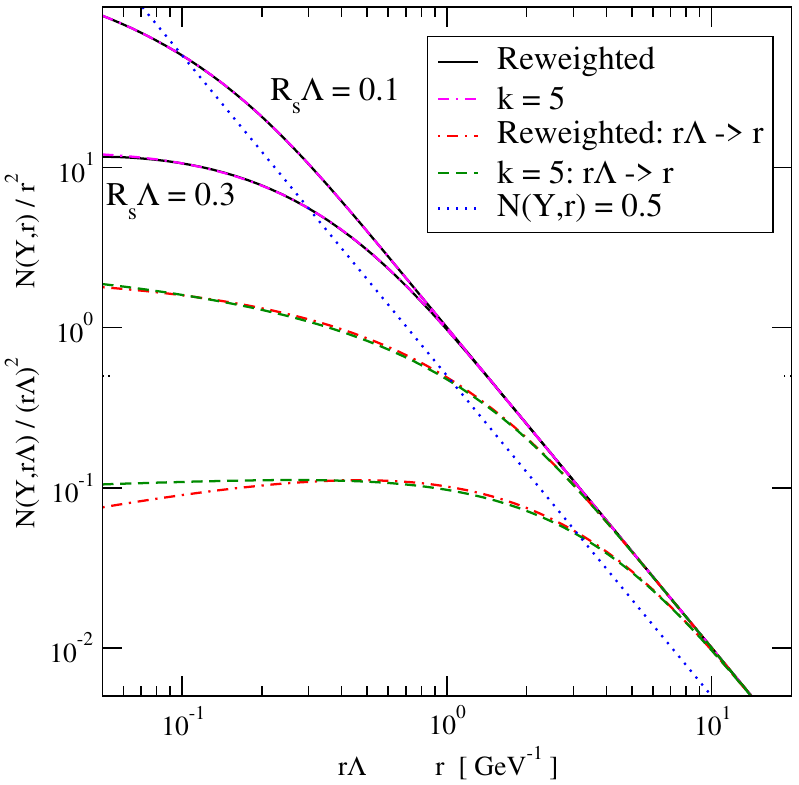}
\hfill
\includegraphics[width=.324\linewidth]{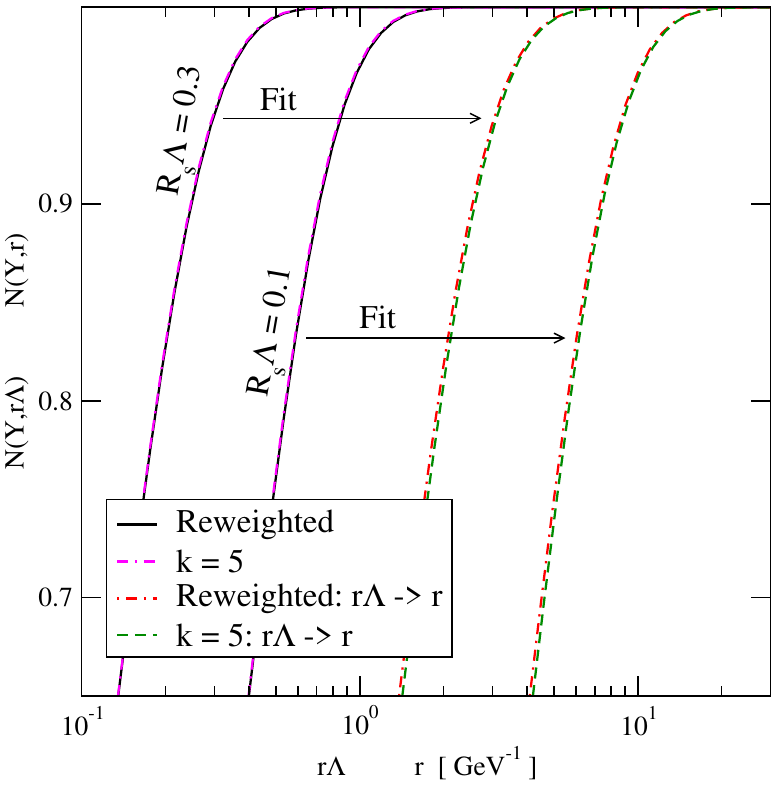}
\hfill
\includegraphics[width=.324\linewidth]{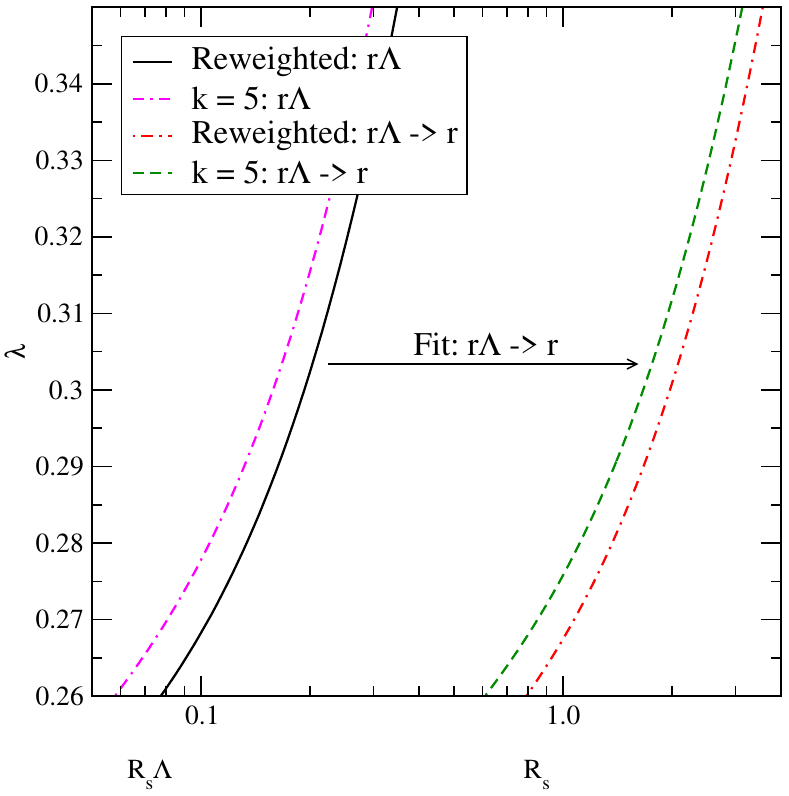}
\caption{\textbf{Left and Middle:} A comparison of $N^{q\Bar q}_Y(r)$ of the
  applied and accurate solution. For a fixed $R_s\Lambda$, there is no
  visible difference between the shapes of the correlators. The same
  feature is approximately preserved after the physical scales are
  determined by the data fits. The correlators are extracted from
  $R_s\Lambda=\{0.1,0.3\}$. The crossing blue dotted line is
  indicating the saturation condition $N^{q\Bar q}_Y(r)=0.5$. \textbf{Right:} The
  deviation between the corresponding evolution speeds remains after
  readjusting $\Lambda$. Thus, the reason for the deviation in
  $\chi^2/\text{dof}$ is a slightly different energy dependencies of
  the saturation scales rather than the actual shapes of the
  correlators.}
 \label{fig:n_compared}
\end{figure}

To summarize, the fits based on the modified BK/GT evolution are
stable against a large variety of modifications. As seen in
Fig.~\ref{fig:lambda_regulators}, the biggest uncertainty to
\emph{evolution} clearly emerges from the infrared regulator.

\bibliography{master,morerefs,all_essential}                   \bibliographystyle{JHEP}  

\end{document}